\documentclass[aps,prb,twocolumn,english,showpacs,superscriptaddress,amssymb,amsfonts]{revtex4}
\usepackage[T1]{fontenc}
\usepackage[latin9]{inputenc}
\usepackage{amsmath}
\usepackage{dsfont}
\usepackage{amstext}

\usepackage{tocvsec2}

\usepackage{amssymb}
\usepackage{amsbsy}
\usepackage{amsthm}
\usepackage{epsfig}
\usepackage{graphicx}
\usepackage{bbm}
\usepackage{hyperref}
\usepackage{color}
\usepackage{multirow}
\usepackage{pdfsync}

\newcommand{\Ref}[1]{Ref.~\onlinecite{#1}}

\newcommand{\bse}{{\boldsymbol{e}}}
\newcommand{\bsg}{{\boldsymbol{g}}}

\newcommand{\ie}{{\emph{i.e.~}}}
\makeatletter

\newcommand{\Rmnum}[1]{\expandafter\@slowromancap\romannumeral #1@}
\makeatother
\newcommand{\imth}{\hspace{1pt}\mathrm{i}\hspace{1pt}}
\newcommand{\alert}[1]{{\color{red}{#1}}}
\newcommand{\eg}{{\emph{e.g.~}}}

\newcommand{\mbz}{{\mathbb{Z}}}
\newcommand{\bea}{\begin{eqnarray}}
\newcommand{\eea}{\end{eqnarray}}
\newcommand{\bpm}{\begin{pmatrix}}
\newcommand{\epm}{\end{pmatrix}}
\newcommand{\bal}{\begin{aligned}}
\newcommand{\eal}{\end{aligned}}

\makeatother

\usepackage{babel}
\begin{document}
\title{ Classification and Properties of Symmetry Enriched Topological Phases:  A Chern-Simons approach with applications to $Z_2$ spin liquids}

\author{Yuan-Ming Lu}%e\footnote{Current address}
\altaffiliation{Current address: \emph{Department of Physics, The Ohio State University, Columbus, Ohio 43210, USA}.}
\affiliation{Department of Physics, University of California, Berkeley, CA 94720, USA}
\affiliation{Materials Science Division, Lawrence Berkeley National Laboratories, Berkeley, CA 94720}

\author{Ashvin Vishwanath}
\affiliation{Department of Physics, University of California, Berkeley, CA 94720, USA}
\affiliation{Materials Science Division, Lawrence Berkeley National Laboratories, Berkeley, CA 94720}

\begin{abstract}
We study 2+1 dimensional phases with topological order, such as fractional quantum Hall states and gapped spin liquids, in the
presence of global symmetries. Phases that share the same topological order can then differ depending on the  action of symmetry, leading to symmetry enriched topological (SET) phases. Here
we present a K-matrix Chern-Simons approach to identify distinct phases with  Abelian topological order, in the presence of unitary or anti-unitary global
 symmetries . A key step is the identification of an smooth edge sewing condition that is used to
check if two putative phases are indeed distinct. We illustrate this method by classifying Z$_2$
topological order (Z$_2$ spin liquids), in the presence of an internal $Z_2$ global
symmetry for which we find six distinct phases. These include two phases with an unconventional action of symmetry that permutes anyons leading to symmetry protected Majorana edge modes. Other routes to realizing protected edge states in SET phases are identified. Symmetry enriched Laughlin states and double semion theories are also discussed. Somewhat surprisingly we observe that  :
(i) gauging the global symmetry of distinct SET phases lead to topological orders with the same total quantum dimension, (ii) a pair of distinct SET phases can yield the same topological order on gauging the symmetry.
\end{abstract}


\maketitle
%\setcounter{tocdepth}{2}
%\begin{widetext}
%\tableofcontents
%\end{widetext}

%\maxsecnumdepth{subsection}

{\small \setcounter{tocdepth}{2} \tableofcontents}

%\twocolumn
\section{Introduction}

It was long believed that phases of matter arose from different patterns of symmetry breaking\cite{Landau1937,Landau1937a}. The discovery of integer\cite{Klitzing1980} and fractional\cite{Tsui1982} quantum Hall (FQH) effects demonstrated however that there exist many different phases of matter which lie outside this paradigm. In particular, the FQH states differ in their internal quasiparticle structure as well as their boundary excitations while preserving all symmetries of the system. This phenomenon is robust against any perturbation and is called\cite{Laughlin1983,Wen2004B} `topological order', which implies ground state degeneracy (GSD) on a closed manifold (a Riemann surface of genus $g$), and emergent anyon excitations\cite{Wilczek1990B} which obey neither bosonic nor fermionic statistics. Another class of topologically ordered phases are gapped quantum spin liquids\cite{Wen2004B}. Recently, several examples of gapped spin liquids have appeared in numerical studies of fairly natural spin-1/2 Heisenberg models, on the kagome\cite{Yan2011} and square lattice (with nearest and next neighbor exchange)\cite{Jiang2012,Wang2011a}. Calculations of topological entanglement entropy\cite{Kitaev2006a,Levin2006} point to $Z_2$ topological order\cite{Jiang2012a,Depenbrock2012}. However, the precise identification of these phases require understanding the interplay between topological order and symmetry in these systems. The symmetries include both on-site global spin rotation and time reversal symmetries, as well as the space group symmetries of the lattice. Kagome lattice antiferromagnets, such as herbertsmithite, may provide experimental realization of this physics, although experimental challenges arising from disorder and residual interactions continue to be actively studied. This motivates the study of distinct topologically ordered phases that may arise in the presence of symmetry\cite{Wen2002,Wang2006,Levin2012a,Essin2013,Mesaros2013,Hung2013a,Hung2013}.

In the presence of symmetry, the structure of topological order is even richer. The microscopic degrees of freedom in the system are either bosons or fermions, and they must form a linear representation of the symmetry group\footnote{It is a subtle issue to define the symmetry group for systems with fermionic microscopic degrees of freedom. The full symmetry group $G_f$ for any fermionic system always contain fermion parity $Z_2^f\equiv\{\bse,(-1)^{N_f}\}$ as an invariant subgroup, since $Z_2^f$ can never be broken by any local interaction\cite{Levin2013}. Usually when we refer to a fermion system with symmetry $G$, it means the full symmetry group $G_f$ is a $Z_2^f$-central extension of group $G$, satisfying $G_f/Z_2^f=G$. The fermionic microscopic d.o.f. form a linear representation of full symmetry group $G_f$, but not of $G$ which is usually referred to as the symmetry group. An example is the fermionic Laughlin state with $G_f=Z_2^f$ (and thus a trivial $G$), as discussed in section \ref{example:fermionic laughlin:Z2f}.} $G_s$. The emergent anyons, however, doesn't need to form a linear representation of $G_s$. Instead they could transform projectively under symmetry operation, \ie each of them can carry a fractional quantum number of symmetry. For example, in Laughlin FQH states\cite{Laughlin1983} at filling fraction $\nu=1/m$, each elementary quasiparticle carries a fraction ($1/m$) of the electron charge. This phenomena is widely known as fractionalization, although a more appropriate name is perhaps \emph{symmetry fractionalization}\cite{Yao2010,Mesaros2013,Essin2013}. The associated symmetry in Laughlin states is the $U(1)$ charge conservation of electrons. While the emergent quasiparticles transform projectively (instead of linearly), the microscopic degrees of freedom always transform linearly under symmetry, simply because each microscopic degrees of freedom can be regarded as a conglomerate of multiple emergent quasiparticles.
%Take the $\nu=1/m$ Laughlin state for example, an electron is equivalent to the bound state of $m$ elementary quasiparticles. Even if each quasiparticle carries a fractional symmetry quantum number, an electron (microscopic degree of freedom) still forms a (linear) representation of the symmetry.

\begin{figure}
 \includegraphics[width=0.45\textwidth]{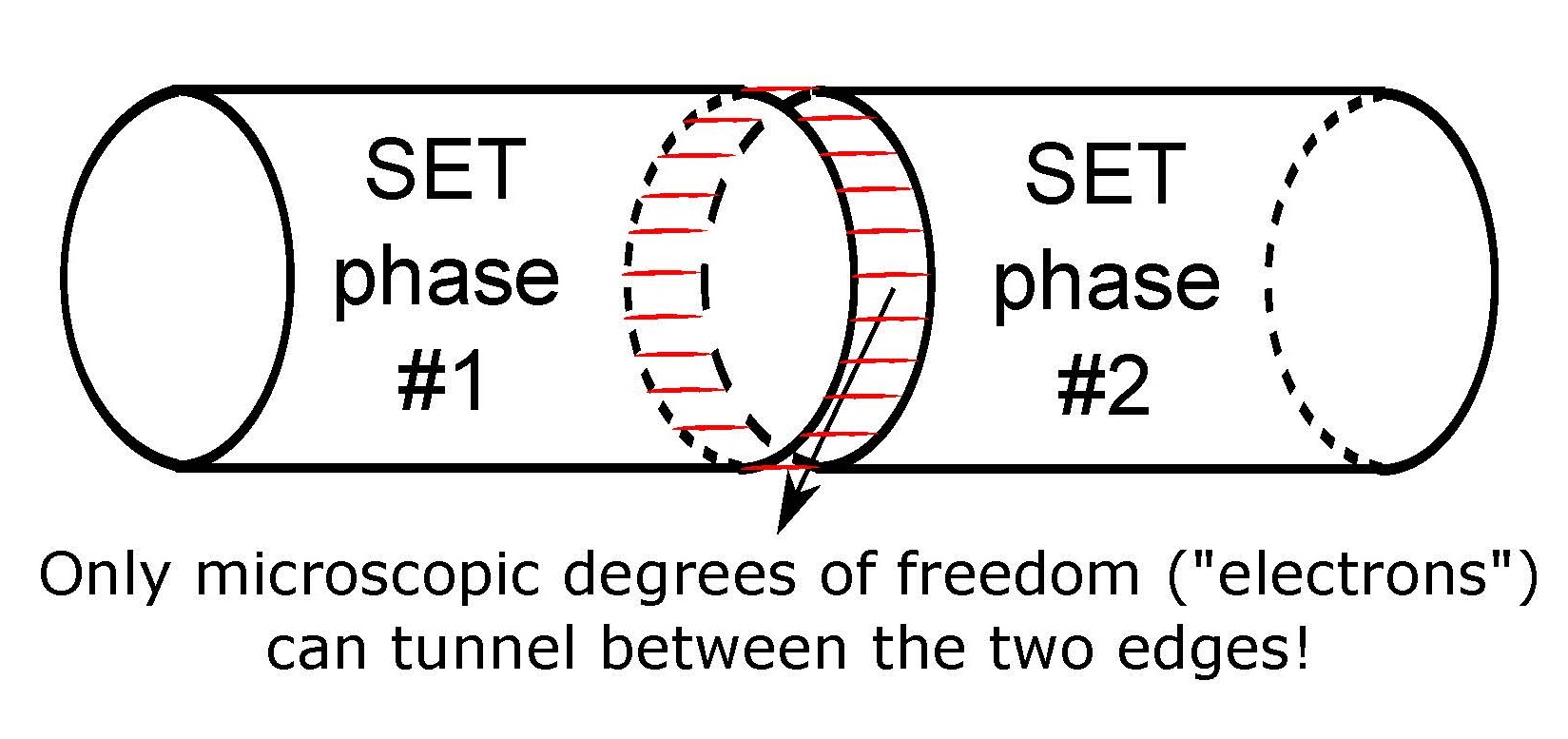}
\caption{(color online) Edge Sewing Criterion to distinguish symmetry enriched topological (SET) phases. Only the microscopic degrees of freedom \ie ``electrons'' (and not gauge charged objects such as anyons/fractionalized quasiparticles) can tunnel between the two edges of a pair of semi-infinite cylinders. If two SET phases can be continuously tuned into one another without a phase transition (while preserving symmetry), there is a ``smooth'' sewing between the two cylinders of SET phases $\#1$ and $\#2$. This implies that all edge excitations are gapped by a few symmetry-allowed terms that tunnel ``electrons'' between the two edges. In the thermodynamic limit these tunneling terms lead to $M$ degenerate ground states, corresponding exactly to the $M$-fold torus degeneracy of the topological order. On the other hand, if the two SET phases are different, there is no such ``smooth'' boundary condition to sew the two edges. A precise version of this statement is formulated in Criterion I in Section \ref{CRITERIA}.}\label{fig:sewing edges}
\end{figure}

Even in the absence of topological order, when symmetry $G_s$ is preserved, different symmetry protected topological (SPT) phases\cite{Chen2013} emerge which are separated from each other through phase transitions. These SPT phases feature symmetry protected boundary states
%. In other words the (energy or entanglement\cite{Li2008}) spectrum of boundary excitations
which will be gapless, unless symmetry $G_s$ is (spontaneously or explicitly) broken on the boundary. Well known examples of SPT phases are topological insulators\cite{Hasan2010,Hasan2011} and superconductors\cite{Qi2011}. In 2+1-D all SPT phases have symmetry protected non-chiral edge modes\cite{Chen2011b,Levin2012,Lu2012a}.
%In a SPT phase, quasiparticle excitations are nothing but microscopic degrees of freedom (bosons or fermions) and they always form a linear (faithful) representation of the symmetry group $G_s$.

The existence of SPT phases further enrich the structure of symmetric topological orders. In other words, topologically ordered phase is not fully determined by how its (anyon) quasiparticles transform (projectively or not) under symmetry: its microscopic degrees of freedom could form a SPT state in parallel with the topological order\cite{Mesaros2013}. The formation of SPT state can \eg bring in new structures to the edge states of the topologically ordered system, and lead to a distinct symmetry enriched topological (SET) order. Therefore, two different SET phases sharing the same topological order can differ by the symmetry transformation on their anyon quasiparticles, or by their distinct boundary excitations. Clearly we have a question here: given two states sharing the same topological order while preserving symmetry $G_s$, can they be continuously connected to each other without a phase transition, if symmetry $G_s$ is preserved?

We address this issue for 2+1-D (Abelian) topological orders. Focusing on on-site (global, instead of spatial) symmetries, we present a universal criterion (Criterion I in section \ref{CRITERIA}) related to the edge states of these 2+1-D SET phases, which works for both unitary and anti-unitary on-site symmetries. The physical picture behind this criterion is demonstrated in FIG. \ref{fig:sewing edges}. Two SET phases $\#1$ and $\#2$ living on the two cylinders are considered the same if they can be smoothly connected together via tunneling of microscopic degrees of freedom between the two edges. Distinct SET phases on the other hand, present an obstruction to such a smooth sewing.

The above criterion allows us to clarify the structure of symmetry enriched topological (SET) orders in 2+1-D. The method we follow is the Chern-Simons approach, which provides a unified description for low-energy bulk and edge properties of a generic Abelian topological order\cite{Read1990,Wen1992,Frohlich1991} in 2+1-D. In particular the bulk-edge correspondence\cite{Witten1988,Wen1995} in Chern-Simons approach enables us to identify all edge excitations with their bulk counterparts, such as the microscopic degrees of freedom (bosons/fermions) and anyons. Therefore the above criterion of smooth sewing boundary conditions for two different SET phases can be made precise within the Chern-Simons approach (see section \ref{CRITERIA}).

More concretely, a 2+1-D Abelian topological phase is fully characterized by a symmetric integer matrix ${\bf K}$ in the Chern-Simons approach. When symmetry $G_s$ is preserved in the system, the anyons could carry a fractional symmetry quantum number (or transform projectively under the symmetry\cite{Wen2002}), while the microscopic degrees of freedom (bosons/fermions) must form linear representations of the symmetry group $G_s=\{\bsg\}$. The relation between microscopic degrees of freedom and fractionalized anyon excitations is especially clear in the Chern-Simons approach.
%Knowing how to implement (on-site) symmetries for an Abelian topological order in the Chern-Simons approach\cite{Lu2012a,Levin2012a}, we can fully characterize a 2+1-D SET phase by a set of data $[{\bf K},\{\eta^\bsg,{\bf W}^\bsg,\delta\vec\phi^\bsg|\bsg\in G_s\}]$, where $\{{\bf W}^\bsg,\delta\vec\phi^\bsg|\bsg\in G_s\}$ labels how (anyon) quasiparticles transform under symmetry $G_s=\{\bsg\}$.

Based on the mentioned Criterion I to differentiate distinct SET phases, we can classify all different SET phases with the same topological order $\{K\}$ and symmetry $G_s$. In this work, we studied various examples: $Z_2$ spin liquids\cite{Kitaev2003}, double semion theory\cite{Freedman2004,Levin2005} and bosonic/fermionic Laughlin states\cite{Laughlin1983} at filling fraction $\nu=1/m$. We consider both anti-unitary time reversal symmetry $G_s=Z_2^T$, and unitary $G_s=Z_2 \,~{\rm or}\,~ Z_2\times Z_2$. These are the analogs of the spin rotation symmetry of Heisenberg magnets. Classification of these SET phases with symmetry $G_s$ are summarized in TABLE \ref{tab:Z2SL:time reversal}-\ref{tab:laughlin1/2k:z2:k=even}. In the case of anti-unitary time-reversal symmetry ($G_s=Z_2^T$) symmetry, our classification based on Chern-Simons approach is unable to capture one extra SET phase, in which distinct anyons are permuted by time reversal operation.

In particular, we highlight the classification of the simplest class of SETs, Z$_2$ (toric code) topological order with a $G_s=Z_2$ onsite symmetry. We find a total of six phases with our method. Of these, only two nontrivial phases are understood in terms of distinct fractional charges. Other phases include combination with SPTs, or unconventional symmetry action that permutes anyons and leads to protected Majorana edge states.

\underline{Unconventional SET Phases:} We divide SET phases into two types - conventional and unconventional. In conventional SET phases, all (anyon) quasiparticles merely obtain a $U(1)$ phase under any symmetry operation.
%More specifically in data (\ref{data of a SET phase}), the matrices ${\bf W}^\bsg=1_{N\times N}$ for any $\bsg\in G_s$ in a SET phase of type-I. We call such a SET phase of type-I a ``conventional'' SET phase.
In contrast, in the more exotic `unconventional' SET phases, certain symmetry operations {\em exchange} two inequivalent anyons,  instead of just acquiring $U(1)$ phases. For example, under on-site unitary $Z_2$ symmetry operation, the two anyons \ie the electric charge $e$ and magnetic vortex $m$ of a $Z_2$ spin liquids are exchanged (see TABLE \ref{tab:Z2SL:z2:unconventional}). Previously, such a transformation law was considered in the Wen `plaquette' model\cite{Wen2003,Kou2008,Cho2012} for translation symmetry, in contrast to the internal $G_s=Z_2$ symmetry considered here.
%Formally in data (\ref{data of a SET phase}), at least one matrice ${\bf W}^\bsg\neq1_{N\times N}$ for certain $\bsg\in G_s$ in a SET phase of type-II. We call these type-II SET phases ``unconventional'' SET phases.
These ``unconventional'' SET phases have some striking properties. First, the edge features gapless Majorana edge modes that are protected by symmetry. Next, if $Z_2$ symmetry is broken at the edge, then a Majorana fermion is trapped at the edge domain wall. Finally, as illustrated in FIG. \ref{fig:boundary fermion}, when a pair of electric charge ($e$) is created at opposite sides  of a sphere, we can divide the system into two subsystems $A$ and $B$, so that there is one electric charge $e$ localized in each subsystem. Now if we perform the $Z_2$ symmetry operation only in subsystem $A$ (flip all the spins), the electric charge $e$ therein will become a magnetic vortex $m$. Since an electric charge $e$ and a magnetic vortex $m$ differs by a fermion $f$ ($e\times f=m$ or $m\times f=e$) in the $Z_2$ spin liquid, this means a fermion mode $f$ must simultaneously appear at the boundary separating subsystem $A$ and $B$, as the Ising symmetry is acted on $A$. This is discussed in Section \ref{UNCONVENTIONAL Z2 SPIN LIQUIDS} and Section \ref{Sec:MeasureMajoranas}.

\underline{Symmetry Protected Edge States:} In general, the non-chiral topological orders, like $Z_2$ topological order and double semion models, do not have gapless excitations at the edge. However, these may appear with additional symmetry. Indeed, the `unconventional' $Z_2$ SET phases have Majorana edge states. Since they are protected by an on-site $Z_2$ symmetry, they are stable even in the presence of disorder that breaks translation symmetry along the edge. Two further mechanisms for gapless edge modes in `conventional' SET phases may be identified. The first is the trivial observation that adding an SPT phase could lead to a corresponding protected edge state. The second mechanism operates when both the electric and magnetic particle of the $Z_2$ gauge theory transforms projectively under symmetry. Then, one cannot condense neither of them at the edge - implying a protected edge.  Details and a sufficient condition (\ref{thm 1:sufficient edge}) for protected edge states will appear in Sec.\ref{EDGE STABILITY}.

\underline{Gauging Symmetry:} A powerful tool in studying the effect of an onsite unitary symmetry $G_s$ is the consequence of gauging it \cite{Levin2012,Hung2013}.  This means the global $G_s$ symmetry is promoted to a local ``gauge symmetry'', which leads to new topological orders. Distinct topological orders can help distinguish different actions of the symmetry in the ungauged theory.
%For example: SET phases reduces to symmetry protected topological (SPT) phases in the limit of no topological order, which are described by nonlinear sigma models with topological terms.
By this procedure in 2+1-D, nonlinear sigma models with topological terms, which describe SPT phases\cite{Chen2013} can be mapped to gauge theories with a topological term \cite{Levin2012,Hung2013}, discussed by Dijkgraaf and Witten \cite{Dijkgraaf1990}. In this work we systematically study the consequences of gauging unitary on-site symmetry in Abelian SET phases. For many ``conventional'' SET phases with Abelian symmetries, the new topological order obtained by gauging symmetry is Abelian, and Chern-Simons theory is a natural framework to derive it. There are some cases of ``conventional'' SET phases that will lead to non-Abelian topological orders by gauging symmetry, though, an example being $Z_2$ spin liquids with $Z_2\times Z_2$ spin rotational symmetry (see section \ref{example:z2sl:Z2xZ2}).

In the examples studied in this work ($Z_2$ spin liquids, double semion theories and $\nu=1/2k$ bosonic Laughlin states with onsite $G_s=Z_2$ symmetry), different SET phases seem to lead to distinct topological orders (with different anyon contents) by gauging the unitary symmetry $G_s=Z_2$. However this doesn't always happen for a general symmetry group $G_s$. Remarkably for all different $G_s$-symmetry-enriched topological phases with the same topological orders (same GSD and anyon statistics), once we gauge the unitary symmetry $G_s$, they lead to distinct intrinsic topological orders with the {\em same} total quantum dimension\cite{Nayak2008} $\mathcal{D}=\sqrt{\sum_\alpha d_\alpha^2}$. Therefore these distinct topological orders obtained by gauging unitary symmetry $G_s$ also shares the same topological entanglement entropy\cite{Kitaev2006a,Levin2006} $\gamma=\log\mathcal{D}$. Although this is an observation from the examples studied in this paper, we conjecture that it generically holds for all SET phases with a finite unitary symmetry group $G_s$.

Somewhat surprisingly, this furnishes examples where two gauge theories with distinct Dijkgraaf-Witten\cite{Dijkgraaf1990} topological terms correspond to the same topological order. Here the topological terms arising for the gauge group  $Z_2\times Z_2$  are obtained by gauging SPT phases and correspond to elements of $H^3(Z_2\times Z_2,\,U(1))$.  Theories for distinct elements are shown to be equivalent on relabeling quasiparticles (an $GL(4,Z)$ transformation). Therefore the distinction between these theories requires additional information such as specification of electric vs. magnetic vortices (Appendix \ref{COMPARISON}).

 For ``unconventional'' SET phases, however, gauging the symmetry always leads to non-Abelian topological orders. A general argument for this conclusion is provided in section \ref{Sec:gauge unconventional SET}. For example the unconventional Ising-symmetry-enriched $Z_2$ spin liquids, after gauging the Ising ($G_s=Z_2$) symmetry, lead to non-Abelian topological orders with 9-fold GSD on a torus. Interestingly, they can be naturally embedded within Kitaev's 16-fold way classification\cite{Kitaev2006} of 2+1-D $Z_2$ gauge theories (see TABLE \ref{tab:Z2SL:z2:unconventional} and \ref{tab:Z2SL:z2:unconventional:vertex algebra}). Notably as mentioned earlier, these non-Abelian topological orders also have total quantum dimension $\mathcal{D}=16$, the same as that of Abelian $Z_2\times Z_2$ (or $Z_4$) gauge theories which are obtained by gauging $Z_2$ symmetry in ``conventional''p SET phases. In this case a vertex algebra approach\cite{Lu2010} can be introduced to extract all information of the non-Abelian topological order (Appendix \ref{app:vertex algebra}). In particular after gauging the on-site Ising ($Z_2$) symmetry, new quasiparticles $\{q_\bsg\}$ (coined $Z_2$ symmetry fluxes) emerge as deconfined excitations. It is a non-Abelian anyon in the unconventional SET case, which corresponds to the edge domain wall bound state in FIG. \ref{fig:edge domain wall}.

\underline{Spin-1/2 From {\bf K}-Matrix CS Theory:} We demonstrate how an emergent `spin 1/2' excitation can be realized in the Chern Simons formalism, by studying $Z_2$ gauge theories with $Z_2\times Z_2$ symmetry.  The latter has a projective representation that can protect a two-fold degenerate state, analogous to spin 1/2. This is accomplished by expanding the 2$\times$2 K-matrix of a $Z_2$ gauge theory to a 4$\times$4 matrix by adding a trivial insulator layer (a 2$\times$2 `trivial' block). Symmetry transformations implemented in this expanded space have the desired properties (in Section \ref{example:z2sl:Z2xZ2}).

\underline{Connection to Other Work:} A symmetry based approach was used to classify $Z_2$ spin liquids in Ref.\onlinecite{Essin2013}. An advantage of that approach is that it treated both internal and space group symmetries. However, topological distinctions and the appearance of edge states are not captured. Also,  the `unconventional' symmetry realizations were not discussed. Finally, as mentioned in Ref.\onlinecite{Essin2013} the symmetry based approach produces forbidden SETs, that cannot be realized in 2+1-D, but only as the surface state of a 3+1-D SPT phase\cite{Vishwanath2013}. Our Chern-Simons approach does not produce such states. A different classification scheme in Ref.\onlinecite{Mesaros2013}, produces a subset of our `conventional' phases although explicit lattice realizations are given for them. Finally, Ref.\onlinecite{Hung2013} gave a classification based on gauging the symmetry, which misses distinctions between phases as discussed previously. In this work we show that plausibly different phases given in \Ref{Mesaros2013} and \Ref{Hung2013} (belonging to distinct Dijkgraaf-Witten topological terms) actually correspond to the same SET phase. Our approach is perhaps closest to that adopted pin Ref.\onlinecite{Levin2012a}, which however was restricted to time reversal symmetric topological states.  Thus the results in this paper go beyond previous classifications of Z$_2$-symmetry-enriched  $Z_2$ gauge theories (including $Z_2$ spin liquid and double semion theory), and a detailed comparison is given in Appendix \ref{COMPARISON}.\\

This paper is organized as follows. In Section \ref{SET-UP} we introduce the Chern-Simons K-matrix approach to (Abelian) symmetry enriched topological (SET) phases in 2+1-D. Rules for implementing on-site symmetry in a topologically ordered phase are discussed in Section \ref{IMPLEMENT SYMMETRY}, with criteria to differentiate distinct SET phases in Section \ref{CRITERIA}. Next, in Section \ref{EXAMPLES}, we demonstrate our approach by classifying SET phases in a few examples. They include:(i)  $Z_2$ spin liquid with (anti-unitary) time reversal symmetry ($G_s=Z_2^T$) symmetry (section \ref{example:z2sl:Z2T}, TABLE \ref{tab:Z2SL:time reversal}),(ii) $Z_2$ spin liquid with unitary Ising ($G_s=Z_2$) symmetry (section \ref{example:z2sl:Z2}, TABLE \ref{tab:Z2SL:z2:conventional} and \ref{tab:Z2SL:z2:unconventional}), (iii) double semion theory with unitary Ising ($G_s=Z_2$) symmetry (Appendix \ref{app:double semion}, TABLE \ref{tab:double semion:z2:conventional}) and (iv) even-denominator bosonic Laughlin state with unitary Ising ($G_s=Z_2$) symmetry (section \ref{example:bosonic laughlin:Z2}, TABLE \ref{tab:laughlin1/2k:z2} and \ref{tab:laughlin1/2k:z2:k=even}).  Appendix \ref{app:gauging} and \ref{app:vertex algebra} provide detailed instructions on how to gauge unitary symmetries in 2+1-D SET phase.

\section{Chern-Simons approach to symmetry enriched Abelian topological orders in 2+1-D}\label{SET-UP}

\subsection{Chern-Simons theory description of 2+1-D Abelian topological orders}

In two spatial dimensions, a generic gapped phase of matter is believed to be described by a low-energy effective Chern-Simons theory in the long-wavelength limit\cite{Dijkgraaf1990,Read1990,Wen1992,Frohlich1991,Fradkin1998}. Both the bulk anyon excitations and the gapless edge states are captured by the effective theory\cite{Wen1995}. Examples include integer and fractional quantum Hall states\cite{Wen2004B}, gapped quantum spin liquids\cite{Kalmeyer1987,Hansson2004,Kou2008} and topological insulators/superconductors. When we restrict ourselves to the case of gapped Abelian phases where all the elementary excitations in the bulk obey Abelian statistics\cite{Wilczek1990B}, a complete description is given in terms of Abelian $U(1)^N$ Chern-Simons theory\cite{Read1990,Wen1992,Frohlich1991,Wen1995}. To be specific, the low-energy effective Lagrangian of $U(1)^N$ Chern-Simons theory has the following generic form
\bea\label{U(1)^N Chern-Simons}
\mathcal{L}_{CS}=\frac{\epsilon_{\mu\nu\lambda}}{4\pi}\sum_{I,J=1}^Na^I_\mu{\bf K}_{I,j}\partial_\nu a^J_\lambda-\sum_{I=1}^N a^I_\mu j_I^\mu+\cdots
\eea
where $\mu,\nu,\lambda=0,1,2$ in 2+1-D and summation over repeated indices are always assumed. Here $\cdots$ represents higher-order terms, such as Maxwell terms $\sim(\epsilon_{\mu\nu}\partial_\mu a^I_\nu)^2$. ${\bf K}$ is a symmetric $N\times N$ matrix with integer entries. Notice that the $U(1)$ gauge fields $a^I_\mu$ are all \emph{compact} in the sense that they are coupled to \emph{quantized} gauge charges with currents $j_I^\mu$. In the first quantized language the quantized quasiparticle currents $j_I^\mu$ are written as
\bea
\notag \forall~I=1,\cdots,N:~~~j^0_I({\bf r})=\sum_nl_I^{(n)}\delta({\bf r}-{\bf r}^{(n)}),\\
\notag j^\alpha_I({\bf r})=\sum_nl_I^{(n)}{{\dot{r}}^{(n)}_\alpha}\delta({\bf r}-{\bf r}^{(n)}),~~~\alpha=1,2.
\eea
where ${\bf r}^{(n)}=(r_1^{(n)},r_2^{(n)})$ denotes the position of the $n$-th quasiparticle, and gauge charges $l_I^{(n)}$ are all quantized as integers. We can simply label the $n$-th quasiparticle by its gauge charge vector ${\bf l}^{(n)}=(l^{(n)}_1,\cdots,l^{(n)}_N)^T$.  The self(exchange) statistics of a quasiparticle ${\bf l}$ is given by its statistical angle
\bea\label{statistics:self}
\theta_{\bf l}=\pi{\bf l}^T{\bf K}^{-1}{\bf l},~~~{\bf l}\in\mbz^N.
\eea
while the mutual(braiding) statistics of a quasiparticle ${\bf l}$ and ${\bf l}^\prime$ is characterized by
\bea\label{statistics:mutual}
\tilde\theta_{{\bf l},{\bf l}^\prime}=2\pi{\bf l}^T{\bf K}^{-1}{\bf l}^\prime,~~~{\bf l},{\bf l}^\prime\in\mbz^N.
\eea
The above statistics comes from the nonlocal Hopf Lagrangian\cite{Wilczek1983} of currents $j^\mu_I$, obtained by integrating out the gauge fields $a_\mu^I$ in (\ref{U(1)^N Chern-Simons}). A simple observation from (\ref{statistics:mutual}) is that for a quasiparticle excitation with gauge charge
\bea\label{vector:electron}
\tilde{\bf l}={\bf K}{\bf l},~~~{\bf l}\in\mbz^N.
\eea
its mutual statistical with any other quasiparticle ${\bf l}^\prime$ is a multiple of $2\pi$. In other words, the quasiparticles $\tilde{\bf l}={\bf K}{\bf l}$ are local\cite{Frohlich1990} with respect to any other quasiparticles ${\bf l}^\prime$. Therefore they are interpreted as the ``gauge-invariant'' microscopic degrees of freedom in the physical system: such as electrons\cite{Wen1992} in a fractional quantum Hall state, and spin-1 magnons in a spin-$1/2$ $Z_2$ spin liquid\cite{Wen2002}. Another direct observation is that when all diagonal elements of matrix ${\bf K}$ are \emph{even} integers, the microscopic degrees of freedom have bosonic statistics $\theta=0\mod2\pi$, and (\ref{U(1)^N Chern-Simons}) describes a bosonic system. When at least one diagonal elements of ${\bf K}$ are \emph{odd} integers, there are fermionic microscopic degrees of freedom in the system.

The ground state degeneracy (GSD), as an important character for the topologically ordered phase described by effective theory (\ref{U(1)^N Chern-Simons}) is\cite{Wen1995}
\bea\notag
\text{GSD}=|\det{\bf K}|^g.
\eea
on a Riemann surface of genus $g$. On the torus with $g=1$, the corresponding GSD$=|\det{\bf K}|$ also equals the numbers of different anyon types (or the number of distinct superselection sectors\cite{Frohlich1990,Kitaev2006}) in the 2+1-D topological ordered system. A simple picture is the following: two anyons differing by a (local) microscopic excitations are the same (or more precisely, belong to the same superselection sector) in the sense that they share the same braiding properties:
\bea\notag
{\bf l}^\prime\simeq{\bf l}^{\prime\prime}\Longleftrightarrow{\bf l}^\prime-{\bf l}^{\prime\prime}={\bf K}{\bf l},~~~~~~{\bf l},{\bf l}^\prime,{\bf l}^{\prime\prime}\in\mbz^N.
\eea
Therefore different quasiparticle types correspond to inequivalent integer vectors ${\bf l}\in\mbz^N$ in a $N$-dimensional lattice, where the Bravais lattice primitive vectors are nothing but the $N$ column vectors of matrix ${\bf K}$. As a result $|\det{\bf K}|$, the volume of the primitive cell in ${\bf l}$-space, counts the number of different quasiparticle types (or superselection sectors) in a topologically ordered system described by effective theory (\ref{U(1)^N Chern-Simons}).

\subsection{Edge excitations of an Abelian topological order}

There is a bulk-edge correspondence\cite{Witten1989,Wen1995} for effective theory (\ref{U(1)^N Chern-Simons}). When put on an open manifold $\mathcal{M}$ with a boundary $\partial\mathcal{M}$, the gauge invariance of effective Lagrangian (\ref{U(1)^N Chern-Simons}) implies the existence of edge states on the boundary $\partial\mathcal{M}$. The $N$ chiral boson fields $\{\phi_I\simeq\phi_I+2\pi|1\leq I\leq N\}$ capture the edge excitations. To be specific, assuming the manifold $\mathcal{M}$ covers the lower half-plane $r_2<0$, then edge excitations localized on the boundary $\partial\mathcal{M}=\{(r_1,r_2)|r_2=0\}$ has the following effective Lagrangian
\bea\label{edge:right}
\mathcal{L}_{rE}=\frac1{4\pi}\sum_{I,J}\big({\bf K}_{I,J}\partial_0\phi_I\partial_1\phi_J-{\bf V}_{I,J}\partial_1\phi_I\partial_1\phi_J\big).
\eea
where $rE$ stands for the \emph{right edge}. On the other hand, if the manifold $\mathcal{M}$ instead covers the upper half-plane $r_2>0$, the corresponding edge theory becomes
\bea\label{edge:left}
\mathcal{L}_{lE}=-\frac1{4\pi}\sum_{I,J}\big({\bf K}_{I,J}\partial_0\phi_I\partial_1\phi_J+{\bf V}_{I,J}\partial_1\phi_I\partial_1\phi_J\big).
\eea
where $lE$ means \emph{left edge} here. ${\bf V}$ is a positive-definite real symmetric $N\times N$ matrix, determined by microscopic details of the system. The edge effective theories (\ref{edge:right})-(\ref{edge:left}) imply the following Kac-Moody algebra\cite{Wen1995} of chiral boson fields:
\bea\label{k-m algebra}
[\phi_I(x),\partial_y\phi_J(y)]=\pm2\pi{\bf K}^{-1}_{I,J}\imth\delta(x-y).
\eea
where $+$($-$) sign corresponds to the right(left) edge. The signature $(n_+,n_-)$ of matrix ${\bf K}$ now has a clear physical meaning from (\ref{edge:right})-(\ref{edge:left}): each positive(negative) eigenvalue of ${\bf K}$ corresponds to a right-mover (left-mover) on the right edge (\ref{edge:right}) and a left-mover (right-mover) on the left edge (\ref{edge:left}).

Similar to the quasiparticle excitations in the bulk labeled by their gauge charge ${\bf l}$, associated quasiparticles on the edge $V_{\bf l}=\exp(\imth\sum_I l_I\phi_I)$ are also labeled by an integer vector ${\bf l}=(l_1,\cdots,l_N)^T$. This identification between bulk quasiparticle ${\bf l}$ and edge excitations $\hat{V}_{\bf l}$ indicates that each (local) microscopic degree of freedom (\ref{vector:electron}) in the bulk also has a correspondent local excitation on the edge: $\hat{V}_{\tilde{\bf l}}=\hat{V}_{{\bf K}{\bf l}}$. For a $N\times N$ matrix ${\bf K}$, all these local excitations on the edge are composed of the following $N$ independent local excitations (microscopic degrees of freedom on the edge):
\bea\notag
&e^{\imth\sum_J{\bf K}_{I,J}\phi_J(x,t)},~~~1\leq I\leq N.\notag
\eea
In the context of fractional quantum Hall states, these local operators on the edge are called\cite{Wen1990,Wen1995} ``electron operators''.

Here let's go over the simplest case with no symmetry, when symmetry group $G_s=\{\bse\}$ and $\bse$ denotes the identity element of a group. In this case all the (local) microscopic boson degrees of freedom can condense in the bulk, and accordingly on the edge the following Higgs terms can be added to Lagrangian (\ref{edge:right})-(\ref{edge:left})
\bea
&\notag\mathcal{L}_{Higgs}=\sum_IC_I\big(e^{\imth\chi_I}\hat{M_I}+~h.c.\big)\\
&\label{higgs term}=\sum_{I=1}^N C_I\cos\big(p_I\sum_J{\bf K}_{I,J}\phi_J(x,t)+\chi_I\big).
\eea
where $C_I$ and $\chi_I$ are all real parameters. Notice that constant factor
\bea
&p_I\equiv \big(3-(-1)^{{\bf K}_{I,I}}\big)/2,~~~\forall~1\leq I\leq N.\notag
\eea
guarantees the self statistics (\ref{statistics:self}) of local quasiparticle
\bea\label{local boson}
&\hat{M}_I(x,t)\equiv e^{\imth p_I\sum_J{\bf K}_{I,J}\phi_J(x,t)},~~~1\leq I\leq N.
\eea
is bosonic, since if $\hat{M}_I$ is fermionic the Higgs term (\ref{higgs term}) will violate locality. The Abelian topological order (featured by GSD on genus-$g$ Riemann surfaces and anyon statistics) will not be affected by these Higgs terms\cite{Wen1990b,Lu2012,Lu2012a}, since all anyon excitations are local with respect to the microscopic boson degrees of freedom. As a result the condensation of local bosonic degrees of freedom $\{\hat{M}_I\}$ will not trigger a phase transition, when there is no symmetry in the Abelian topological order. Hence in a general ground these Higgs terms (\ref{higgs term}) should be include in the low-energy effective theory (\ref{U(1)^N Chern-Simons}),(\ref{edge:right})-(\ref{edge:left}) of an Abelian topological order, in the absence of any symmetry. Although naively effective theory (\ref{U(1)^N Chern-Simons}) and (\ref{edge:right})-(\ref{edge:left}) seems to have $N$ conserved $U(1)$ currents, this $U(1)^N$ symmetry will disappear when these Higgs terms are considered.

Since Higgs terms (\ref{higgs term}) are generally present in the edge effective theory (when there is no symmetry), they will introduce backscattering processes on the edge. A natural question is the stability of gapless edge excitations\cite{Haldane1995}. When $n_+\neq n_-$ for the p signature $(n_+,n_-)$ of matrix ${\bf K}$, there is a net chirality for edge states (\ref{edge:right})-(\ref{edge:left}) and they cannot be fully gapped out by the Higgs terms (\ref{higgs term}). A physical consequence is a nonzero thermal Hall conductance in the system\cite{Kane1996}. If $n_+=n_-$ on the other hand, there is no net chirality on the edge. But this doesn't mean the edge states can be gapped out by Higgs term (\ref{higgs term}): the simplest counterexample is ${\bf K}=\bpm3&0\\0&-5\epm$, whose edge cannot be gapped out even in the absence of any symmetry\cite{Levin2013}. When the system preserves symmetry $G_s$, the structure of edge states is richer. Typically some Higgs terms in (\ref{higgs term}) will be forbidden by symmetry, and there will be symmetry-protected edge excitations\cite{Levin2009,Cho2012,Lu2012a} in the Abelian topological order. In other words certain branches of edge excitations will either remain gapless when symmetry $G_s$ is preserved, or become gapped out when symmetry $G_s$ is spontaneously broken on the edge. For a general discussion on the stability of edge modes in an Abelian topological order we refer the readers to section III of \Ref{Lu2012a}. For the SET phases studied in this work, their edge stabilities are briefly discussed in section \ref{EDGE STABILITY}.

\subsection{Different Chern-Simons theories can describe the same topological order}

For symmetric unimodular ${\bf K}$ matrix with $\det{\bf K}=\pm1$, the ground state of system (\ref{U(1)^N Chern-Simons}) is unique on any closed manifold. Consistent with the nondegenerate ground state on torus, any quasiparticle ${\bf l}$ is either bosonic or fermionic with trivial mutual statistics with each other. Hence there is no topological order in the system\cite{Lu2012a,Levin2012a} when $\det{\bf K}=\pm1$. However the corresponding gapped phase can still have gapless chiral edge modes on its boundary, which are stable against any perturbations. Well-known examples are the integer quantum Hall effects where ${\bf K}$ is an $N\times N$ identity matrix. On the other hand, if ${\bf K}$ matrix satisfies the following ``trivial" condition:
\bea\label{trivial phase}
&\notag\text{Trivial phase}:~\text{for}~N=\text{dim}{\bf K}=~\text{even},\\
&\det{\bf K}=(-1)^{N/2},~(n_+,n_-)=(N/2,N/2).
\eea
the edge excitations will be non-chiral (the same number of right- and left-movers) and are generally gapped in the absence of symmetry\cite{Lu2012a}. In these cases we call the corresponding phase a \emph{trivial phase} in 2+1-D, since it's featureless both in the bulk and on the edge and can be continuously connected to a trivial product state without any phase transition\cite{Chen2013}.\\

One key point we want to emphasize is that the Chern-Simons theory description for a certain topologically ordered phase is \emph{not unique}. In other words, two different ${\bf K}$ matrices for effective theory (\ref{U(1)^N Chern-Simons}) can correspond to the same topological phase, with the same set of quasiparticle (anyon) excitations. The two features described below are crucial for the classification of symmetry enriched topological orders.

First of all, the following $GL(N,\mbz)$ transformation on the ${\bf K}$ matrix yields an equivalent description for the same phase
\bea\label{GL(N,Z) transformation}
{\bf K}\simeq\tilde{\bf K}={\bf X}^T{\bf K}{\bf X},~~~{\bf X}\in GL(N,\mbz).
\eea
where $GL(N,\mbz)$ represents the group of $N\times N$ unimodular matrices. This $GL(N,\mbz)$ transformation ${\bf X}$ merely relabels the quasiparticle (anyon) excitations so that ${\bf l}\rightarrow\tilde{\bf l}={\bf X}^{-1}{\bf l}$. It's straightforward to see that all the topological properties, such as quasiparticle statistics and GSD are invariant under such a $GL(N,\mbz)$ transformation. A brief introduction to $GL(N,\mbz)$ group is given in Appendix \ref{app:gl(n,z)}.

Secondly, notice that a trivial phase satisfying (\ref{trivial phase}) can always be added to a topologically ordered phase without changing any topological properties (such as quasiparticle statistics, GSD and chiral central charge of edge excitations\cite{Kitaev2006}). One just needs to enlarge the Hilbert space to include some new microscopic degrees of freedom, which form a trivial phase. Mathematically addition of a topologically ordered phase with matrix ${\bf K}$ and a trivial phase with matrix ${\bf K}_t$ satisfying (\ref{trivial phase}) is carried out by the matrix direct sum\cite{Lu2012a}:
\bea\label{stable equivalency}
&{\bf K}\simeq{\bf K}^\prime={\bf K}\oplus{\bf K}_t,\\
&\notag\det{\bf K}_t=(-1)^{N_t/2},~~~N_t=\text{dim}{\bf K}_t=~\text{even}.
\eea
Therefore two ${\bf K}$ matrices of different dimensions can describe the same topologically ordered phase. Typically in a bosonic system (where the microscopic degrees of freedom are all bosons) the generic trivial phase is represented by\cite{Lu2012a,Levin2012a}
\bea\label{trivial phase:boson}
{\bf K}_t=\bpm0&1\\1&0\epm\oplus\bpm0&1\\1&0\epm\oplus\cdots\oplus\bpm0&1\\1&0\epm.
\eea
Meanwhile in a fermionic system, both (\ref{trivial phase:boson}) and
\bea\label{trivial phase:fermion}
{\bf K}_t=\bpm1&0\\0&-1\epm\oplus\bpm1&0\\0&-1\epm\oplus\cdots\oplus\bpm1&0\\0&-1\epm.
\eea
together represent a generic trivial phase.

\subsection{Implementing symmetries in Abelian topological orders}\label{IMPLEMENT SYMMETRY}

Our discussions in the previous section didn't assume any symmetry\footnote{Although the formulation of $U(1)^N$ Chern-Simons theory in (\ref{U(1)^N Chern-Simons}) seems to suggest existence of $N$ conserved $U(1)$-currents, they can actually be explicitly broken by \eg introducing Higgs terms\cite{Lu2012a}. These Higgs terms merely condense local \emph{bosonic} excitations (instead of nonlocal anyonic excitations) and hence don't change the superselection sectors or any topological properties of the phase.} in the topologically ordered phase. Without any symmetry, an Abelian topological order is fully characterized by its ${\bf K}$ matrix. In the presence of symmetry, however, ${\bf K}$ matrix alone is not enough to describe a symmetry enriched topological (SET) phase: \eg distinct SET phases that are separated from each other by phase transitions can share the same ${\bf K}$ matrix. The missing information is how the bulk quasiparticles (with currents $j^\mu_I$) in effective theory (\ref{U(1)^N Chern-Simons}) transform under the symmetry. The corresponding information in the edge states (\ref{edge:right})-(\ref{edge:left}) is how the chiral boson fields $\{\phi_I,1\leq I\leq N\}$ transform under symmetry.

We will restrict to unitary and anti-unitary \emph{onsite} (or global) symmetries in this work. By onsite symmetries we mean the local Hilbert space is mapped to itself\cite{Chen2011b} under the symmetry transformation, so that the symmetries act in a ``on-site'' fashion. In this case studying the symmetry transformations of bulk quasiparticles (with currents $j^\mu_I$) is equivalent to\cite{Lu2012a} studying the symmetry transformations of edge chiral bosons $\{\phi_I,1\leq I\leq N\}$. Henceforth we'll focus on the chiral boson variables on the edge to study their transformation rules under symmetry operations, in the presence of a symmetry group $G_s$.

Most generally, under the operation of symmetry group element $\bsg\in G_s$, the chiral boson fields $\{\phi_I\}$ transform in the following way\cite{Lu2012a}:
\bea\label{sym transf}
&\phi_I(x,t)\rightarrow\sum_{J}\eta^\bsg{\bf W}^\bsg_{I,J}\phi_J(x,t)+\delta\phi_I^\bsg,\\
&\notag\eta^\bsg{\bf K}=\big({\bf W}^\bsg\big)^T{\bf K}{\bf W}^\bsg,~~~{\bf W}^\bsg\in GL(N,\mbz).
\eea
where $\eta^\bsg=+1(-1)$ for a unitary (anti-unitary) on-site symmetry. This is simply because under an anti-unitary symmetry operation (such as time reversal $t\rightarrow-t$) the Chern-Simons term $\epsilon^{\mu\nu\lambda}a^I_\mu\partial_\nu a^J_\lambda$ changes sign, and in order to keep the Lagrangian (\ref{U(1)^N Chern-Simons}) in the bulk or (\ref{edge:right})-(\ref{edge:left}) on the edge invariant, ${\bf K}$ must change sign under the $GL(N,\mbz)$ rotation ${\bf W}^\bsg$.

Notice that the above symmetry transformations $\{{\bf W}^\bsg,\delta\phi^\bsg|\bsg\in G_s\}$ must be compatible with group structure of symmetry group $G_s$. This provides a strong constraint on the allowed choices of $GL(N,\mbz)$ rotations $\{{\bf W}^\bsg\}$ and $U(1)$ phase shifts $\{\delta\phi^\bsg_I\simeq\delta\phi^\bsg_I+2\pi\}$. To be precise, the consistent conditions for symmetry transformations $\{{\bf W}^\bsg,\delta\phi^\bsg|\bsg\in G_s\}$ on an Abelian topological order characterized by matrix ${\bf K}$ is summarized in the following statement:\\

The \emph{(nonlocal) quasiparticle} excitations $\{\hat{Q}_I(x,t)\equiv e^{\imth\phi_I(x,t)}\}$ transform \emph{projectively} under symmetry group $G_s$, while the \emph{(local) microscopic boson degrees of freedom} $\{\hat{M}_I(x,t)\equiv e^{\imth p_I\sum_J{\bf K}_{I,J}\phi_J(x,t)}\}$ in (\ref{local boson}) must form a \emph{linear representation} of symmetry group $G_s$. Here constant factor $p_I=1$ if ${\bf K}_{I,I}=$~even, or $p_I=2$ if ${\bf K}_{I,I}=$~odd.
%\bea\label{group compatibility condition}
%\eea
\\

In the following we'll discuss why (local) microscopic boson degrees of freedom must form a linear representation of symmetry group $G_s$. Imagine an Abelian topolgical ordered phase preserves symmetry $G_s$. For simplicity let's consider $G_s=Z_2=\{\bsg,\bse\}$ for an illustration. We denote the generator of the $Z_2$ group as $\bsg$. It satisfies the following $Z_2$ multiplication rule:
\bea\label{Z2 group}
\bsg\cdot\bsg\equiv\bsg^2=\bse.
\eea
And under this $Z_2$ symmetry operation $\bsg$ the edge chiral bosons transform as (\ref{sym transf}).~% with $\eta^\bsg=+1$, since $\bsg$ is a unitary symmetry.
Consider we weakly break the $Z_2$ symmetry without closing the bulk energy gap (no phase transition). Now $Z_2$ operation $\bsg$ is not a symmetry anymore and there is no symmetry in the system. Therefore all the local bosonic degrees of freedom $\{\hat{M}_I(x,t)\equiv e^{\imth p_I\sum_J{\bf K}_{I,J}\phi_J(x,t)}|1\leq I\leq N\}$ can condensed and Higgs term (\ref{higgs term}) should be allowed. At the same time, notice that $\bsg^2=\bse$ is still a ``symmetry'' of the system. When symmetry operation $\bsg$ act twice, its transformations on chiral bosons $\vec{\phi}(x,t)\equiv\big(\phi_1(x,t),\cdots,\phi_N(x,t)\big)^T$ become
\bea
&\notag\vec{\phi}(x,t)\overset{\bsg}\longrightarrow\eta^\bsg{\bf W}^\bsg\vec{\phi}(x,t)+\delta\vec{\phi}^\bsg\overset{\bsg}\longrightarrow\\
&\big({\bf W}^\bsg\big)^2\vec{\phi}(x,t)+\big(1_{N\times N}+\eta^\bsg{\bf W}^\bsg\big)\delta\vec{\phi}^\bsg.\label{sym transf: Z2 twice}
\eea
where $1_{N\times N}$ denotes an $N\times N$ identity matrix. And we must require all Higgs terms (\ref{higgs term}) with arbitrary parameters $\{C_I,\chi_I\}$ are allowed by ``symmetry'' $\bsg^2=\bse$ in (\ref{sym transf: Z2 twice}). In other words all the Higgs terms in (\ref{higgs term}) should remain invariant when $Z_2$ operation $\bsg$ act twice as in (\ref{sym transf: Z2 twice})! This means the argument of any cosine (Higgs) terms in (\ref{higgs term}) must be invariant up to a $2\pi$ phase, leading to the following relation:
\bea
&\notag{\bf P}{\bf K}\big({\bf W}^\bsg\big)^2={\bf P}{\bf K},~~~{\bf P}{\bf K}\big(1_{N\times N}+\eta^\bsg{\bf W}^\bsg\big)\delta\vec{\phi}^\bsg=2\pi{\bf n}.
\eea
where we defined $N\times N$ diagonal matrix ${\bf P}_{I,J}=p_I\delta_{I,J}$ and ${\bf n}=(n_1,\cdots,n_N)^T\in\mbz^N$ is an integer vector. The above relation can be rewritten as
\bea
&\big({\bf W}^\bsg\big)^2=1_{N\times N},~~\eta^\bsg{\bf K}=\big({\bf W}^\bsg\big)^T{\bf K}{\bf W}^\bsg,\label{constraint: sym transf Z2}\\
&\notag\big(1_{N\times N}+\eta^\bsg{\bf W}^\bsg\big)\delta\vec{\phi}^\bsg=2\pi({\bf P}{\bf K})^{-1}{\bf n},~~~{\bf n}\in\mbz^N.
\eea
These are the \emph{group compatibility conditions} on the symmetry transformation (\ref{sym transf}) for a $Z_2$ symmetry group $G_s=\{\bsg,\bse=\bsg^2\}$. These conditions will be applied in the examples later.

In a generic case, symmetry group $G_s$ (and its multiplication table) is fully determined by a set of algebraic relations
\bea
\notag\mathcal{A}_{m_1,\cdots,m_{N_g}}\equiv \bsg_1^{m_1}\cdot\bsg_2^{m_2}\cdots\bsg_{N_g}^{m_{N_g}}=\bse.
\eea
where $\{\bsg_1,\cdots,\bsg_{N_g}\}$ is a set of generators in group $G_s$. Each algebraic relation $\mathcal{A}_{m_1,\cdots,m_{N_g}}$ gives rise to a consistent condition with an integer vector ${\bf n}_{m_1,\cdots,m_{N_g}}$, just like (\ref{constraint: sym transf Z2}) in the $G_s=Z_2$ case. When all these group compatibility conditions are satisfied, any local bosonic degrees of freedom
\bea%\label{operator:local boson}
&\hat{B}_{\bf l}(x,t)\equiv\exp\Big(\imth{\bf l}^T{\bf P}{\bf K}\vec{\phi}(x,t)\Big),~~~{\bf l}\in\mbz^N.\notag
\eea
is invariant under symmetry operation $\mathcal{A}_{m_1,\cdots,m_{N_g}}$. By definition they form a linear representation of the symmetry group $G_s$. On the other hand, a generic quasiparticle excitation
\bea%\label{operator:nonlocal fermion/anyon}
&\notag\hat{V}_{\bf l}(x,t)\equiv\exp\Big(\imth{\bf l}^T\vec{\phi}(x,t)\Big),~~~{\bf l}\in\mbz^N.
\eea
could still transform nontrivially under consecutive symmetry operation $\mathcal{A}_{m_1,\cdots,m_{N_g}}$ (which equals identity $\bse$ in symmetry group $G_s$). Therefore these (fermionic or anyonic) excitations transform projectively\cite{Wen2002,Lu2012a} under symmetry group $G_s$.

\subsection{Criteria for different symmetry enriched topological orders}\label{CRITERIA}

In the previous section we discussed the consistent conditions on the symmetry transformations on the quasiparticle excitations in an Abelian topological order. In terms of chiral boson fields $\vec{\phi}(x,t)$ which captures the quasiparticle contents in an Abelian topological order, under symmetry transformations (\ref{sym transf}) (labeled by $\{{\bf W}^\bsg,\delta\vec\phi^\bsg|\bsg\in G_s\}$ for symmetry group $G_s$), the (local) bosonic degrees of freedom (\ref{local boson}) transform linearly while (nonlocal) anyonic degrees of freedom $\{e^{\imth\phi_I}\}$ can transform projectively. Together with matrix ${\bf K}$ which contains all the topological properties, the following set of data
\bea\label{data of a SET phase}
[{\bf K},\{\eta^\bsg,{\bf W}^\bsg,\delta\vec\phi^\bsg|\bsg\in G_s\}]
\eea
fully characterizes a symmetry enriched topological (SET) phase in the presence of symmetry group $G_s$.

A natural question is: is such a data a unique fingerprint for a SET phase? Can two different sets of data describe the same SET phase? Not surprisingly the answer is yes. A trivial example is discussed earlier when symmetry group is trivial $G_s=\{\bse\}$ (\ie no symmetry) and two different ${\bf K}$ matrices corresponds to the same Abelian topological order. So how can we tell whether two sets of data (\ref{data of a SET phase}) describe the same SET phase or not? In the following we'll propose a few criteria, which thoroughly address this issue.

The first criterion comes from the physical picture that there is no ``smooth'' boundary condition under which we can sew two different SET phases with the same topological order and symmetry group $G_s$. This is rooted in the fact that two different SET phases cannot be continuously (no phase transitions in between) connected to each other without breaking the symmetry. On the other hand, if two symmetric states belong to the same SET phase, there must exist a ``transparent'' smooth boundary between these two states that preserves symmetry. In the following we'll establish the above physical picture in a more precise mathematical setup.

Consider two symmetric Abelian states described by K-matrix ${\bf K}^L$ and ${\bf K}^R$. First we require ${\bf K}^L\simeq{\bf K}^R$ describe the same Abelian topological order in the absence of symmetry, \ie they have the same topological properties such as GSD ($|\det{\bf K}^L|=|\det{\bf K}^R|$) and quasiparticle statistics. This is because two SET phases are certainly different if they correspond to different topological orders when symmetry is broken.

Consider a left edge (\ref{edge:left}) of SET phase $[{\bf K}^L,\{\eta^\bsg,{\bf W}^\bsg_L,\delta\vec\phi^\bsg_L|\bsg\in G_s\}]$ and the right edge (\ref{edge:right}) of SET phase $[{\bf K}^R,\{\eta^\bsg,{\bf W}^\bsg_R,\delta\vec\phi^\bsg_R|\bsg\in G_s\}]$ are sewed together by introducing tunneling terms between the two edges (see FIG. \ref{fig:sewing edges}). We denote the chiral boson fields as $\{\phi^L_I\}$ on the left edge and $\{\phi^R_J\}$ on the right edge. Notice that only microscopic degrees of freedom (\ref{vector:electron}) whose mutual statistics (\ref{statistics:mutual}) with any quasiparticle are multiples of $2\pi$, can appear in the tunneling term between the right and left edges\cite{Wen1990} as shown in FIG. \ref{fig:sewing edges}. Therefore a general tunneling term has the following Lagrangian density
\bea
&\notag\mathcal{H}_{tunnel}=\sum_{\alpha}T_\alpha\cos\Big(({\bf l}^L_\alpha)^T{\bf K}^L\vec\phi^L-({\bf l}^R_\alpha)^T{\bf K}^R\vec\phi^R+\varphi_\alpha\Big),\\
&{\bf l}^L_\alpha,{\bf l}^R_\alpha\neq0,~~~({\bf l}^L_\alpha)^T{\bf K}^L{\bf l}^L_\alpha-({\bf l}^R_\alpha)^T{\bf K}^R{\bf l}^R_\alpha=0,~~\forall~\alpha.\label{tunneling term}
\eea
where $T_\alpha,\varphi_\alpha$ are real parameters. According to Kac-Moody algebra (\ref{k-m algebra}) for the chiral bosons, the condition on the 2nd line means the variables in each cosine term of (\ref{tunneling term}) commute with itself and can be localized at a classical value $\langle({\bf l}^L_\alpha)^T{\bf K}^L\vec\phi^L-({\bf l}^R_\alpha)^T{\bf K}^R\vec\phi^R\rangle$. Of course every tunneling term in (\ref{tunneling term}) must be allowed by symmetry, \ie they remain invariant under symmetry transformation (\ref{sym transf}). The edge states is fully gapped\cite{Lu2012a} if each chiral boson field $\phi^{L/R}_I$ is either pinned at a classical value or doesn't commute with at least one variable of the cosine terms in (\ref{tunneling term}). Notice that each cosine term in (\ref{tunneling term}) must contain local operators from both edges.

If the set of symmetric tunneling terms (\ref{tunneling term}) cannot fully gap out the boundary between the two states, the two symmetric states clearly cannot belong to the same SET phase. On the contrary, even if the boundary can be fully gapped out by (\ref{tunneling term}), the two states on both sides of the boundary may still correspond to different SET phases, as elaborated below.

For an Abelian topological order characterized by $N\times N$ matrix ${\bf K}$, its Abelian anyons $e^{\imth\vec\mu\cdot\vec\phi}$ are labeled by vectors $\vec\mu\in\mbz^N$ which form a $N$-dimensional integer lattice\cite{Frohlich1991,Wen1992,Cano2014}. The primitive vectors $\{{\bf b}^J\}$ are given by rows of ${\bf K}$-matrix \ie ${\bf b}^J_I={\bf K}_{I,J}$, which forms a Bravais lattice. This defines an equivalence relation between anyon vectors $\vec l$: two anyon vectors differing by a Bravais lattice vector correspond to the same sector of Abelian anyon. Distinct anyons in an Abelian topological order thus correspond to different ``sublattices'' of the anyon lattice. Fusion rule of distinct anyon sectors corresponds to addition of the anyon vectors in the anyon lattice $\Lambda_{\bf K}$.

The effective K-matrix for the boundary between ${\bf K}^L$ and ${\bf K}^R$ is ${\bf K}={\bf K}^L\oplus(-{\bf K}^R)$. If a set of tunneling terms (\ref{tunneling term}) fully gap out the boundary, tunneling an anyon $\vec\mu_R$ from the right side through the boundary would inject a different anyon $\vec\mu_L$ into the left side in a way compatible with the boundary terms (\ref{tunneling term}). Mathematically we can define the vector product between two anyons $\vec\mu$ and $\vec\mu^\prime$ in terms of bilinear form
$(\vec\mu,\vec\mu^\prime)\equiv\vec\mu^T{\bf K}^{-1}\vec\mu^\prime$, and the anyons tunnelled through the boundary $(\vec\mu_L^T,-\vec\mu_R^T)^T$ is compatible with tunneling terms if and only if they are orthogonal to all tunneling vectors $\big(({\bf l}^L)^T{\bf K}^L,-({\bf l}^R)^T{\bf K}^R\big)^T$ \ie
\bea
\vec\mu^T_L{\bf l}^L_\alpha-\vec\mu^T_R{\bf l}^R_\alpha=0,~~~\forall~\alpha.\label{tunneling condition}
\eea
This establishes a mapping between anyons $\vec\mu_L$ in the left state ${\bf K}^L$ and anyons $\vec\mu_R$ in the right state ${\bf K}^R$, once tunnelling terms (\ref{tunneling term}) fully gap out the boundary.\\

{\bf Definition of smooth edge sewing condition}: \emph{Symmetric tunneling terms (\ref{tunneling term}) provides a smoothing sewing between the left and right side, if and only if (i) they fully gap out the boundary without breaking symmetry, (ii) the mapping (\ref{tunneling condition}) from anyons $\vec\mu_L\in\Lambda_{L}$ on the left side to anyons $\vec\mu_R\in\Lambda_{R}$ on the right side realizes an isomorphism between their anyon lattices $\Lambda_{L}$ and $\Lambda_{R}$.}\\

Since an isomorphism between the two anyon lattices preserves all the fusion rules (vector addition) and anyon statistics (vector product) of all anyon sectors. Therefore the smooth sewed boundary would be transparent for the two sides: tunneling an anyon $\vec\mu_R$ from the right side through the boundary would inject an anyon $\vec\mu_L$ with the same fractional statistics and fusion rules into the left side, as if the anyon $\vec\mu_R$ passes through the boundary. This provides a one-to-one correspondence between anyons on both sides in a way compatible with the symmetry. This motivates our criterion for different Abelian SET phases:\\

{\bf Criterion I}: \emph{Two sets of data $[{\bf K}^L,\{\eta^\bsg,{\bf W}^\bsg_L,\delta\vec\phi^\bsg_L|\bsg\in G_s\}]$ (for the left edge) and $[{\bf K}^R,\{\eta^\bsg,{\bf W}^\bsg_R,\delta\vec\phi^\bsg_R|\bsg\in G_s\}]$ (for the right edge) belong to the same SET phase if and only if there exists a set of tunneling terms (\ref{tunneling term}) connecting the two edges, which satisfy the smooth sewing condition defined above.}\\

We've argued the sufficiency of this criterion, below we show that it's also a necessary condition. Assume that smooth sewing condition cannot be achieved for the boundary between two states. Since the distinct anyon sectors are the same for ${\bf K}^L$ and ${\bf K}^R$, we only need to consider the case that the mapping from $\Lambda_L$ to $\Lambda_R$ is not injective without loss of generality\footnote{In the case when the mapping from $\Lambda_L$ to $\Lambda_R$ is not surjective, the map from $\Lambda_R$ to $\Lambda_L$ will not be injective.}. This means tunneling a nontrivial anyon $\vec\mu_{L,0}\neq0$ from the left side can simply inject a local boson (vacuum sector) $\vec\mu_{R,0}\sim0$ into the right side, via the symmetric tunneling terms (\ref{tunneling term}). This will lead to significant observable effects. Consider a hybrid sphere where the ``equator belt'' region is our symmetric state ${\bf K}^L$, while the rest region (near north and south poles) is occupied by symmetric state ${\bf K}^R$. We can immediately show the hybrid system actually have ground state degeneracy\cite{Lu2014c,Wang2013b} on a sphere, since we can create quasiparticles $e^{\imth(\vec\mu_{L,0}^T\cdot\vec\phi_L-\vec\mu_{R,0}^T\cdot\vec\phi_R)}$ on the pair of boundaries, and drag the pair of nontrivial anyon $e^{\imth(\vec\mu_{L,0}^T\cdot\vec\phi_L}$ towards the equator before annihilating them on the equator. Such a string operator gives rise to a degenerate ground state. Such degenerate gapped ground states are impossible on a sphere (no non-contractible loops) if the two regions host the same SET phase.

Let's take a look at the simplest case, when the two SET states share exactly the same set of data (\ref{data of a SET phase}). In this case the tunneling term (\ref{tunneling term}) essentially sews the left and right edge of the same SET phase. The smooth sewing between left and right edge basically tunnels the same local microscopic degrees of freedom $\hat{V}_{{\bf K}{\bf l}}^L$ of one edge with its counterpart $\hat{V}_{{\bf K}{\bf l}}^R$ on the other edge. The following tunneling term
\bea\label{tunneling:smooth}
\mathcal{H}_{tunnel}^0=\sum_{I=1}^N T_I\cos\Big(\sum_J{\bf K}_{I,J}(\phi^L_J-\phi^R_J)+\varphi_I\Big).
\eea
is allowed by symmetry $G_s$ and will gap out the edge states. Notice that all cosine terms commute with each other, so they can be minimized simultaneously. One important feature of the above tunneling terms is that there are $|\det{K}|$ inequivalent classical minima\cite{Wen1990} for the $\{\phi^L_J-\phi^R_J\}$ variables of the cosine terms. In other words, the chiral bosons will be pinned at one of the $|\det{K}|$ classical values by the above tunneling terms. In this case the one-to-one correspondence between left and right anyons is simply $\vec\mu_L=\vec\mu_R$.

%
%
%When the two sets of data, $[{\bf K}^L,\{\eta^\bsg,{\bf W}^\bsg_L,\delta\vec\phi^\bsg_L|\bsg\in G_s\}]$ for the left edge and $[{\bf K}^R,\{\eta^\bsg,{\bf W}^\bsg_R,\delta\vec\phi^\bsg_R|\bsg\in G_s\}]$ for the right edge correspond to two different SET phases, on the other hand, there is no way to smoothly sew the left and right edges together. In this case when tunneling term (\ref{tunneling term}) allowed by symmetry $G_s$ is added, either certain chiral boson modes remain gapless or the number of classical minima is more than $|\det{\bf K}^L|=|\det{\bf K}^R|$. And our first criterion is\\
%
%{\bf Criterion I}: \emph{Two sets of data $[{\bf K}^L,\{\eta^\bsg,{\bf W}^\bsg_L,\delta\vec\phi^\bsg_L|\bsg\in G_s\}]$ (for the left edge) and $[{\bf K}^R,\{\eta^\bsg,{\bf W}^\bsg_R,\delta\vec\phi^\bsg_R|\bsg\in G_s\}]$ (for the right edge) belong to the same SET phase if and only if there exists a tunneling term (\ref{tunneling term}) connecting the two edges, which gaps out all chiral boson fields and has $|\det{\bf K}^{L/R}|$-fold degenerate classical minima.}\\

This criterion applies universally to both unitary and anti-unitary symmetries (such as time reversal symmetry). When $\det{\bf K}^{L/R}=\pm1$ it automatically reduces to the criterion for different symmetry protected topological (SPT) phases in the Chern-Simons approach\cite{Lu2012a}. A direct consequence of Criterion I are the following two corollaries {\bf Corollary I}:

\emph{If the two sets of data share the same matrix ${\bf K}^L={\bf K}^R$, and all their local microscopic degrees of freedom (\ref{vector:electron}) transform in the same way under symmetry $G_s$, then they belong to the same SET phase since their edges can be sewed together smoothly by term (\ref{tunneling:smooth}).}
\\

Next, notice that a $GL(N,\mbz)$ transformation (\ref{GL(N,Z) transformation}) can always be performed on a ${\bf K}$ matrix without changing the topological order. It simply relabels different quasiparticles. Besides, $U(1)$ gauge transformations can always be performed on gauge fields $a^I_\mu$ and chiral bosons $\{\phi_I\}$. The most general gauge transformations that relabel quasiparticles have the following form
\bea\label{gauge transf}
\phi_I(x,t)\rightarrow\sum_J{\bf X}_{I,J}\phi_J(x,t)+\Delta\phi_I,~~~{\bf X}\in GL(N,\mbz).
\eea
where $\Delta\phi_I\in[0,2\pi)$ are constants. We denote such a gauge transformation as $\{{\bf X},\Delta\vec{\phi}\}$. Under such a gauge transformation, the set of data (\ref{data of a SET phase}) changes as
\bea\label{gauge equivalency}
&{\bf K}\overset{\{{\bf X},\Delta\vec{\phi}\}}\longrightarrow{\bf X}^T{\bf K}{\bf X},\\
&\notag\forall~\bsg\in G_s,~~~{\bf W}^\bsg\overset{\{{\bf X},\Delta\vec{\phi}\}}\longrightarrow{\bf X}^{-1}{\bf W}^\bsg{\bf X},\\
&\notag\delta\vec\phi^\bsg\overset{\{{\bf X},\Delta\vec\phi\}}\longrightarrow{\bf X}^{-1}\Big(\delta\vec\phi^\bsg+(\eta^\bsg{\bf W}^\bsg-1_{N\times N})\Delta\vec\phi\Big).
\eea
and $\eta^\bsg$ remains invariant. Here comes the second corollary\\

{\bf Corollary II}: \emph{any two sets of data (\ref{data of a SET phase}) that can be related to each other by a gauge transformation (\ref{gauge equivalency}) correspond to the same SET phase.}\\

Last but not least, an important lesson from studying SPT phases is that there is a duality\cite{Dijkgraaf1990,Levin2012} between SPT phases and gauge theories (or intrinsic topological orders). This duality is established by gauging the (unitary) symmetry $G_s$ in the SPT phase, \ie coupling the physical degrees of freedom (which transform under symmetry $G_s$) to a gauge field\cite{Levin2012} (with gauge group $G_s$). One conjecture is that different SPT phases with $G_s$ symmetry always leads to distinct $G_s$ gauge theories. Naively one can ask the same question for SET phases\cite{Hung2013a}: \ie will two different SET phases (with $G_s$ symmetry) always lead to distinct intrinsic topological orders, when the symmetry $G_s$ is gauged? In the examples studied in this work, by gauging unitary symmetry $G_s=Z_2$ different SET phases do lead to distinct topological orders, with different quasiparticle statistics. Besides these distinct topological orders all share the same total quantum dimension\cite{Nayak2008} $\mathcal{D}$ (and hence the same topological entanglement entropy\cite{Kitaev2006a,Levin2006} $\gamma=\log\mathcal{D}$). However one can show that for a general symmetry group this not true: \ie different SET phases can result in the same topological order by gauging the symmetry. For a counterexample let's consider two different SET phases with $G_s=Z_2\times Z_2$ symmetry.  They are constructed by stacking a topologically-ordered layer, which doesn't transform under symmetry at all, with two different $Z_2\times Z_2$-SPT layers respectively\footnote{The $Z_2\times Z_2$-SPT phases in two dimensions (2d) have a $\mbz_2^3$ classification, labeled by three binary numbers $\{n_{i}=0,1|i=1,2,3\}$.}. Clearly they cannot be smoothly connected to each other without phase transitions while preserving $Z_2\times Z_2$ symmetry, thus are distinct SET phases. However after gauging the global $Z_2\times Z_2$ symmetry, they can lead to the same intrinsic topological order \eg
\bea
\bpm0&2\\2&0\epm\oplus{\bf K}(010)\simeq\bpm0&2\\2&0\epm\oplus{\bf K}(110)\notag
\eea
where ${\bf K}(n_1n_2n_3)$ are intrinsic topological orders obtained by gauging $Z_2\times Z_2$-SPT phases in 2d, as defined in (\ref{kmat:spt:Z2xZ2}).

Although the above statement is not always true, the converse (or inverse) statement is necessarily true:\\

{\bf Criterion II}: \emph{After gauging the unitary symmetry $G_s$, if two SET phases (with symmetry $G_s$) lead to two different topological orders, they must belong to two distinct SET phases}.\\

Criterion II only applies to unitary symmetries. As will become clear in the examples, in certain cases (including those which we call ``unconventional'' SET phases), gauging an Abelian symmetry in an Abelian topological order will lead to non-Abelian topological orders\cite{Barkeshli2010b}.\\

In the following sections we will demonstrate these criteria by classifying different SET phases with (anti-)unitary $Z_2$ symmetries. The Abelian topological orders that will be studied include $Z_2$ spin liquid\cite{Hansson2004,Kou2008} with ${\bf K}\simeq\bpm0&2\\2&0\epm$, double semion theory\cite{Freedman2004,Levin2005} with ${\bf K}\simeq\bpm2&0\\0&-2\epm$ and Laughlin states\cite{Laughlin1983} (${\bf K}\simeq m$) at different filling fractions $\nu=1/m$. Among them $Z_2$ spin liquid and double semion theory are non-chiral Abelian phases, in the sense that their edge excitations have no net chirality. And in the absence of symmetry their edge excitations will generically be gapped. On the other had Laughlin states are chiral Abelian phases with quantized thermal Hall conductance.

\section{Examples}\label{EXAMPLES}

In this section we'll apply the Chern-Simons approach discussed in previous sections to various Abelian topological orders. We start by classifying $Z_2$ spin liquids with time reversal symmetry $G_s=Z_2^T$ and with a unitary $Z_2$ symmetry $G_s=Z_2$. Usually by $Z_2$ spin liquids people refer to gapped many-spin ground states supporting fractionalized spin-carrying quasiparticles, coined ``spinons'' and other fractionalized quasiparticles carrying no spin quantum numbers, coined ``visons''. The mutual (braiding) statistics of a spinon and a vison is semionic ($\theta_{s,v}=\pi$), while the self statistics of a spinon/vison is bosonic. A $Z_2$ spin liquid has 4-fold GSD on a torus. All these topological properties are captured by the Chern-Simons theory (\ref{U(1)^N Chern-Simons}) with
\bea\label{K mat:Z2 spin liquid}
{\bf K}\simeq\bpm0&2\\2&0\epm.
\eea
In the context of this work, we don't assume spin rotational symmetry and hence visons/spinons generally cannot be distinguished from their spin quantum numbers. Despite this fact we still use the name ``$Z_2$ spin liquid'' to label this Abelian topological order. The 4 degenerate ground states on a torus correspond to the 4 superselection sectors, which are associated with the 4 inequivalent quasiparticles:
\bea\label{qp contents:Z2 spin liquid}
&1\simeq\bpm0\\0\epm\simeq\bpm2\\0\epm\simeq\bpm0\\2\epm,\\
&\notag e\simeq\bpm1\\0\epm,~~~m\simeq\bpm0\\1\epm,~~~f\simeq\bpm1\\1\epm.
\eea
where both $e$ and $m$ have bosonic (self)statistics and they correspond to electric charge and magnetic vortex in a $Z_2$ gauge theory\cite{Kitaev2003} respectively. $f$ is the bound state of an electric charge and a magnetic vortex, with fermionic statistic. $0$ corresponds to any local excitations (\ref{vector:electron}) with no fractional statistics, belonging to the vacuum sector. In the folklore of $Z_2$ spin liquid, a vison is $e$ (or $m$), and accordingly a bosonic spinon is $m$ (or $e$).

\subsection{Classifying $Z_2$ spin liquids with time reversal symmetry}\label{example:z2sl:Z2T}

As a warmup we consider Abelian topological order (\ref{K mat:Z2 spin liquid}) with symmetry group $Z_2^T=\{\bsg,\bse=\bsg^2\}$ with algebra (\ref{Z2 group}). Notice that the generator of $Z_2^T$ group, $\bsg$ is an anti-unitary operation with $\eta^\bsg=-1$ in (\ref{sym transf}). In this case we rely on Criterion I and its corollaries to differentiate various $Z_2^T$-SET phases.

The associated group compatibility condition (\ref{constraint: sym transf Z2}) for $G_s=Z_2^T$ in Abelian topological order (\ref{K mat:Z2 spin liquid}) is
\bea
&\notag\big({\bf W}^\bsg\big)^2=1_{2\times 2},~~-\bpm0&2\\2&0\epm=\big({\bf W}^\bsg\big)^T\bpm0&2\\2&0\epm{\bf W}^\bsg,\\
&\big(1_{2\times 2}-{\bf W}^\bsg\big)\delta\vec{\phi}^\bsg=\pi\bpm0&1\\1&0\epm{\bf n},~~~{\bf n}\in\mbz^2.\label{condition:Z2SL:time reversal}
\eea
The ${\bf W}^\bsg\in GL(2,\mbz)$ solution to the above conditions is ${\bf W}^\bsg=\pm\bpm1&0\\0&-1\epm$. However notice that the following $GL(2,\mbz)$ gauge transformations (\ref{gauge equivalency}) keep the ${\bf K}$ matrix (\ref{K mat:Z2 spin liquid}) invariant:
\bea%\label{gauge redundancy:Z2 spin liquid}
\notag{\bf X}=\pm1_{2\times2},~~~\pm\bpm0&1\\1&0\epm.
\eea
Therefore ${\bf W}^\bsg=\bpm1&0\\0&-1\epm$ and ${\bf W}^\bsg=\bpm-1&0\\0&1\epm$ are equivalent, related by gauge transformation ${\bf X}=\pm\bpm0&1\\1&0\epm$ in (\ref{gauge equivalency}). And we can fix the gauge by choosing
\bea
\notag{\bf W}^\bsg=\bpm1&0\\0&-1\epm
\eea
for time reversal operation $\bsg$. Then solving the 2nd line of conditions (\ref{condition:Z2SL:time reversal}) we obtain
\bea\notag
n_2=0,~~~\delta\vec{\phi}=\bpm\delta\phi_1\\ \frac{n_1}{2}\pi\epm\mod2\pi.
\eea
We can always choose a gauge transformation $\{{\bf X}=1_{2\times2},\Delta\vec\phi\}$ in (\ref{gauge equivalency}) so that $\delta\phi_1=0$. Meanwhile since ${\bf l}=(2,0)^T$ and ${\bf l}=(0,2)^T$ are the local excitations in the system, according to Corollary II, $n_1=$~even all corresponds to the same SET phase. Meanwhile $n_1=$~odd leads to another SET phase, which is distinct from the $n_1=$~even SET phase. This is because the magnetic vortex $m\simeq(0,1)^T$ transforms projectively in $n_1=$~odd phase, but transforms linearly in the $n_2=$~even phase under time reversal symmetry. It's straightforward to check that there is no way to smoothly sew the two edges of $n_1=$~even and $n_1=$~odd SET phases by a time-reversal-invariant tunneling term (\ref{tunneling term}), which has 4-fold degenerate classical minima. Therefore according to Criterion I they belong to two different SET phases. Hence Chern-Simons approach produces two different classes of $Z_2^T$-symmetry-enriched $Z_2$ spin liquids, as summarized in TABLE \ref{tab:Z2SL:time reversal}.

\begin{table}[tb]
\centering
\begin{tabular}{ |c||c|c| }
\hline
\multicolumn{3}{|c|}{${\bf K}\simeq\bpm0&2\\2&0\epm$ with symmetry $G_s=Z_2^T=\{\bsg,\bse=\bsg^2\}$} \\
\hline
\multicolumn{3}{|c|}{Data set in (\ref{data of a SET phase}): $[{\bf K}=\bpm0&2\\2&0\epm,\{\eta^\bsg=-1,{\bf W}^\bsg,\delta\vec\phi^\bsg\}]$}\\
\hline
Label&$\#1$&$\#2$\\
\hline
${\bf W}^\bsg$&$\bpm1&0\\0&-1\epm$&$\bpm1&0\\0&-1\epm$\\
\hline$\delta\vec\phi^\bsg$&$(0,n\pi)^T$&$(0,\pi/2+n\pi)^T$\\
\hline Proj. Sym. ($m=(0,1)^T$)&No&Yes\\
\hline Gapless edges&No&No\\
\hline
 \end{tabular}
\caption{Two different $Z_2$ spin liquids with (anti-unitary) time reversal symmetry $Z_2^T$ classified by Abelian Chern-Simons theory. In SET phase $\#1$ all quasiparticles in (\ref{qp contents:Z2 spin liquid}) transform linearly under $Z_2^T$ symmetry, while in SET phase $\#2$ quasiparticle $m$ transforms projectively under $Z_2^T$ symmetry. The data set in the 2nd line completely characterizes these SET phases. ``Proj. Sym.'' is short for ``projective realization of symmetry'' in the table.}
\label{tab:Z2SL:time reversal}
\end{table}

Since the previous calculations are based on $2\times2$ ${\bf K}$ matrix (\ref{K mat:Z2 spin liquid}), it is natural to ask: what if we enlarge the dimension of ${\bf K}$ matrix (\ref{stable equivalency}) by introducing the trivial part  with (\ref{trivial phase:boson})? In this new representation of the same Abelian topological order, will we get more SET phases or not? Notice that the trivial part (\ref{trivial phase:boson}) is nothing but the ${\bf K}$ matrix for a bosonic SPT phase\cite{Levin2012a,Lu2012a} in 2+1-D. For anti-unitary $Z_2^T$ symmetry, there is no nontrivial bosonic SPT phase\cite{Chen2013,Lu2012a} in 2+1-D. This means the edge chiral bosons for the trivial parts can always be gapped out by introducing symmetry-allowed backscattering cosine (Higgs) terms, whose classical minima is pinned at a unique classical value since $|\det{\bf K}_t|=1$. According to Criterion I, in the presence of $Z_2^T$ symmetry, when the dimension of ${\bf K}$ is enlarged by adding the trivial parts, it will not introduce any new SET phases.

At the end we discuss the stability of edge excitations in the two SET phases. Notice that the chiral bosons $\{\phi_{1,2}\}$ transform as
\bea\notag
\bpm\phi_1(x,t)\\ \phi_2(x,t)\epm\overset{\bsg}\longrightarrow\bpm-\phi_1(x,t)\\ \phi_2(x,t)+\frac{n_1}{2}\pi\epm.
\eea
under time reversal operation $\bsg$. As a result the edges can be completely gapped by introducing Higgs terms
\bea\notag
\mathcal{H}_{higgs}=C\cos\big(2\phi_1(x,t)\big).
\eea
which pins chiral boson field $\phi_1(x,t)$ to a classical value $\langle\phi_1(x,t)\rangle=0$ or $\pi$, without breaking the time reversal symmetry. Therefore in general there are no gapless edge states for the two SET phases with $G_s=Z_2^T$.

Potentially, one could conceive of a phase where both electric and magnetic vortices transform projectively under time reversal symmetry. However, such a phase is only possible as the surface state of a 3+1-D SPT phase with time reversal symmetry\cite{Vishwanath2013}. The {\bf K}-matrix classification correctly reproduces the fact that this phase is forbidden.

Meanwhile, in the presence of time reversal symmetry only, there exists one $Z_2$ spin liquid phase which cannot be described by Abelian Chern-Simons theory with a ${\bf K}$ matrix. This SET phase can be constructed \eg in terms of Abrikosov-fermion\cite{Abrikosov1965,Affleck1988b,Baskaran1988} representation of a spin-$1/2$ system, where the spin-$1/2$ fermionic spinons form a topological superconductor in class DIII\cite{Schnyder2008,Kitaev2009}. Roughly speaking, such a topological superconductor consists of a $p+\imth p$ chiral superconductor of spin-$\uparrow$ fermions, and a $p-\imth p$ chiral superconductor of spin-$\downarrow$ fermions so that time reversal symmetry is preserved. In contrast, SET phase \#2 in TABLE \ref{tab:Z2SL:time reversal} corresponds to a non-topological $s$-wave singlet superconductor of fermionic spinons. On the other hand, phase \#1 in TABLE \ref{tab:Z2SL:time reversal} can be realized in a spin-1 system, where fermionic spinons (carrying integer spins) again form a trivial $s$-wave superconductor in the fermion representation\cite{Liu2010a} of spin-1. These three time-reversal-enriched $Z_2$ spin liquids cannot be adiabatically tuned into each other without a phase transition. A full classification of $Z_2$ spin liquids with time reversal symmetry is constituted of these three distinct phases.

\begin{table*}[tb]
\centering
\begin{ruledtabular}
\begin{tabular}{ |c||c|c|c|c|}
\hline
\multicolumn{5}{|c|}{${\bf K}\simeq\bpm0&2\\2&0\epm$ with unitary symmetry $G_s=Z_2=\{\bsg,\bse=\bsg^2\}$} \\
\hline
\multicolumn{5}{|c|}{Data set in (\ref{data of a SET phase}): $[{\bf K}=\bpm0&2\\2&0\epm\oplus\bpm0&1\\1&0\epm,\{\eta^\bsg=+1,{\bf W}^\bsg=1_{4\times4},\delta\vec\phi^\bsg\}]$}\\
\hline
Label&$\#1$&$\#2$&$\#3$&$\#4$\\
\hline$\delta\vec\phi^\bsg$&$\bpm0\\0\\ \pi\\0\epm$&$\bpm0\\0\\ \pi\\ \pi\epm$&$\bpm\pi/2\\0\\ \pi\\0\epm\simeq\bpm\pi/2\\0\\ \pi\\ \pi\epm$&$\bpm\pi/2\\ \pi/2\\ \pi\\0\epm\simeq\bpm\pi/2\\ \pi/2\\ \pi\\ \pi\epm$\\
\hline Proj. Sym. ($e\simeq(1,0,0,0)^T$)&No&No&Yes&Yes\\
\hline Proj. Sym. ($m\simeq(0,1,0,0)^T$)&No&No&No&Yes\\
\hline Proj. Sym. ($f\simeq(1,1,0,0)^T$)&No&No&Yes&No\\
\hline Symmetry protected edge states&No&Yes&No&Yes\\
\hline Central charge $c$ of edge states&0&1&0&1\\
\hline {\color{blue} After gauging symmetry $\bsg$: }&&&&
\\{\color{blue} }~${\bf K}_g\simeq$&$\bpm0&2&0&0\\2&0&0&0\\0&0&0&2\\0&0&2&0\epm$&
$\bpm0&2&0&0\\2&0&0&0\\0&0&2&0\\0&0&0&-2\epm$&
$\bpm0&4\\4&0\epm$&
$\bpm4&0\\0&-4\epm$\\
\hline${\color{blue}\theta_{q_\bsg}/2\pi\equiv h_{q_\bsg}\mod1}$&$0, 1/2$&$\pm1/4$&$0,\pm1/4,1/2$&$\pm1/8,\pm3/8$\\
\hline$\color{blue}\tilde\theta_{q_\bsg,e}/2\pi\mod1$&$0,1/2$&$0,1/2$&$\pm1/4$&$\pm1/4$\\
\hline$\color{blue}\tilde\theta_{q_\bsg,m}/2\pi\mod1$&$0,1/2$&$0,1/2$&$0,1/2$&$\pm1/4$\\
%\hline$\color{blue}\tilde\theta_{q_\bsg,f}\mod2\pi$&$0$&$0$&$\pi/2$&$\pi/2$&$\pi$&$\pi$\\
\hline \color{green} Comparison to \Ref{Hung2013}&$(000)$&$(100)$&$(010)~\&~(110)~\&~m_1=0$&$m_1=2$\\
\hline
 \end{tabular}
\caption{Classification of ``conventional'' $Z_2$ spin liquids enriched by onsite (unitary) $G_s=Z_2$ symmetry. There are 4 different ``conventional'' SET phases, where under $Z_2$ symmetry all quasiparticles ($e,m,f$) merely obtain a $U(1)$ phase factor. The data set in the 2nd line completely characterizes these SET phases. ${\bf K}_g$ denotes the topological order, which is obtained by gauging the unitary $G_s=Z_2$ symmetry in the $Z_2$ spin liquid. Some of these SET phases have $Z_2$ symmetry protected edge states, which will be gapless unless $Z_2$ symmetry is spontaneously broken. On gauging the $Z_2$ symmetry (blue entries), new quasiparticle excitations (coined ``$Z_2$ symmetry fluxes'')  $\{q_\bsg\}$ are obtained, as described in Appendix \ref{app:gauging}. Their statistics (\ref{new qp:z2 spin liquid:self})-(\ref{new qp:z2 spin liquid:m}) are also summarized in the table: its self statistics $\theta_{q_\bsg}=2\pi h_{q_\bsg}$ has a one-to-one correspondence with its topological spin
 $\Theta_{q_\bsg}=\exp(\imth\theta_{q_\bsg})=\exp(2\pi\imth h_{q_\bsg})$.}
\label{tab:Z2SL:z2:conventional}
\end{ruledtabular}
\end{table*}

\subsection{Classifying $Z_2$ spin liquids with onsite $Z_2$ symmetry}\label{example:z2sl:Z2}

As discussed earlier, the reason why $2\times2$ matrix ${\bf K}=\bpm0&2\\2&0\epm$ is enough to describe $Z_2^T$-symmetric $Z_2$ spin liquids is that there is no nontrivial $Z_2^T$-SPT phases of bosons in 2+1-D. In other words the possible trivial part (\ref{trivial phase:boson}) that can be added to ${\bf K}$ in (\ref{stable equivalency}) doesn't bring in new structure to SET phases. However, for a unitary $G_s=Z_2$ symmetry, as will become clear later, there is a nontrivial bosonic SPT phase\cite{Chen2013,Chen2011b,Levin2012} whose edge cannot be gapped without breaking the $Z_2$ symmetry. This $Z_2$-SPT phase can be understood in Chern-Simons approach with ${\bf K}=\bpm0&1\\1&0\epm$. Therefore we need to consider $4\times4$ matrix
\bea
{\bf K}=\bpm0&2\\2&0\epm\oplus\bpm0&1\\1&0\epm=\bpm0&2&0&0\\2&0&0&0\\0&0&0&1\\0&0&1&0\epm\notag
\eea
to represent a generic $Z_2$ spin liquid enriched by a unitary $Z_2$ symmetry. The group compatibility condition (\ref{constraint: sym transf Z2}) for the unitary $Z_2$ symmetry ($\eta^\bsg=1$) transformation (\ref{sym transf}) becomes
\bea
&\label{condition:Z2SL:z2}\big({\bf W}^\bsg\big)^2=1_{4\times 4},\\
&\notag\bpm0&2\\2&0\epm\oplus\bpm0&1\\1&0\epm=\big({\bf W}^\bsg\big)^T\bpm0&2\\2&0\epm\oplus\bpm0&1\\1&0\epm{\bf W}^\bsg,\\
&\notag\big(1_{4\times 4}+{\bf W}^\bsg\big)\delta\vec{\phi}^\bsg=\pi\bpm0&1\\1&0\epm\oplus\bpm0&2\\2&0\epm{\bf n},~~~{\bf n}\in\mbz^4.
\eea
The gauge inequivalent solutions to ${\bf W}^\bsg$ is the following\footnote{In fact there are other solutions to conditions (\ref{condition:Z2SL:z2}), such as ${\bf W}^\bsg=-1_{4\times4}$, ${\bf W}^\bsg=-\sigma_x\oplus1_{2\times2}$, ${\bf W}^\bsg=\pm1_{2\times2}\oplus\sigma_x$ and ${\bf W}^\bsg=\pm\sigma_x\oplus\sigma_x$, where $\sigma_{x,y,z}$ are the three Pauli matrices. Whether they may lead to new SET phases are is clear to us at this moment, and we don't consider them in this work.}
\bea\notag
&{\bf W}^\bsg=1_{4\times4},~~~\bpm0&1\\1&0\epm\oplus1_{2\times2}.
\eea
We will discuss these two cases separately in the following. In particular we'll call the 1st case (${\bf W}^\bsg=1_{4\times4}$) ``conventional'' $Z_2$ spin liquids. In contrast we'll call the 2nd case (${\bf W}^\bsg=\bpm0&1\\1&0\epm\oplus1_{2\times2}$) ``unconventional'' $Z_2$ spin liquids, in the sense that distinct quasiparticles ($e$ and $m$) are exchanged under $Z_2$ symmetry operation $\bsg$.

\subsubsection{``Conventional'' $Z_2$-symmetry-enriched $Z_2$ spin liquids}

First we discuss the solution ${\bf W}^\bsg=1_{4\times4}$. In this case each anyon quasiparticle (\ref{qp contents:Z2 spin liquid}) in the $Z_2$ spin liquids merely obtains a $U(1)$ phase factor under the $Z_2$ symmetry operation $\bsg$, and we call them ``conventional'' SET phases. Due to the gauge transformations ${\bf X}=\bpm0&1\\1&0\epm\oplus1_{2\times2},1_{2\times2}\oplus\bpm0&1\\1&0\epm$ which leave the ${\bf K}$ matrix invariant, we know that the integer vector ${\bf n}$ in (\ref{condition:Z2SL:z2}) has the following equivalency relation
\bea\notag
{\bf n}=\bpm n_1\\n_2\\n_3\\n_4\epm\simeq\bpm n_2\\n_1\\n_3\\n_4\epm\simeq\bpm n_1\\n_2\\n_4\\n_3\epm.
\eea
Moreover Corollary II tells us
\bea\notag
{\bf n}\simeq\bpm n_1+2\\n_2\\n_3\\n_4\epm\simeq\bpm n_1\\n_2+2\\n_3\\n_4\epm\simeq\bpm n_1\\n_2\\n_3+2\\n_4\epm\simeq\bpm n_1\\n_2\\n_3\\n_4+2\epm.
\eea
and
\bea\notag
\bpm n_1\\n_2\\0\\0\epm\simeq\bpm n_1\\n_2\\1\\0\epm\simeq\bpm n_1\\n_2\\0\\1\epm.
\eea

\begin{table*}[tb]
\centering
%\begin{ruledtabular}
\begin{tabular}{ |c||c|c|}
\hline
\multicolumn{3}{|c|}{${\bf K}\simeq\bpm0&2\\2&0\epm$ with unitary symmetry $G_s=Z_2=\{\bsg,\bse=\bsg^2\}$} \\
\hline
\multicolumn{3}{|c|}{Data set in (\ref{data of a SET phase}): $[{\bf K}=\bpm0&2\\2&0\epm\oplus\bpm0&1\\1&0\epm,\{\eta^\bsg=+1,{\bf W}^\bsg=\bpm0&1\\1&0\epm\oplus1_{2\times2},\delta\vec\phi^\bsg\}]$}\\
\hline
Label&$\#5$&$\#6$\\
\hline$\delta\vec\phi^\bsg$&$(0,0,\pi,0)^T\simeq(\pi/2,\pi/2,\pi,0)^T$&$(0,0,\pi,\pi)^T\simeq(\pi/2,\pi/2,\pi,\pi)^T$\\%$\bpm0\\0\\ \pi\\0\epm$&$\bpm0\\0\\ \pi\\ \pi\epm$&$\bpm\pi/2\\ \pi/2\\ \pi\\0\epm$&$\bpm\pi/2\\ \pi/2\\ \pi\\ \pi\epm$\\
%\hline Proj. Sym. ($e\simeq\bpm1\\0\\0\\0\epm,m\simeq\bpm0\\1\\0\\0\epm$)&No&No&Yes&Yes\\
\hline Proj. Sym. ($f\simeq(1,1,0,0)^T$%\bpm1\\1\\0\\0\epm$
)&No&No\\
\hline Symmetry protected edge states&Yes&Yes\\
\hline Central charge $c$ of gapless edge states&$1/2$&$3/2$\\
\hline\color{blue} $ h_{q_\bsg}\mod1$&$\pm\frac1{16},\pm\frac9{16}$&
$\pm\frac3{16},\pm\frac5{16}$\\
%\hline$\tilde\theta_{q_\bsg,e}\mod2\pi$&$x$&$x$&$x$&$x$\\
%\hline$\tilde\theta_{q_\bsg,m}\mod2\pi$&$x$&$x$&$x$&$x$\\
%\hline$\tilde\theta_{q_\bsg,f}\mod2\pi$&$x$&$x$&$x$&$x$\\
\hline
\color{blue} {Relation to Kitaev's 16-fold way\cite{Kitaev2006}}&{$(\nu=1)\otimes(\nu=15)\simeq(\nu=7)\otimes(\nu=9)$}&{$(\nu=5)\otimes(\nu=11)\simeq(\nu=3)\otimes(\nu=13)$}\\
\hline
 \end{tabular}
\caption{Classification of ``unconventional'' $Z_2$ spin liquids enriched by onsite (unitary) $G_s=Z_2$ symmetry. There are 2 different ``unconventional'' SET phases, where under $Z_2$ symmetry quasiparticles $e$ and $m$ will exchange. The data set in the 2nd line completely characterizes these SET phases. Both SET phases have $Z_2$ symmetry protected edge states, which will be gapless unless $Z_2$ symmetry is broken. The central charges of these gapless edge states are half integers, in contrast to ``conventional'' $Z_2$ spin liquids where $c\in\mbz$ (see TABLE \ref{tab:Z2SL:z2:conventional}). As described in Appendix \ref{app:vertex algebra}, gauging this ``unconventional'' $Z_2$ symmetry leads to new non-Abelian quasiparticles (blue entries), which has quantum dimension $d_{q_\bsg}=\sqrt2$ and topological spin $\Theta_{q_\bsg}=\exp(2\pi\imth h_{q_\bsg})$. These new quasiparticle excitations are $Z_2$ symmetry fluxes, labeled by $\{q_\bsg\}$. All quasiparticle contents of the non-Abelian topological orders obtained by gauging $Z_2$ symmetry are summarized in TABLE \ref{tab:Z2SL:z2:unconventional:vertex algebra}. The ``gauged'' non-Abelian topological orders for both SET phases have 9-fold GSD on a torus, corresponding to 9 different superselection sectors.}
\label{tab:Z2SL:z2:unconventional}
%\end{ruledtabular}
\end{table*}

Naively there are 6 inequivalent solutions with ${\bf W}^\bsg=1_{4\times4}$ under $Z_2$ symmetry:
\bea
&\notag {\bf n}=\bpm0\\0\\0\\1\epm,~\bpm0\\0\\1\\1\epm,~\bpm0\\1\\0\\1\epm,~\bpm0\\1\\1\\1\epm,~\bpm1\\1\\0\\1\epm,~\bpm1\\1\\1\\1\epm\Longleftrightarrow\\
&\delta\vec\phi^\bsg=\bpm0\\0\\ \pi\\0\epm,\bpm0\\0\\ \pi\\ \pi\epm,\bpm\pi/2\\0\\ \pi\\0\epm,\bpm\pi/2\\0\\ \pi\\ \pi\epm,\bpm\pi/2\\ \pi/2\\ \pi\\0\epm,\bpm\pi/2\\ \pi/2\\ \pi\\ \pi\epm.\notag
\eea
However these 6 SET phases are not all different. According to Criterion I, one can show that
\bea
\delta\vec\phi^\bsg=\bpm\pi/2\\0\\ \pi\\0\epm\simeq\bpm\pi/2\\0\\ \pi\\ \pi\epm\simeq\bpm\pi/2\\-\pi\\ \pi\\ \pi\epm={\bf X}^{-1}\delta\vec\phi^\bsg\notag
\eea
by a $GL(4,\mbz)$ gauge transformation (\ref{gauge equivalency}) with
\bea
{\bf X}=\bpm1&0&0&0\\0&1&1&0\\0&0&1&0\\-2&0&0&1\epm,~~~{\bf X}^{T}{\bf K}{\bf X}={\bf K}.\notag
\eea
Similarly another two states belong to the same SET phase by the gauge transformation ${\bf X}$:
\bea
\delta\vec\phi^\bsg=\bpm\pi/2\\ \pi/2\\ \pi\\0\epm\simeq\bpm\pi/2\\ \pi/2\\ \pi\\ \pi\epm\simeq\bpm\pi/2\\-\pi/2\\ \pi\\ \pi\epm={\bf X}^{-1}\delta\vec\phi^\bsg\notag
\eea

As a result we obtain 4 inequivalent SET phases with ${\bf W}^\bsg=1_{4\times4}$ under $Z_2$ symmetry, with their symmetry transformations $\delta\vec\phi^\bsg$ summarized in TABLE \ref{tab:Z2SL:z2:conventional}. We require $\delta\vec\phi\neq0$ so that the local excitations (\ref{vector:electron}) form a faithful representation\cite{Lu2012a} of symmetry group $G_s$. In the following we briefly discuss the consequence of gauging the unitary $Z_2$ symmetry. A detailed prescription of gauging a unitary symmetry in the Chern-Simons approach is given in Appendix \ref{app:gauging}, where we've shown that gauging $Z_2$ symmetry $\{{\bf W}^\bsg=1_{4\times4},\delta\vec\phi^\bsg=\pi(i_1/2,i_2/2,1,i_4)^T\}$ yields an Abelian topological order described by matrix ${\bf K}_g$ in (\ref{K mat:Z2 spin liquid:gauing z2 sym}). Take $\#4$ for an example, after gauging $Z_2$ symmetry we have
\bea\notag
&{\bf K}_g=\bpm0&0&0&2\\0&0&2&-1\\0&2&0&-1\\2&-1&-1&-2\epm\simeq\bpm4&0&0&0\\0&-4&0&0\\0&0&0&1\\0&0&1&0\epm\simeq\bpm4&0\\0&-4\epm.
\eea
where the first equivalency $\simeq$ is realized by gauge transformation (\ref{gauge equivalency}) with
\bea\notag
{\bf X}=\bpm2&-1&0&0\\1&-1&0&0\\-1&3&1&1\\2&-2&0&-1\epm,~~~\det{\bf X}=1.
\eea
From TABLE \ref{tab:Z2SL:z2:conventional} one can see that different SET phases lead to distinct (intrinsic) topological orders by gauging their $Z_2$ symmetry. New quasiparticles $\{q_\bsg\}$ emerge when we gauge the symmetry, whose topological spins $\Theta_{q_\bsg}\equiv\exp(2\pi\imth h_{q_\bsg})$ and mutual statistics $\tilde\theta_{q_\bsg,e},~\tilde\theta_{q_\bsg,m}$ with original anyons ($e$ and $m$) serves as important characters of the SET phase (see TABLE \ref{tab:Z2SL:z2:conventional}).

The stability of edge excitations is also summarized in TABLE \ref{tab:Z2SL:z2:conventional}. For SET phases $\#2$ the gapless edge excitations come from the trivial $\bpm0&1\\1&0\epm$ part (lower $2\times2$ part) of ${\bf K}$ matrix, which corresponds to the symmetry protected edge modes of bosonic $Z_2$-SPT phases. In fact \#2 phase corresponds to nothing but stacking a layer of bosonic $Z_2$-SPT phase with a layer of $Z_2$ spin liquid which doesn't transform under $Z_2$ symmetry operation. On the other hand for $\#4$, the topologically ordered $\bpm0&2\\2&0\epm$ part (upper $2\times2$) of ${\bf K}$ matrix contributes to $c=1$ gapless edge states. In other words in a $Z_2$ spin liquid, if both $e$ and $m$ transform projectively under $Z_2$ symmetry, the edge excitations is protected to be gapless unless symmetry is broken. The edge chiral bosons $\{\phi_{1,2}\}$ for ${\bf K}=\bpm0&2\\2&0\epm$ can be refermionized as right-moving branch $\psi_R\sim\exp\big[\imth(\phi_1+\phi_2)\big]$ and left-moving branch $\psi_L\sim\exp\big[\imth(\phi_1-\phi_2)\big]$. The edge effective theory (\ref{edge:right}) can be rewritten as
\bea\label{edge:refermionize}
\mathcal{L}_{rE}=\imth\psi^\dagger_R(\partial_t-v_+\partial_x)\psi_R-\imth\psi^\dagger_L(\partial_t-v_-\partial_x)\psi_L.
\eea
where $v_{\pm}=({\bf V}_{1,2}\pm\sqrt{{\bf V}_{1,2}^2+{\bf V}_{1,1}{\bf V}_{2,2}})/2$ are the velocities of edge modes. It's easy to see that under $Z_2$ symmetry $\bsg$ the chiral fermions transform as $\psi_R\rightarrow-\psi_R$ and $\psi_L\rightarrow\psi_L$ for SET phase $\#4$, and hence backscattering term $\psi_L\psi_R+h.c.$ and $\psi_L^\dagger\psi_R+h.c.$ are forbidden by $Z_2$ symmetry. As a result there are gapless edge states in $\#4$ SET phase, protected by $Z_2$ symmetry.

%From TABLE \ref{tab:Z2SL:z2:conventional} it seems that in SET phases $\#5$ and $\#6$, anyonic quasiparticles ($e,m,f$) transform in the same fashion. They both have symmetry protected edge states, and gauge into the same intrinsic $Z_2$ topological order ${\bf K}_g\simeq\bpm0&4\\4&0\epm$. It is Criterion I that dictates they are different SET phases. Physically they do have different edge state structures: both in $\#5$ and $\#6$ a pair of counter-propagating chiral fermions from topological order $\bpm0&2\\2&0\epm$ part are protected by symmetry, while in $\#6$ there are also gapless edge excitations contributed by the $\bpm0&1\\1&0\epm$ part. Meanwhile, the $Z_2$ flux $q_\bsg$ obtained by gauging $Z_2$ symmetry $\bsg$ has different topological spins $\exp(2\pi\imth h_{q_\bsg})$ in SET phases $\#5$ and $\#6$, as shown in TABLE \ref{tab:Z2SL:z2:conventional}.
%
%

\subsubsection{``Unconventional'' $Z_2$-symmetry-enriched $Z_2$ spin liquids}\label{UNCONVENTIONAL Z2 SPIN LIQUIDS}

Now we turn to solutions to (\ref{condition:Z2SL:z2}) with ${\bf W}^\bsg=\bpm0&1\\1&0\epm\oplus1_{2\times2}$. Notice that one can always choose a proper gauge $\Delta\vec\phi$ in (\ref{gauge equivalency}) so that $\delta\phi^\bsg_1=\delta\phi^\bsg_2$, and hence $n_1=n_2$ in (\ref{condition:Z2SL:z2}). Naively there are 4 different solutions of this type to (\ref{condition:Z2SL:z2}): they are $\delta\vec\phi^\bsg=\pi(n_1/2,n_1/2,1,n_3)$ with $n_{1,3}=0,1$. However one can show that these 4 solutions are related by a gauge transformation $\Delta\vec\phi=(\pi/2,0,0,0)$ and ${\bf X}=1_{4\times4}$ in (\ref{gauge equivalency}):
\bea
&\notag\delta\vec\phi^\bsg=\bpm\pi/2\\ \pi/2\\ \pi\\n_3\pi\epm\simeq\bpm0\\0\\ \pi\\n_3\pi\epm\simeq\bpm0\\ \pi\\ \pi\\n_3\pi\epm\\
&={\bf X}^{-1}\Big(\delta\vec\phi^\bsg+({\bf W}^\bsg-1_{4\times4})\Delta\vec\phi\Big).\notag
\eea
where the 2nd equivalency is due to Criterion I. Consequently there are only two different ``unconventional'' $Z_2$ spin liquids enriched by unitary onsite $Z_2$ symmetry, as summarized in TABLE \ref{tab:Z2SL:z2:unconventional}. We can see that two distinct superselection sectors, \ie the electric charge $e$ and magnetic vortex $m$ exchange under onsite $Z_2$ symmetry $\bsg$, hence we call them ``unconventional'' SET phases. Notice that unlike ``conventional'' SET phases in TABLE \ref{tab:Z2SL:z2:conventional}, we cannot determine whether $e$ or $m$ transform projectively under unitary $Z_2$ symmetry $\bsg$ or not, since they transform into each other under $\bsg$. Both SET phases in TABLE \ref{tab:Z2SL:z2:unconventional} host symmetry protected edge excitations on the boundary, but the central charge of the gapless edge states is $c=1/2$, different from $c=1$ in ``conventional'' SET phases (see TABLE \ref{tab:Z2SL:z2:conventional}).

For SET phase $\#5$, the edge chiral bosons $\phi_{3,4}$ for the trivial $\bpm0&1\\1&0\epm$ part of ${\bf K}$ matrix can be fully gapped by a $\cos\phi_4$ term. However, the chiral bosons $\phi_{1,2}$  for topologically ordered $\bpm0&2\\2&0\epm$ part of ${\bf K}$ are protected by $Z_2$ symmetry. To be precise, in the refermionized description (\ref{edge:refermionize}) for edge states, the chiral fermions transform as
\bea
&\psi_R\overset{\bsg}\longrightarrow(-1)^{n_1}\psi_R,~~~\psi_L\overset{\bsg}\longrightarrow\psi_L^\dagger.
\eea
for $\delta\vec\phi^\bsg=\pi(n_1/2,n_1/2,1,n_3)$ with $n_{1,3}=0,1$ in TABLE \ref{tab:Z2SL:z2:unconventional}. We can rewrite each chiral fermion in terms of two Majorana fermion\cite{Kou2008} $\xi_{R/L}$ and $\eta_{R/L}$
\bea\notag
\psi_{R/L}\equiv\xi_{R/L}+\imth\eta_{R/L}
\eea
When $n_1=0$, the following backscattering term is allowed by $Z_2$ symmetry
\bea\notag
\mathcal{H}_{bs}\propto\psi_R(\psi_L+\psi_L^\dagger)+h.c.=4\xi_R\xi_L.
\eea
Therefore the $\xi_{R/L}$ branch of Majorana fermions are gapped, while the $\eta_{R/L}$ branch is protected by $Z_2$ symmetry. When $n_1=1$, similarly the following backscattering term
\bea\notag
\mathcal{H}_{bs}\propto\psi_R(\psi_L-\psi_L^\dagger)+h.c.=4\eta_R\eta_L.
\eea
is allowed by $Z_2$ symmetry. It will gap out $\eta_{R/L}$ modes and leave Majorana modes $\xi_{R/L}$ gapless. As a consequence a $c=1/2$ branch of Majorana fermions will remain gapless, unless $Z_2$ symmetry is broken on the edge. Together with the $c=1$ $Z_2$-symmetry-protected chiral boson edge\cite{Lu2012a} from $\bpm0&1\\1&0\epm$ part for SET phase $\#6$, we obtain the total central charge $c$ for both ``unconventional'' SET phases as summarized in TABLE \ref{tab:Z2SL:z2:unconventional}.\\

Aside from gapless edge states, another important feature for these ``unconventional'' SET phases is that they lead to non-Abelian topological orders once the unitary $Z_2$ symmetry is gauged. For these unconventional SET phases, a vertex algebra approach is introduced in Appendix \ref{app:vertex algebra} to gauge the unitary symmetry. The quasiparticle contents of the ``gauged'' non-Abelian topological orders for both SET phases are summarized in TABLE \ref{tab:Z2SL:z2:unconventional:vertex algebra}. The ``gauged'' topological orders are related to the ``unconventional'' $Z_2$ gauge theories describing fermions with \emph{odd} Chern number $\nu$ coupled to a $Z_2$ gauge field, as Kitaev described in his 16-fold way classification of 2+1-D $Z_2$ gauge theories\cite{Kitaev2006}. More specifically, these non-Abelian topological orders are $Z_2\times Z_2$ gauge theories, the direct product of $\nu=$odd $Z_2$ gauge theory and its time reversal counterpart $\bar\nu=16-\nu$, as summarized in TABLE \ref{tab:Z2SL:z2:unconventional}. Hence these ``gauged'' topological orders both have non-chiral edge states (chiral central charge $c_-=0$), which will generally be gapped in the absence of extra symmetry.

After gauging the symmetry, new quasiparticles with quantum dimension $d_{q_\bsg}=\sqrt2$~emerge as deconfined excitations, called $Z_2$ symmetry fluxes $\{q_\bsg\}$. When any quasiparticle $q$ in the original SET phase is moved adiabatically around a $Z_2$ symmetry flux, it becomes its image $\hat{\bsg}q$ under $Z_2$ symmetry operation. These $Z_2$ symmetry fluxes are similar to a Majorana bound state in the vortex core of a spinless $p+\imth p$ superconductor\cite{Read2000} in 2+1-D. However, they have different topological spin than those in $p+\imth p$ superconductors. To be specific, there are 4 inequivalent $Z_2$ symmetry fluxes with topological spin $\exp(\pm\frac{\nu}{16}2\pi\imth)$ and $\exp(\pm\frac{8+\nu}{16}2\pi\imth)$, as shown in TABLE \ref{tab:Z2SL:z2:unconventional}. All these non-Abelian topological orders have 9-fold GSD on a torus, corresponding to 9 different superselection sectors shown in TABLE \ref{tab:Z2SL:z2:unconventional:vertex algebra}. It's not hard to see that SET phases $\#5$ and $\#6$ do lead to different non-Abelian topological orders after $Z_2$ symmetry is gauged. In particular, their $9\times9$ modular $\mathcal{S}$ and $\mathcal{T}$ matrices in the basis of TABLE \ref{tab:Z2SL:z2:unconventional:vertex algebra} are shown in the end of Appendix \ref{app:vertex algebra}. Combining the edge states and ``gauged'' topological orders summarized in TABLE \ref{tab:Z2SL:z2:unconventional}, indeed there are 2 distinct ``unconventional'' $Z_2$-symmetry-enriched $Z_2$ spin liquids.\\

~\\

Therefore a full classification of $Z_2$ spin liquids (or $Z_2$ toric codes) enriched by unitary onsite $Z_2$ symmetry include 6 different SET phases. 2 of these 6 SET phases are ``unconventional'' (TALBE \ref{tab:Z2SL:z2:unconventional}), in the sense that distinct superselction sectors are exchanged under onsite $Z_2$ symmetry, while the other 4 SET phases are ``conventional'' (TABLE \ref{tab:Z2SL:z2:conventional}). Here many `conventional' SET phases lead to Abelian topological orders by gauging the unitary symmetry, while `unconventional' SET phases always lead to non-Abelian topological orders. In spite of these diversities, a general rule seems to apply to all SET phases:\\

{\bf Conjecture I}: \emph{All different SET phases (with the same `ungauged' topological order and onsite unitary symmetry $G_s$) always lead to topological orders with the same total quantum dimension\cite{Kitaev2006,Nayak2008} $\mathcal{D}$ (and hence the same topological entanglement entropy\cite{Kitaev2006a,Levin2006} $\gamma=\log\mathcal{D}$) , after finite unitary symmetry $G_s$ is gauged.}\\

One can easily see this conjecture holds for all examples studied in this paper, such as $Z_2$ spin liquids in TABLE \ref{tab:Z2SL:z2:conventional}-\ref{tab:Z2SL:z2:unconventional}, double semion theories in TABLE \ref{tab:double semion:z2:conventional} and $\nu=1/2k$ Laughlin states in TABLE \ref{tab:laughlin1/2k:z2}-\ref{tab:laughlin1/2k:z2:k=even}. For example all ``gauged'' topological orders from $Z_2$-symmetry enriched $Z_2$ spin liquids have $\mathcal{D}=16$, no matter Abelian (`conventional') or non-Abelian (`unconventional'). Similarly $\nu=1/2k$ Laughlin states enriched by $Z_2$ symmetry has $\mathcal{D}=8k$.

This conclusion however doesn't apply to continuous (unitary) symmetries. For a simplest example let's consider bosonic SPT phases protected by $G_s=U(1)$ symmetry\cite{Lu2012a} as a special case of $G_s=U(1)$ SET phases. They are featured by \emph{even integer} Hall conductance $\sigma_{xy}=2q,~q\in\mbz$ in units of $e^2/h$ where $e$ is the fundamental charge of bosons. Gauging the unitary $G_s=U(1)$ symmetry leads to Abelian $U(1)_{2q}$ Chern-Simons theories, whose total quantum dimension $\mathcal{D}=|2q|$ clearly differs for different $U(1)$-SPT phases.

\subsubsection{Measurable effects of ``unconventional'' $Z_2$ symmetry realizations}
\label{Sec:MeasureMajoranas}

Suppose there is a $Z_2$ spin liquid which preserves $Z_2$ spin rotational symmetry (for integer spins), are there measurable effects for these SET phases? More specifically, what are the distinctive measurable features of the ``unconventional'' $Z_2$-SET phases? In this section we'll try to answer these questions in two aspects, \ie measurements in the bulk and on the edge. We'll focus on unconventional SET phases in this section.\\

\begin{figure}
 \includegraphics[width=0.45\textwidth]{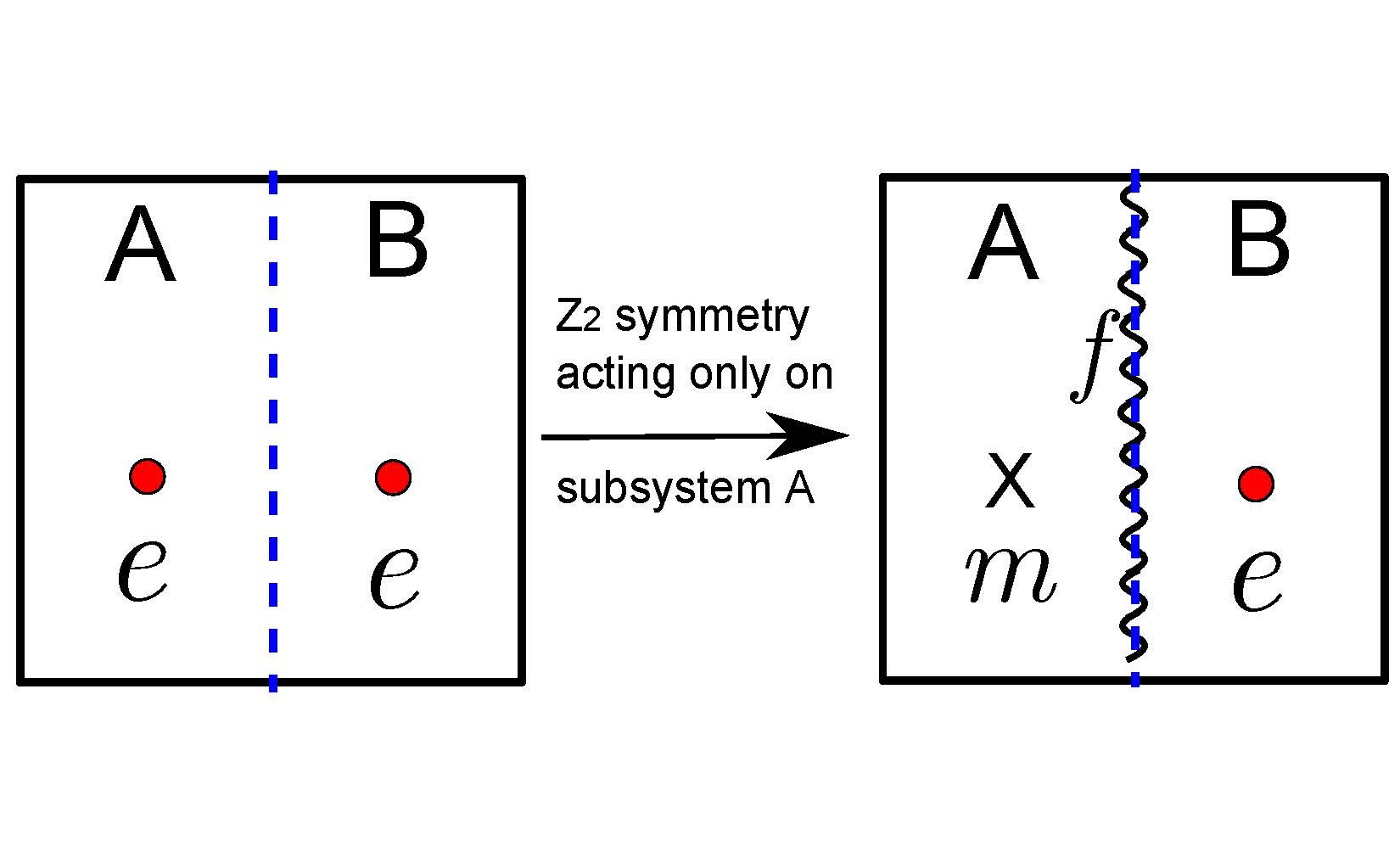}
\caption{(color online) A fermion mode ($f$) localized at the boundary between two subsystem $A$ and $B$ which from a bipartition of the on a sphere, where the ``unconventional'' Ising-symmetry-enriched $Z_2$ spin liquid resides. Under the Ising symmetry operation, an electric charge $e$ will transform into a magnetic vortex $m$. Consider one electric charge is created in each subsystem. If we perform Ising ($Z_2$) symmetry only on subsystem $A$, a fermion mode will emerge on the boundary, as the electric $e$ charge turns into a magnetic vortex $m$ in $A$.}\label{fig:boundary fermion}
\end{figure}

First of all, an important ingredient of SET phases is how their quasiparticles transform under symmetry $G_s$. This gives us one way to measure an SET phase: to create quasiparticle excitations and apply symmetry operation on them. For example for a $Z_2$ spin liquid on a closed manifold (a sphere, say), a pair of electric charges $e\simeq(1,0,0,0)^T$  (or magnetic vortices $m\simeq(0,1,0,0)^T$) can be created on top of the groundstate. For an on-site unitary symmetry (such as $Z_2$ spin-flip symmetry $\bsg$), one can choose to perform the symmetry operation only on a part of the whole system. For example, FIG. \ref{fig:boundary fermion} shows such a striking measurable effects on the unconventional $Z_2$-symmetry-enriched $Z_2$ spin liquids. Assume in the SET phase a pair of electric charges $e$ are created, one in subsystem $A$ and the other in subsystem $B$. The whole system $A\bigcup B$ lives on a closed manifold, say a sphere on which the groundstate is unique. If we only perform the ``unconventional'' $Z_2$ symmetry operation $\bsg$ in subsystem $A$ (but not in $B$), the electric charge $e\simeq(1,0,0,0)^T$ in subsystem $A$ will become a magnetic vortex $m\simeq(0,1,0,0)^T$. However one electric charge and one magnetic vortex cannot be created simultaneously ($e\times m=f\neq1$) on top of the groundstate: the conservation of ``topological charge'' requires the existence of an extra fermion $f$ in the system! Such a fermion mode $f$ lives on the boundary (dashed line in FIG. \ref{fig:boundary fermion}) between subsystem $A$ and $B$, as shown by the wavy line in FIG. \ref{fig:boundary fermion}. This effect happens in all 4 SET phases in TABLE \ref{tab:Z2SL:z2:unconventional}.\\
\begin{figure}
 \includegraphics[width=0.45\textwidth]{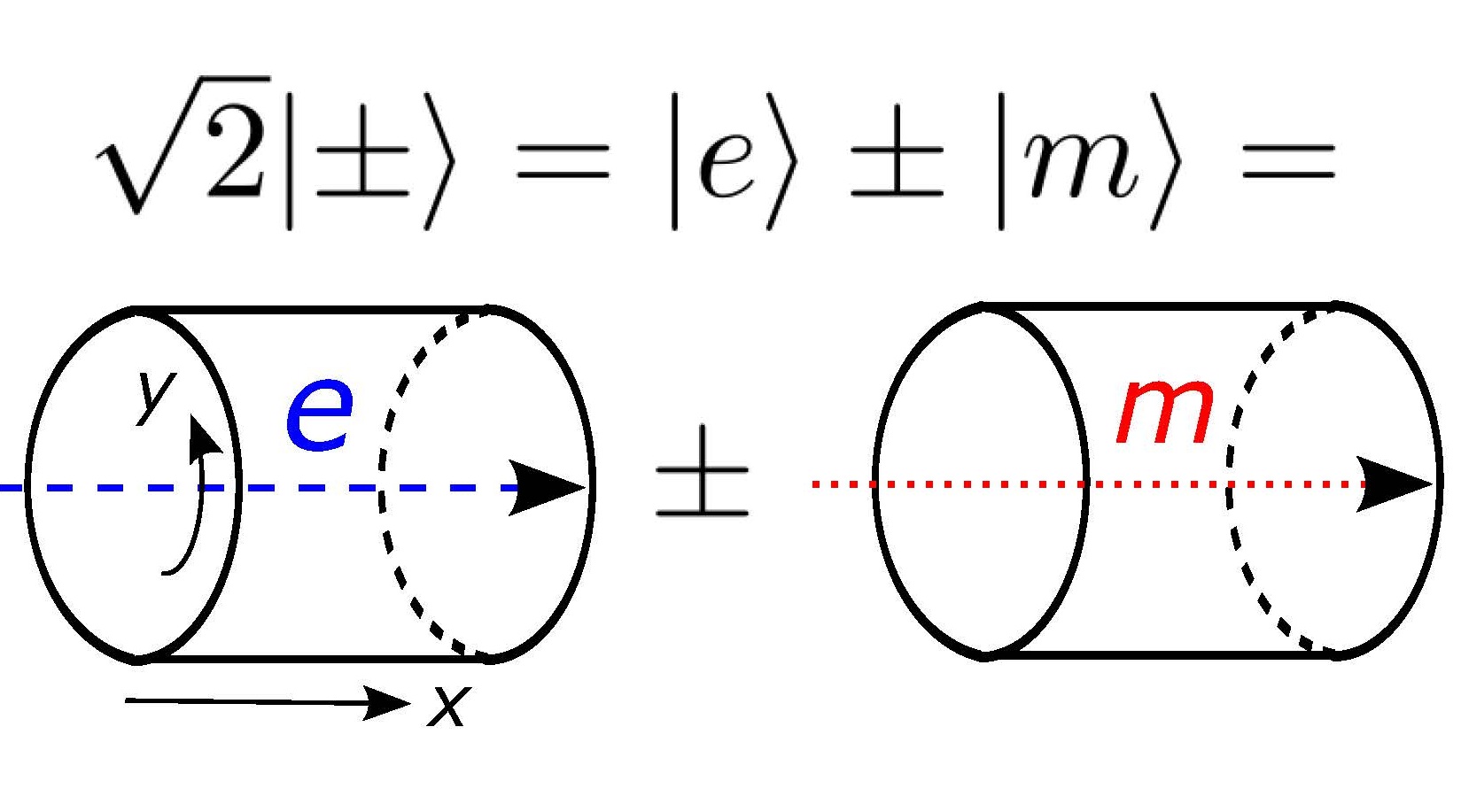}
\caption{(color online) The Ising symmetry eigenstates are linear combinations of minimal entropy states (MESs) for a ``unconventional'' Ising-symmetry-enriched $Z_2$ spin liquid, since one MES $|e\rangle$ transforms into another MES $|m\rangle$ under Ising symmetry operation.}\label{fig:cylinder ground states}
\end{figure}

Secondly, there are degenerate ground states once we put the SET phases on a closed manifold with nontrivial topology (with nonzero genus). For example, they have 4-fold GSD on a torus (or infinite cylinder). How these ground states transform under symmetry is a reflection of how anyon quasiparticles transform under symmetry. Specifically, one can always choose a set of basis called the minimal entropy states (MESs)\cite{Zhang2012}. As the name implies, each MES is a superposition of degenerate ground states which minimizes the bipartite entanglement entropy\cite{Eisert2010}, once a certain entanglement cut is chosen: \eg along the $y$-direction in the middle of the infinite cylinder as shown in FIG. \ref{fig:cylinder ground states}. The MESs are flux eigenstates\cite{Dong2008}, which keeps maximum knowledge of the states after the entanglement cut. Specifically for a $Z_2$ spin liquid, we can label the 4 MESs as $|1\rangle,~|e\rangle,~|m\rangle,~|f\rangle$ on an infinite cylinder. Remarkably under the ``unconventional'' Ising symmetry operation, two MESs ($|e\rangle$ and $|m\rangle$ exchanges) and their linear combinations $|\pm\rangle=(|e\rangle\pm|m\rangle)/\sqrt2$ are the Ising symmetry eigenstates. Therefore the MESs necessarily breaks Ising symmetry in such an unconventional SET phase! This phenomena can be measured in numerical studies\cite{Jiang2012a}.\\

\begin{figure}
 \includegraphics[width=0.35\textwidth]{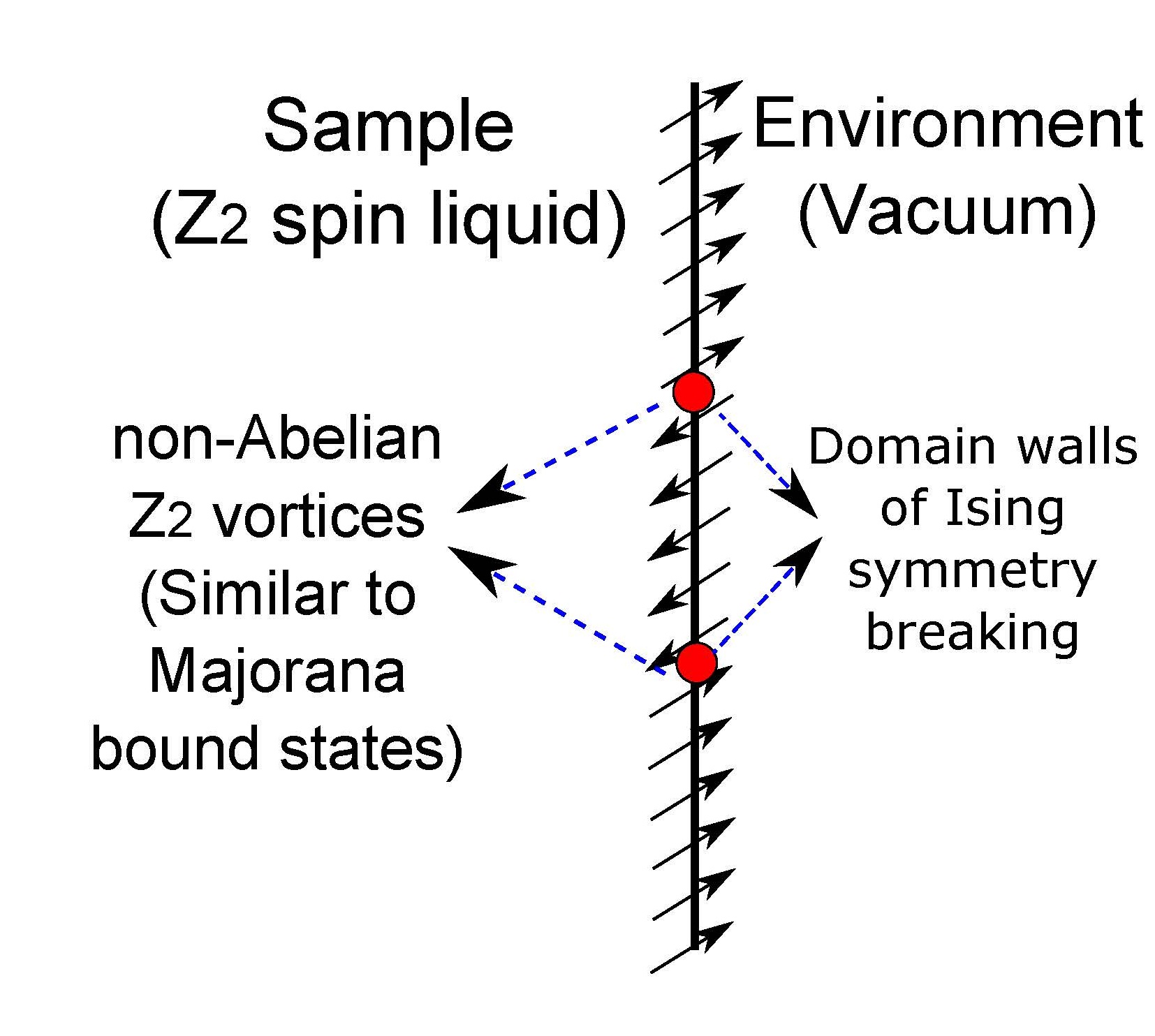}
\caption{(color online) Domain wall bound state on the edge of ``unconventional'' Ising-symmetry-enriched $Z_2$ spin liquids (see TABLE \ref{tab:Z2SL:z2:unconventional}). In these SET phases, under $Z_2$ symmetry operation, one electric charge will transform into a magnetic vortex and vice versa. The on-site unitary $Z_2$ (Ising) symmetry can be, \eg a spin-flip symmetry. On the two sides of the Ising-symmetry domain wall, two different backscattering ``mass'' terms related by spin-flip Ising symmetry are added to gap out the edge states. These two mass terms break $Z_2$ symmetry in opposite ways. A non-Abelian bound state with quantum dimension $d_{q_\bsg}=\sqrt2$ is localized at each Ising domain wall. For a ``conventional'' $Z_2$-SET phase, such a Ising mass domain wall will trap an Abelian bound state with quantum dimension $1$.}\label{fig:edge domain wall}
\end{figure}
Thirdly, the edge state structure encodes many information of a SET phase, when it supports symmetry protected edge modes. For unconventional SET phases, there are always gapless edge excitations protected by symmetry, unless the symmetry is spontaneously broken on the boundary. In the specific case of unconventional $Z_2$-symmetry-enriched $Z_2$ spin liquids summarized in TABLE \ref{tab:Z2SL:z2:unconventional}, one important feature is that SET phases $\#1$ and $\#2$ supports gapless (non-chiral) Majorana fermion excitations on the boundary, with central charge $c=1/2\mod1$. However this is not universal for all unconventional SET phases. A more interesting effect comes from the bound state localized at a $Z_2$ domain wall on the edge. Take SET phase $\#1$ for example, a perturbation on the edge
\bea
\mathcal{H}_1=A_1\cos(2\phi_1)+A_2\cos(2\phi_2)+A_4\cos\phi_4.\label{edge perturbation:z2 spin liquid}
\eea
can fully gap out the edge excitations in (\ref{edge:right})-(\ref{edge:left}) with ${\bf K}=\bpm0&2\\2&0\epm\oplus\bpm0&1\\1&0\epm$, if $A_{1,4}\neq0$ (or $A_{2,4}\neq0$). On one side of the $Z_2$ domain wall on the edge of SET phase $\#1$, we break $Z_2$ symmetry with $A_2=0$ and $A_{1,4}\neq0$. On the other side of the $Z_2$ domain wall, $Z_2$ symmetry is broken in the opposite way so that $A_1=0$ and $A_{2,4}\neq0$. Physically the electric charges are condensed on one side of the domain wall, while magnetic vortices condense on the other side. At the domain wall a non-Abelian (Majorana) bound state\cite{Wen2003,Bombin2010} is localized, which has quantum dimension $\sqrt2$, as illustrated in FIG. \ref{fig:edge domain wall}. Such a domain-wall-bound state is similar to those localized at the (ferromagnetism/superconductivity) mass domain walls of a quantum spin Hall insulator\cite{Fu2009}. In the vertex algebra context, these domain-wall-bound-states correspond to quasiparticle $q_6$ (in the 8th role) in TABLE \ref{tab:Z2SL:z2:unconventional:vertex algebra}. In SET phase $\#5$ in TABLE \ref{tab:Z2SL:z2:unconventional} \eg it has topological spin $\exp(-\imth\pi/8)$. In the bulk-edge correspondence of SET phases, such a bound state on the edge is related to the $Z_2$ symmetry flux $q_\bsg$ in the bulk.\\

\subsubsection{Gauging symmetry in ``unconventional'' SET phases}
\label{Sec:gauge unconventional SET}

In this section we provide a simple pictorial argument, which shows that gauging symmetry in an Abelian ``unconventional'' SET phase will lead to a non-Abelian topological order. In particular we'd like to show that in the new topological order obtained by gauging the symmetry, the quantum dimensions of certain quasiparticle excitations are larger than 1. Therefore the topological order obtained by gauging symmetry must be non-Abelian. Thought we'll use $G_s=Z_2$ symmetry as an illustration, this argument naturally generalizes to other finite unitary symmetry groups.

\begin{figure}
 \includegraphics[width=0.5\textwidth]{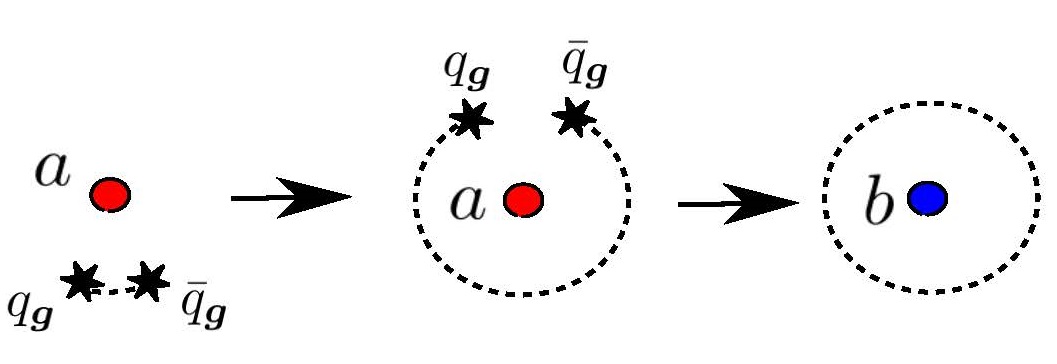}
\caption{(color online) The process in which a symmetry flux $q_\bsg$ and its antiparticle $\bar{q}_\bsg$ are created out of the vaccum, dragged around a quasiparticle $a$ and then annihilated. The dashed line denotes the trajectory of $q_\bsg$ and $\bar{q}_\bsg$ from their creation to annihilation. After this process, symmetry $\bsg$ is implemented in the region inside the closed loop (dashed line), and hence quasiparticle $a$ is transformed into $b$ under $\bsg$ symmetry operation.}\label{fig:gauging}
\end{figure}

Let's assume two distinct quasiparticle types $a$ and $b$ in an Abelian SET phase are exchanged under unitary $Z_2$ symmetry operation $\bsg$. Once $Z_2$ symmetry $\bsg$ is gauged, $Z_2$ symmetry flux $q_\bsg$ becomes deconfined excitations in the system. Hence the new groundstate after gauging symmetry $\bsg$ is a condensate of closed loops (or sting-net condensate\cite{Levin2005}), which are trajectories of a symmetry flux $q_\bsg$ and its anti-particle $\bar{q}_\bsg$ before they annihilate each other (see FIG. \ref{fig:gauging}). Whenever such a closed loop is created, symmetry $\bsg$ is implemented in the region enclosed by the loop which will transform a quasiparticle $a$ inside the loop into $b$. After gauging symmetry $\bsg$, since these closed loops (string-nets) are condensed in the new ground state, quasipaticles $a$ and $b$ are not separately deconfined excitations anymore. Instead their quantum superposition
\bea
\alpha=a+b\notag
\eea
becomes well-defined finite-energy excitation after gauging $Z_2$ symmetry $\bsg$.

More precisely, \eg the excited state wavefunction with spatially separated quasipaticles $\{\alpha({\bf r}_1),\alpha({\bf r}_2),\cdots\}$ (we use $\cdots$ to denote other quasiparticles) is a quantum superposition of four amplitudes
\bea
&\notag|\alpha({\bf r}_1),\alpha({\bf r}_2),\cdots\rangle=|a({\bf r}_1),a({\bf r}_2),\cdots\rangle+|a({\bf r}_1),b({\bf r}_2),\cdots\rangle\\
&\notag+|b({\bf r}_1),a({\bf r}_2),\cdots\rangle+|b({\bf r}_1),b({\bf r}_2),\cdots\rangle.
\eea
Therefore by definition the quantum dimension of new quasiparticle $\alpha$ is $\mathcal{D}_\alpha=\sqrt4=2$, after gauging unitary $Z_2$ symmetry $\bsg$. In the example we studied in this paper, this new quasiparticle is listed on the fifth row of TABLE \ref{tab:Z2SL:z2:unconventional:vertex algebra}.

Similarly, it's straightforward to show that after gauging ``unconventional'' $Z_N$ symmetry which permutes $N$ distinct anyons, a new quasiparticle $\alpha$ with quantum dimension $\mathcal{D}_\alpha=N>1$ will emerge. This demonstrates that gauging an ``unconventional'' SET phase will inevitably lead to non-Abelian topological orders.

%\section{Classifying Laughlin state with onsite $Z_2$ symmetry}
~\\

In previous sections we use the Chern-Simons approach to study SET phase which are non-chiral, \ie there are no gapless edge excitations in the absence symmetry. Chern-Simons approach also applies to chiral topological phases, which has gapless edge modes even in the absence of symmetry. These chiral phases have a nonzero chiral central charge\cite{Kitaev2006} $c_-$ and quantized thermal Hall effect\cite{Kane1996}, which necessarily breaks time reversal symmetry. In the following we'll use Laughlin states as illustrative examples of chiral SET phases.

\subsection{Classifying bosonic Laughlin state at $\nu=\frac1{2k},~(k\in\mbz)$ with onsite $Z_2$ symmetry}\label{example:bosonic laughlin:Z2}

\begin{table*}[tb]
\centering
%\begin{ruledtabular}
\begin{tabular}{ |c||c|c|c|c|}
\hline
\multicolumn{5}{|c|}{${\bf K}\simeq2k$~~\alert{($k=$~odd)}~~with unitary symmetry $G_s=Z_2=\{\bsg,\bse=\bsg^2\}$} \\
\hline
\multicolumn{5}{|c|}{Data set in (\ref{data of a SET phase}): $[{\bf K}=(2k)\oplus\bpm0&1\\1&0\epm,\{\eta^\bsg=+1,{\bf W}^\bsg=1_{3\times3},\delta\vec\phi^\bsg\}]$}\\
\hline
Label&$\#1$&$\#2$&$\#3$&$\#4$\\
\hline$\delta\vec\phi^\bsg$&$(0,\pi,0)^T$&$(0,\pi,\pi)^T$&$(\pi/2k,\pi,0)^T$&$(\pi/2k,\pi,\pi)^T$\\
%\hline Proj. Sym. ($s\simeq(1,0,0)^T$)&No&No&Yes&Yes\\
%\hline Symmetry protected edge&No&Yes&No&Yes&Yes&Yes\\
\hline \color{blue} After gauging symmetry:&&&&\\${\bf K}_g\simeq$&$\bpm2k&0&0\\0&0&2\\0&2&0\epm$&
$\bpm2k&0&0\\0&2&0\\0&0&-2\epm$&
$\bpm8k&0&0\\0&0&1\\0&1&0\epm\simeq8k$&
$\bpm2k&-1&0\\-1&-2&2\\0&2&0\epm$\\
\hline \color{blue} $h_{q_\bsg}=\theta_{q_\bsg}/2\pi\mod1$&$\frac{n^2}{4k},\frac{n^2}{4k}+\frac12~(n\in\mbz)$&$\frac{n^2}{4k}\pm\frac14~(n\in\mbz)$&$\frac{4k+1+4n(n+1)}{16k}\pm\frac14~(n\in\mbz)$&$\frac{1+4n(n+1)}{16k}\pm\frac14~(n\in\mbz)$\\
\hline \color{blue}$\tilde\theta_{q_\bsg,p}/2\pi\mod1$&$\frac{n}{2k},\frac{n}{2k}+\frac12~(n\in\mbz)$&$\frac{n}{2k},\frac{n}{2k}+\frac12~(n\in\mbz)$
&$\frac{2n+1}{4k},\frac{2n+1}{4k}+\frac12~(n\in\mbz)$&$\frac{2n+1}{4k},\frac{2n+1}{4k}+\frac12~(n\in\mbz)$\\
\hline
 \end{tabular}
\caption{Classification of ``conventional'' $\nu=\frac1{2k},~k=$~odd bosonic Laughlin states (or chiral spin liquids with $2k$-fold GSD on torus) enriched by onsite (unitary) $G_s=Z_2$ symmetry. There are 4 different conventional SET phases, where under $Z_2$ symmetry all quasiparticles merely obtain a $U(1)$ phase factor. The data set in the 2nd line completely characterizes these SET phases. ${\bf K}_g$ denotes the topological order, which is obtained by gauging the unitary $G_s=Z_2$ symmetry in the $Z_2$ spin liquid. On gauging the $Z_2$ symmetry (blue entries) new quasiparticle excitations $\{q_\bsg\}$ ($Z_2$ symmetry fluxes) are obtained. Their statistics (\ref{new qp:laughlin:self})-(\ref{new qp:laughlin:mutual}) are also summarized in the table.}
\label{tab:laughlin1/2k:z2}
%\end{ruledtabular}
\end{table*}

A Laughlin state\cite{Laughlin1983} at filling fraction $\nu=1/m$ is described by ${\bf K}\simeq m$ in effective theory (\ref{U(1)^N Chern-Simons}). When $m=$even it describes a bosonic topological order, while $m=$odd corresponds to a fermionic state. Such an effective theory also describes chiral spin liquids\cite{Kalmeyer1987,Wen1989}. Here we start from the simplest case \ie $m=2$. It has 2-fold GSD on torus, corresponding to two different types of quasiparticles (or two superselection sectors): boson 1 and semion $s$. Under a unitary $Z_2$ symmetry a semion always transforms into a semion, hence we don't expect any unconventional SET phases where two inequivalent quasiparticles exchange under $Z_2$ operation. Again due to the existence of nontrivial $Z_2$-SPT phase of bosons in 2+1-D, we use the following $3\times3$ matrix
\bea\label{K mat:1/2 laughlin}
{\bf K}=\bpm2&0&0\\0&0&1\\0&1&0\epm
\eea
in (\ref{U(1)^N Chern-Simons}) to represent a generic $\nu=1/2$ Laughlin state with $Z_2$ symmetry. The group compatibility conditions (\ref{constraint: sym transf Z2}) for symmetry transformations (\ref{sym transf}) are ($\eta^\bsg=1$ for unitary $Z_2$ symmetry)
\bea
&\notag\big({\bf W}^\bsg\big)^2=1_{3\times 3},~~\bpm2&0&0\\0&0&1\\0&1&0\epm=\big({\bf W}^\bsg\big)^T\bpm2&0&0\\0&0&1\\0&1&0\epm{\bf W}^\bsg,\\
&\big(1_{3\times 3}+{\bf W}^\bsg\big)\delta\vec{\phi}^\bsg=\bpm{\frac12}&0&0\\0&0&2\\0&2&0\epm{\bf n}\pi,~~~{\bf n}\in\mbz^3.\label{condition:laughlin1/2}
\eea
The inequivalent solutions to the above conditions are ${\bf W}^\bsg=1_{3\times3}$ and
\bea
\delta\vec\phi^\bsg=(\frac{i_1\pi}{2},\pi,i_3\pi)^T.
\eea
In this case they correspond to 4 different SET phases with $i_1=0,1$ and $i_3=0,1$, as summarized in TABLE \ref{tab:laughlin1/2k:z2} with $k=1$. This can be easily understood according to Criterion II, since they lead to 4 distinct topological orders when we gauge the unitary $Z_2$ symmetry.

Following the Chern-Simons approach to gauge the unitary $Z_2$ symmetry described in Appendix \ref{app:gauging}, we obtain the following ``gauged'' topological order
\bea
&\notag{\bf K}_g^{-1}={\bf M}^T{\bf K}^{-1}{\bf M},~~{\bf M}=\bpm1&0&\frac{i_1}{2}\\0&1&\frac{i_3}2\\0&0&\frac12\epm.
\eea
and hence
\bea\label{K mat:1/2 laughlin:gauging z2 sym}
{\bf K}_g=\bpm2&-i_1&0\\-i_1&-2i_3&2\\0&2&0\epm.
\eea

Take SET phase $\#3~(i_1=1,i_3=0)$ in TABLE \ref{tab:laughlin1/2k:z2} for example, its ``gauged'' topological order contains the following quasiparticles in (\ref{K mat:1/2 laughlin:gauging z2 sym}):
\bea\label{qp:gauged:laughlin:3}
\gamma\equiv\bpm0\\0\\ \gamma\epm,~~~\theta_\gamma=\frac{\gamma^2}{8}\pi,~~~\gamma=0,1,\cdots,7.
\eea
with $(1,0,0)^T\simeq(0,0,2)^T$ and $(0,1,0)^T\simeq(0,0,4)^T$. The new quasiparticles, \ie $Z_2$ symmetry fluxes with $\gamma=$~odd in (\ref{qp:gauged:laughlin:3}) have topological spins $\pm e^{\imth\pi/8}$.

Meanwhile for SET phase $\#4$ in TABLE \ref{tab:laughlin1/2k:z2}, its ``gauged'' theory has the following quasiparticle contents in (\ref{K mat:1/2 laughlin:gauging z2 sym}):
\bea\label{qp:gauged:laughlin:4}
\gamma\equiv\bpm0\\0\\ \gamma\epm,~~~\theta_\gamma=\frac{5\gamma^2}{8}\pi,~~~\gamma=0,1,\cdots,7.
\eea
where $(1,0,0)^T\simeq(0,0,2)^T$ and $(0,1,0)^T\simeq(0,0,4)^T$. One can easily show that $i_3=0,1$ in (\ref{K mat:1/2 laughlin:gauging z2 sym}) correspond to distinct topological orders: \eg the new quasiparticles or $Z_2$ symmetry fluxes with $\gamma=$~odd in (\ref{qp:gauged:laughlin:4}) in SET phase $\#4$ have topological spins $\pm e^{\imth5\pi/8}$, in contrast to $\pm e^{\imth\pi/8}$ in phase $\#3$.\\

In a complete parallel fashion we can study generic ``conventional'' $Z_2$-symmetry-enriched even-denominator Laughlin state at $\nu=1/(2k),~k\in\mbz$. Without loss of generality, a $\nu=1/(2k)$ Laughlin state with unitary $Z_2$ symmetry is represented by
\bea\label{K mat:1/2k laughlin}
{\bf K}=2k\oplus\bpm0&1\\1&0\epm=\bpm2k&0&0\\0&0&1\\0&1&0\epm.
\eea
It has $2k$ different quasiparticles (or superselection sectors) labeled as
\bea%\label{qp contents:1/2k laughlin}
\notag Q_a=\bpm a\\0\\0\epm,~~~h_{Q_a}=\frac{\theta_{Q_a}}{2\pi}=\frac{a^2}{2k},~~~a=0,1,\cdots,2k-1.
\eea

The group compatibility conditions (\ref{constraint: sym transf Z2}) for symmetry transformations (\ref{sym transf}) have the following inequivalent solutions
\bea
{\bf W}^\bsg=1_{3\times3},~~~\delta\vec\phi^\bsg=(\frac{i_1\pi}{2k},\pi,i_3\pi)^T.
\eea
where $i_{1,3}=0,1$. The solution $i_1=2$ represents the same SET phase as $i_1=0$, according to Corollary II in the criterions. Comparing with the $\nu=1/2$ bosonic Laughlin state case, we can see there is a universal structure for all bosonic Lauglin state with ${\bf K}\simeq2k,~k=$~odd. To be specific, for a $\nu=1/2k~(k=~\text{odd})$ bosonic Laughlin state, there are 4 different $Z_2$-SET phases as summarized in TABLE \ref{tab:laughlin1/2k:z2}. The quasiparticles (or edge chiral bosons) transform as
\bea
\phi\overset{\bsg}\longrightarrow\phi+(\frac{i_1\pi}{2k},\pi,i_3\pi)^T,~~~i_{1,3}=0,1.
\eea\label{sym transf:laughlin:k=even}
under ``conventional'' $Z_2$ operation.

Again following the Chern-Simons approach to gauge the unitary $Z_2$ symmetry described in Appendix \ref{app:gauging}, we obtain the following topological order
\bea
&\notag{\bf K}_g^{-1}={\bf M}^T{\bf K}^{-1}{\bf M},~~{\bf M}=\bpm1&0&\frac{i_1}{2}\\0&1&\frac{i_3}2\\0&0&\frac12\epm.
\eea
and hence
\bea\label{K mat:1/2k laughlin:gauging z2 sym}
{\bf K}_g=\bpm2k&-i_1&0\\-i_1&-2i_2&2\\0&2&0\epm,~~~i_{1,3}=0,1.
\eea
All these 4 Abelian topological orders obtained by gauging symmetry has $|\det{{\bf K}_g}|=8k$ fold GSD on a torus.

After gauging the $Z_2$ symmetry, we obtain new quasiparticles which are the $Z_2$ symmetry fluxes $\{q_\bsg\}$. A generic $Z_2$ symmetry flux is represented by $q_\bsg=(i_1,i_3,1)^T/2+{\bf l},~{\bf l}\in\mbz$. Its topological spin is given by $\exp(2\pi\imth h_{q_\bsg})$, where
\bea
&\notag h_{q_\bsg}=\frac{\theta_{q_\bsg}}{2\pi}=\frac12q_\bsg^T{\bf K}^{-1}q_\bsg\\
&\notag=\frac12\big(\frac{\delta\vec\phi^\bsg}{2\pi}\big)^T{\bf K}\frac{\delta\vec\phi^\bsg}{2\pi}+\frac12{\bf l}^T{\bf K}^{-1}{\bf l}+\frac1{2\pi}{\bf l}^T\delta\vec\phi^\bsg\\
&\label{new qp:laughlin:self}=\frac{i_1^2}{16k}+\frac{l_1^2+i_1l_1}{4k}+\frac{i_3}4+\frac{l_2+i_3l_3}2\mod1.
\eea
Its mutual statistics with the fundamental Laughlin quasihole $p\equiv(1,p_2,p_3)^T$ is
\bea\label{new qp:laughlin:mutual}
\tilde\theta_{q_\bsg,p}=\frac{\pi(i_1+2l_1)}{2k}+\pi(p_2+i_3p_3)\mod2\pi.
\eea
as summarized in TABLE \ref{tab:laughlin1/2k:z2}.

It's straightforward to check that $\#3$ and $\#4$ are different SET phases, since they lead to distinct topological orders after gauging $Z_2$ symmetry (Corollary II). For example certain $Z_2$ symmetry flux in phase $\#3$ has topological spin $e^{\imth\pi/8k}$, while such $Z_2$ symmetry flux don't exist in phase $\#4$ when $k=$~odd.

\begin{table*}[tb]
\centering
%\begin{ruledtabular}
\begin{tabular}{ |c||c|c|c|}
\hline
\multicolumn{4}{|c|}{${\bf K}\simeq2k$~~\alert{($k=$~even)}~~with unitary symmetry $G_s=Z_2=\{\bsg,\bse=\bsg^2\}$} \\
\hline
\multicolumn{4}{|c|}{Data set in (\ref{data of a SET phase}): $[{\bf K}=(2k)\oplus\bpm0&1\\1&0\epm,\{\eta^\bsg=+1,{\bf W}^\bsg=1_{3\times3},\delta\vec\phi^\bsg\}]$}\\
\hline
Label&$\#1$&$\#2$&$\#3$\\
\hline$\delta\vec\phi^\bsg$&$(0,\pi,0)^T$&$(0,\pi,\pi)^T$&$(\pi/2k,\pi,0)^T\simeq(\pi/2k,\pi,\pi)^T$\\
%\hline Proj. Sym. ($s\simeq(1,0,0)^T$)&No&No&Yes&Yes\\
%\hline Symmetry protected edge&No&Yes&No&Yes&Yes&Yes\\
\hline \color{blue} After gauging symmetry:&&&\\${\bf K}_g\simeq$&$\bpm2k&0&0\\0&0&2\\0&2&0\epm$&
$\bpm2k&0&0\\0&2&0\\0&0&-2\epm$&
$\bpm8k&0&0\\0&0&1\\0&1&0\epm\simeq\bpm2k&-1&0\\-1&-2&2\\0&2&0\epm$\\
\hline \color{blue} $h_{q_\bsg}=\theta_{q_\bsg}/2\pi\mod1$&$\frac{n^2}{4k},\frac{n^2}{4k}+\frac12~(n\in\mbz)$&$\frac{n^2}{4k}\pm\frac14~(n\in\mbz)$&$\frac{1+4n(n+1)}{16k}\pm\frac14~(n\in\mbz)$\\
\hline \color{blue}$\tilde\theta_{q_\bsg,p}/2\pi\mod1$&$\frac{n}{2k},\frac{n}{2k}+\frac12~(n\in\mbz)$&$\frac{n}{2k},\frac{n}{2k}+\frac12~(n\in\mbz)$
&$\frac{2n+1}{4k},\frac{2n+1}{4k}+\frac12~(n\in\mbz)$\\
\hline
 \end{tabular}
\caption{Classification of ``conventional'' $\nu=\frac1{2k},~k=$~even bosonic Laughlin states (or chiral spin liquids with $2k$-fold GSD on torus) enriched by onsite (unitary) $G_s=Z_2$ symmetry. There are 3 different conventional SET phases, where under $Z_2$ symmetry all quasiparticles merely obtain a $U(1)$ phase factor. This is in contrast to 4 distinct SET phases when $k=$~odd (see TABLE \ref{tab:laughlin1/2k:z2}). The data set in the 2nd line completely characterizes these SET phases. ${\bf K}_g$ denotes the topological order, which is obtained by gauging the unitary $G_s=Z_2$ symmetry in the $Z_2$ spin liquid. On gauging the $Z_2$ symmetry (blue entries) new quasiparticle excitations $\{q_\bsg\}$ ($Z_2$ symmetry fluxes) are obtained. Their statistics (\ref{new qp:laughlin:self})-(\ref{new qp:laughlin:mutual}) are also summarized in the table: its self statistics $\theta_{q_\bsg}=2\pi h_{q_\bsg}$ has a one-to-one correspondence with its topological spin $\Theta_{q_\bsg}=\exp(2\pi\imth h_{q_\bsg})$.}
\label{tab:laughlin1/2k:z2:k=even}
%\end{ruledtabular}
\end{table*}

When $k=$~even, on the other hand, $\#3$ and $\#4$ states actually lead to the same $Z_2$-SET phase. A careful analysis reveals that $i_3=0$ and $i_3=1$ actually belong to the same SET phase when $i_1=1$ in (\ref{sym transf:laughlin:k=even}). They are related by gauge transformation (\ref{gauge equivalency}) as follows:
\bea
&\notag{\bf X}=\bpm1&0&-1\\2k&1&-k\\0&0&1\epm,~~~{\bf X}^T{\bf K}{\bf X}={\bf K},\\
&\delta\vec{\phi}^\bsg=\bpm\frac\pi{2k}\\ \pi\\ \pi\epm\simeq\bpm\frac\pi{2k}\\0\\ \pi\epm\simeq\bpm\frac\pi{2k}+\pi\\-k\pi\\ \pi\epm={\bf X}^{-1}\delta\vec{\phi}^\bsg.\notag
\eea
when $k=$~even. Therefore with $k=$~even there are only 3 different ``conventional'' $\nu=\frac1{2k}$ Laughlin states enriched by a unitary $Z_2$ symmetry, as summarized in TABLE \ref{tab:laughlin1/2k:z2:k=even}.

In general there are also many ``unconventional'' $\nu=1/2k$ Laughlin states, where distinct superselection sectors exchange under $Z_2$ symmetry operation. We leave the classification of these ``unconventional'' Laughlin states to future study. A few interesting examples are discussed in \Ref{Lu2014b}.

\subsection{Fermionic Laughlin state at $\nu=\frac1{2k+1},~(k\in\mbz)$ with $Z_2^f$ symmetry is unique}\label{example:fermionic laughlin:Z2f}

Now let's turn to the simplest fermionic Laughlin state with ${\bf K}\simeq3$. It has 3-fold GSD on a torus and anyon excitations with statistics $\theta=\pi/3,~-2\pi/3$. We consider the following matrix in effective theory (\ref{U(1)^N Chern-Simons})
\bea\label{K mat:1/3 laughlin}
{\bf K}=3\oplus{\bf K}_t.
\eea
where ${\bf K}_t$ generically take the form of (\ref{trivial phase:boson}) and (\ref{trivial phase:fermion}). Notice that for a fermion system with only $Z_2^f=\{\bse,\bsg=(-1)^{N_f}\}$ (fermion number parity) symmetry, there is no nontrivial SPT phases\cite{Gu2012,Lu2012a} in 2+1-D, which hosts gapless edge excitations protected by $Z_2^f$ symmetry. This fact suggests that ${\bf K}=3$ is enough to describe a $\nu=1/3$ fermionic Laughlin state with only $Z_2^f$ symmetry. Now for such a $1\times1$ matrix ${\bf K}=3$, the group compatibility conditions (\ref{constraint: sym transf Z2}) becomes (note that ${\bf P}=2$ for fermions)
\bea\label{condition:laughlin1/3:z2f}
{\bf W}^\bsg=1,~~~2\delta\phi^\bsg=\frac{2\pi}6n,~~~n\in\mbz.
\eea
for a unitary $Z_2$ symmetry $\bsg$. The gauge inequivalent solutions are $\delta\phi^\bsg=\frac{n}6\pi$ with $n\in\mbz$. However notice a fermions in this system have gauge charge $3$ in (\ref{U(1)^N Chern-Simons}), or alternatively it's represented by $e^{3\imth\phi}$ on the edge (\ref{edge:right})-(\ref{edge:left}). Under the $Z_2^f$ operation $\bsg$ each fermion obtains a $-1$ sign, which means $3\delta\phi^\bsg=\pi\mod2\pi$ and $n$ must be even in (\ref{condition:laughlin1/3:z2f}). Therefore the quasiparticle (chiral boson) transform as
\bea
\phi\overset{\bsg=(-1)^{N_f}}\longrightarrow\phi+\frac{2n+1}{3}\pi,~~~n\in\mbz.
\eea
According to Corollary II on smooth sewing condition between edges, we know different integer $n\in\mbz$ above correspond to the same $Z_2^f$-SET phase. As a result when only fermion number parity ($Z_2^f$ symmetry) is conserved, the Laughlin $\nu=1/3$ state of fermions is unique.

It's straightforward to see that after gauging the $Z_2^f$ symmetry, we obtain an Abelian topological order
\bea
{\bf K}_g=3\times4=12.
\eea

In fact, the above conclusion is true for any fermionic Laughlin state at $\nu=1/(2k+1),~k\in\mbz$ with conserved fermion number parity. In the presence of $Z_2^f$ symmetry, it is unique with ${\bf K}=2k+1$. After gauging the $Z_2^f$ symmetry, we obtain an Abelian topological order ${\bf K}_g=4(2k+1)$.

\subsection{$Z_2$ spin liquids with onsite $Z_2\times Z_2$ symmetry}\label{example:z2sl:Z2xZ2}

In the end we turn to a $Z_2$ spin liquid ${\bf K}\simeq\bpm0&2\\2&0\epm$ in the presence of $Z_2\times Z_2$ symmetry. The symmetry group $G_s=Z_2\times Z_2=\{\bse,\bsg_1,\bsg_2,\bsg_1\bsg_2\}$ consists of two generators $\bsg_1$ and $\bsg_2$, satisfying the following algebra:
\bea\label{z2 x z2 group}
\bsg_1^2=\bsg_2^2=(\bsg_1\bsg_2)^2=\bse.
\eea
In an integer-spin\footnote{In a half-integer spin system, on the other hand, spin rotations by an angle of $2\pi$ will lead to a phase $-1$. Therefore the group structure generated by $\pi$-spin-rotations along $\hat{x}$ and $\hat{z}$ directions is not $Z_2\times Z_2$ as in (\ref{z2 x z2 group}).} system these two generators $\bsg_{1,2}$ can correspond to \eg spin rotations along $\hat{x}~(\bsg_1)$ and $\hat{z}~(\bsg_2)$ direction by an angle of $\pi$. Naturally the $\pi$-spin-rotation along $\hat{y}$ direction corresponds to group element $\bsg_1\bsg_2$.

Here we'll not attempt to fully classify all $Z_2\times Z_2$-symmetry-enriched $Z_2$ spin liquids. Instead, we focus on one nontrivial example, where spinons transforms projectively under $Z_2\times Z_2$ symmetry, in the sense that under $2\pi$-spin-rotation along any ($\hat{x},\hat{y},\hat{z}$) direction the spinon (or electric charge $e$) obtains a phase $-1$, just like a half-integer spin. On the other hand, the vison (or magnetic vortex $m$) transforms trivially under the spin rotations. Such a SET phase can be easily realized by \eg Schwinger boson\cite{Auerbach1994B} representation of $Z_2$ spin liquids, for integer spin-$S$ ($S=0,1,2,\cdots$)
\bea
{\bf S}=\frac12\bpm b^\dagger_\uparrow\\b^\dagger_\downarrow\epm^T\vec\sigma
\bpm b_\uparrow\\b_\downarrow\epm
\eea
where $\vec\sigma$ are Pauli matrices. The following constraint needs to be enforced for each spin
\bea\label{constraint:schwinger boson}
b^\dagger_\uparrow b_\uparrow+b^\dagger_\downarrow b_\downarrow=2S
\eea
to guarantee ${\bf S}^2=S(S+1)$ for a spin-$S$ system. Once the bosons $b_{\uparrow/\downarrow}$ form a pair superfluid (but not a superfluid) with $\langle bb\rangle\neq0$ (but $\langle b\rangle=0$), the resulting spin-$S$ state after projection into the physical Hilbert space (\ref{constraint:schwinger boson}) is a $Z_2$ spin liquid\cite{Sachdev1992,Wang2006}. Its spinon excitations $b_{\uparrow/\downarrow}$ carry half-spin each, hence transforming projectively under $SO(3)$ (and hence $Z_2\times Z_2$) spin rotations. By ``transforming projectively" we simply mean that after all three symmetry operations in (\ref{z2 x z2 group}) which equals identity operation $\bse$, all spinons obtain $-1$ phase instead of remaining invariant (or transforming linearly). In the following we'll show such a $Z_2\times Z_2$ SET phase can be captured in the Chern-Simons approach.

Starting from a $4\times4$ matrix ${\bf K}_0=\bpm0&2\\2&0\epm\oplus\bpm0&1\\1&0\epm$ to describe $Z_2$ spin liquid, for clarity we first perform a $GL(4,\mbz)$ gauge transformation (\ref{gauge equivalency}) on ${\bf K}_0$
\bea\label{K mat:z2 spin liquid:z2 x z2}
&{\bf K}={\bf X}^T{\bf K}_0{\bf X}=\bpm0&0&-1&1\\0&0&1&1\\-1&1&0&0\\1&1&0&0\epm,\\
&\notag{\bf X}=\bpm1&0&0&0\\0&0&-1&0\\1&1&0&0\\0&0&1&1\epm,~~~\det{\bf X}=1.
\eea
We study $Z_2$ spin liquid with $Z_2\times Z_2$ spin rotational symmetry in the above representation ${\bf K}$. Notice that
\bea
{\bf K}^{-1}=\frac12{\bf K}=\bpm0&0&-1/2&1/2\\0&0&1/2&1/2\\-1/2&1/2&0&0\\1/2&1/2&0&0\epm\notag
\eea
Apparently the first two components ($\phi_{1,2}$ in the edge chiral boson context) can be regarded as spinons, which obey semionic mutual statistics with the last two components ($\phi_{3,4}$) \ie the visons. Two spinons (visons) are mutually local w.r.t. each other as indicated by (\ref{statistics:mutual}).

In a $Z_2\times Z_2$ symmetry group (\ref{z2 x z2 group}), the group compatibility conditions for symmetry transformations $\{{\bf W}^{\bsg_{1,2}},\delta\vec\phi^{\bsg_{1,2}}\}$ in (\ref{sym transf}) are
\bea\notag
&\label{condition:Z2SL:z2 x z2}\big({\bf W}^{\bsg_{1,2}}\big)^2=\big({\bf W}^{\bsg_{1}}{\bf W}^{\bsg_{2}}\big)^{2}=1_{4\times4};\\
&\notag(1_{4\times4}+{\bf W}^{\bsg_{1,2}})\delta\vec\phi^{\bsg_{1,2}}=2\pi{\bf K}^{-1}{\bf n}_{1,2};\\
&\notag\big(1_{4\times4}+{\bf W}^{\bsg_{1}}{\bf W}^{\bsg_{2}}\big)\big(\delta\vec\phi^{\bsg_2}+{\bf W}^{\bsg_2}\delta\vec\phi^{\bsg_1}\big)=2\pi{\bf K}^{-1}{\bf n}.
\eea
where ${\bf n}_{1},{\bf n}_2,{\bf n}\in\mbz^4$. Among all inequivalent solutions to these group compatibility conditions (\ref{z2 x z2 group}), the following one
\bea
&\notag{\bf W}^{\bsg_1}=\bpm0&1&0&0\\1&0&0&0\\0&0&-1&0\\0&0&0&1\epm,~~~\delta\vec\phi^{\bsg_1}=\bpm\pi/2\\ \pi/2\\0\\0\epm,\\
&\notag{\bf W}^{\bsg_2}=1_{4\times4},~~~\delta\vec\phi^{\bsg_2}=(\pi/2,-\pi/2,0,0)^T.\\
&\notag{\bf n}_1={\bf n}=(0,0,0,1)^T,~{\bf n}_2=(0,0,-1,0)^T.
\eea
corresponds to such a integer-spin $Z_2$ spin liquid where spinons transform projectively under the $Z_2\times Z_2$ spin rotations. To be specific, the quasiparticles transform as
\bea
&\notag\vec\phi=\bpm\phi_1\\ \phi_2\\ \phi_3\\ \phi_4\epm\overset{\bsg_1}\longrightarrow\bpm\phi_2+\pi/2\\ \phi_1+\pi/2\\-\phi_3\\ \phi_4\epm,\\
&\vec\phi=\bpm\phi_1\\ \phi_2\\ \phi_3\\ \phi_4\epm\overset{\bsg_2}\longrightarrow\bpm\phi_1+\pi/2\\ \phi_2-\pi/2\\ \phi_3\\ \phi_4\epm.
\eea
Indeed each spinon ($\phi_{1,2}$) acquires -1 phase after every $2\pi$-spin-rotation, while visons ($\phi_{3,4}$) transform trivially.

Notice that in such a SET phase there is no symmetry protected gapless edge states, \ie generically all edge states are gapped in the presence of $Z_2\times Z_2$ symmetry. Specifically the following backscattering terms can be added to the edge action (\ref{edge:right})-(\ref{edge:left}) without breaking symmetry
\bea
\notag\mathcal{L}_{Higgs}=C_3\cos(2\phi_3)+C_4\cos(2\phi_4)
\eea
All the chiral boson modes $\{\phi_i|i=1,2,3,4\}$ on the edge will be gapped out by this term.

%\section{Classifying $Z_2$ spin liquids with translational symmetries}

\subsection{Condition for symmetry protected edge states in SET phases}\label{EDGE STABILITY}

In this section we briefly comments on the symmetry protected edge states in all SET phases discussed above. First of all for a chiral topological order, such as Laughlin state\cite{Laughlin1983} at filling fraction $\nu=1/m$, its edge excitations have net chirality $|n_+-n_-|\neq0$ and hence cannot be destroyed even in the absence of any symmetry. These chiral topological orders are featured by quantized thermal Hall transport\cite{Kane1996}.

On the other hand, the edge excitations of a non-chiral topological order have both right and left movers and can be fully gapped out in the absence of any symmetry\cite{Wang2012a}, such as in $Z_2$ spin liquids ${\bf K}\simeq\bpm0&2\\2&0\epm$ and double semion theory ${\bf K}\simeq\bpm2&0\\0&-2\epm$. In the presence of global symmetry $G_s$, they might have symmetry protected gapless edge modes, if the edge backscattering terms are forbidden by symmetry. To be specific, the backscattering terms are typically Higgs terms (\ref{higgs term}). The edge states will be fully gapped, if and only if each chiral boson fields $\phi_i$ is either pinned at a classical minimal by the Higgs terms or doesn't commute with at least one Higgs term\cite{Lu2012a}.

Take $Z_2$ spin liquids with ${\bf K}=\bpm0&2\\2&0\epm\oplus\bpm0&1\\1&0\epm$ for example, either $A_1\cos(2\phi_1+\alpha_1)$ or $A_2\cos(2\phi_2+\alpha_2)$ could gap out chiral boson fields $\phi_{1,2}$, since $[\phi_1(x),\phi_2(y)]\neq0$. Similarly either $A_3\cos(\phi_3+\alpha_3)$ or $A_4\cos(\phi_4+\alpha_4)$ could gap out chiral bosons $\phi_{3,4}$. When these terms are not allowed by symmetry, there could be gapless excitations on the edge protected by symmetry. The symmetry protected edge modes in ``conventional'' Ising-symmetry-enriched $Z_2$ spin liquids are summarized in TABLE \ref{tab:Z2SL:z2:conventional}. Among the 4 different conventional SET phases, $\#2$ and $\#4$ upport symmetry protected gapless edge modes.

SET phase $\#4$ provides an interesting example. Here, both electric and magnetic particles transform projectively under the global $Z_2$ symmetry. Hence the edge perturbations $A_1\cos(2\phi_1+\alpha_1)$ and $A_2\cos(2\phi_2+\alpha_2)$ which attempt to condense them, are both disallowed, leading to a symmetry protected edge state. Although symmetry does allow one backscattering term, such as $\cos(2\phi_1+\phi_3+\alpha_{13})$ in SET phase $\#4$, the rest part (generated by $\phi_2-\phi_4$ and $\phi_1$) still remains gapless and is responsible for central charge $c=1$.

For ``unconventional'' Ising-symmetry-enriched $Z_2$ spin liquids summarized in TABLE \ref{tab:Z2SL:z2:unconventional}, there are always Ising-symmetry protected Majorana edge modes with central charge $c=1/2\mod1$.

For double semion theory with ${\bf K}=\bpm2&0\\0&-2\epm\oplus\bpm0&1\\1&0\epm$ on the other hand, either $A_+\cos(2\phi_1+2\phi_2+\alpha_+)$ or $A_-\cos(2\phi_1-2\phi_2+\alpha_-)$ can fully gap out chiral bosons $\phi_{1,2}$. Meanwhile again either $A_3\cos(\phi_3+\alpha_3)$ or $A_4\cos(\phi_4+\alpha_4)$ could gap out chiral bosons $\phi_{3,4}$. When symmetry forbids these terms on the edge, there will be gapless edge excitations. The results are summarized in TABLE \ref{tab:double semion:z2:conventional} for Ising-symmetry-enriched double semion theories. Among the 6 different SET phases, only $\#3,~\#4$ and $\#6$ host symmetry protected gapless edge states. They all have central charge $c=1$.

An immediate observation from results in TABLE \ref{tab:Z2SL:z2:conventional}-\ref{tab:double semion:z2:conventional} is summarized as follows:

{\bf Conjecture II}: \emph{If every $\bsg$-symmetry fluxes $\{q_\bsg\}$ in a non-chiral SET phase carries nontrivial topological spin ($h_{q_\bsg}\neq0\mod1$) after gauging the symmetry, this SET phase must support $\bsg$-symmetry protected gapless edge states.}
\bea\label{thm 1:sufficient edge}~\eea

There is a natural physical picture behind this conclusion. Symmetric edge states can always be obtained when we start from a symmetry-breaking edge. By proliferating or condensing the defects of the broken symmetry, such as domain walls for unitary $Z_2$ symmetry $\bsg$ here, one can restore symmetry on the edge\cite{Lu2014c} and obtain a gapped symmetric edge. The symmetry defects or domain walls on the edge is nothing but $\bsg$-symmetry flux discussed earlier, as illustrated by FIG. \ref{fig:edge domain wall}. If all $\bsg$-symmetry fluxes carry nontrivial topological spin, \ie none of the symmetry defects on the edge obey bosonic statistics\cite{Lu2014c}, we cannot condense them to restore the symmetry. Hence a gapped symmetric edge is not possible in this situation, and there must be $\bsg$-symmetry protected gapless edge states.

\section{Concluding remarks}\label{CONCLUSION}

In summary, we have presented a general framework to study 2+1-D symmetry enriched topological phases with Abelian topological order, using the Chern-Simons approach. It allows us to implement generic on-site unitary (or anti-unitary) symmetry in an Abelian topologically ordered phase in 2+1-D, to differentiate whether two states belong to the same SET phase or not, and to gauge a unitary (discrete) Abelian symmetry and extract the resultant topological order. Symmetry protected edge states are also easily captured in this framework. Based on this general formulation, we classify all different SET phases in a series of examples, including $Z_2$ spin liquids with time reversal symmetry (TABLE \ref{tab:Z2SL:time reversal}), $Z_2$ spin liquids with unitary Ising symmetry (TABLE \ref{tab:Z2SL:z2:conventional} and \ref{tab:Z2SL:z2:unconventional}), double semion theory with unitary Ising symmetry (TABLE \ref{tab:double semion:z2:conventional}), bosonic Laughlin states with unitary Ising symmetry (TABLE \ref{tab:laughlin1/2k:z2} and \ref{tab:laughlin1/2k:z2:k=even}) and others. We also show that (odd-denominator) fermionic Laughlin states with only conserved fermion number parity ($Z_2^f$ symmetry) is \emph{unique}. Consequences of gauging symmetries and measurable effects, such as gapless edge states, are also discussed for these SET phases.

A number of directions remain. Can the approach applied be extended to spatial symmetries? Can we extend this framework to symmetry enriched non-Abelian topological orders in 2+1-D and SET phases in 3+1-D? While SPT phases form an Abelian group, it is presently unclear if the set of SET phases have additional structure. We leave these questions to future work.

\acknowledgements

We thank Xie Chen, T. Senthil, X. G. Wen and especially Lukasz Fidkowski for numerous discussions. YML thanks Janet Ling-Yan Hung for helpful discussions on \Ref{Dijkgraaf1990}. We are particularly indebted to Max Metlitski for penetrating comments. This work is supported by Office of BES, Materials Sciences Division of the U.S. DOE under contract No. DE-AC02-05CH11231(YML,AV) and by NSF DMR 0645691(AV).

\appendix

\section{Introduction to $GL(N,\mbz)$}\label{app:gl(n,z)}

$GL(N,\mbz)$ is the group of all $N\times N$ unimodular matrices. All $GL(N,\mbz)$ matrices can be generated by the following basic transformations ($i\neq j$):
\begin{eqnarray}
&\notag T^{(i,j)}_{a,b}=\delta_{a,b}+\delta_{a,i}\delta_{b,j},\\
&\notag S^{(i,j)}_{a,b}=\delta_{a,b}(1-\delta_{a,i})(1-\delta_{a,j})+\delta_{a,j}\delta_{b,i}-\delta_{a,i}\delta_{b,j},\\
& D_{a,b}=\delta_{a,b}-2\delta_{a,N}\delta_{b,N}.\label{gl(n,z):generator}
\end{eqnarray}
$T^{(i,j)}{\bf K}$ will add the $j$-th row of matrix ${\bf K}$ to the $i$-th row of ${\bf K}$, while $S^{(i,j)}{\bf K}$ will exchange the $i$-th and $j$-th row of ${\bf K}$ with a factor of $-1$ multiplied on the $i$-th row. $D{\bf K}$ will just multiply the $N$-th row of ${\bf K}$ by a factor of $-1$. ${\bf K}T^{(i,j)},~{\bf K}S^{(i,j)}$ and ${\bf K}D$ correspond to similar operations to columns (instead of rows). A subgroup of $GL(N,\mbz)$ with determinant $+1$ is called $SL(N,\mbz)$ and it's generated by $\{T^{(i,j)},~S^{(i,j)}\}$.

As a simple example when $N=2$, group $GL(2,\mbz)$ is generated by the following basic transformations:
\begin{eqnarray}\label{gl(2,z):generator}
&S=\begin{pmatrix}0&-1\\1&0\end{pmatrix},~~T=\begin{pmatrix}1&1\\0&1\end{pmatrix},~~D=\begin{pmatrix}1&0\\0&-1\end{pmatrix}.
\end{eqnarray}
The following results will be useful
\begin{eqnarray}
&\notag T^n=\begin{pmatrix}1&n\\0&1\end{pmatrix},~(-STS)^n=\begin{pmatrix}1&0\\-n&1\end{pmatrix},~n\in\mathbb{Z}.
\end{eqnarray}

\section{Chern-Simons approach to gauge a unitary symmetry}\label{app:gauging}

In this section we discuss how to obtain the (intrinsic) topological order by gauging the unitary symmetry\cite{Levin2012} in an Abelian SET phase. We'll restrict ourselves to ``conventional'' SET phases, characterized by data $[{\bf K},\{\eta^\bsg=+1,{\bf W}^\bsg=1_{N\times N},\delta\vec\phi^\bsg|\bsg\in G_s\}]$. In these cases the chiral bosons $\phi_I$ only acquire $U(1)$ phase factors $\phi_I\rightarrow\phi_I+\delta\phi^\bsg_I$ after unitary symmetry operation $\bsg\in G_s$. When we couple the quasiparticles to a gauge field (with gauge group $G_s$), the following gauge flux
\bea\notag
&\epsilon^{0\mu\nu}\partial_\mu a^I_\nu({\bf r},t)=\delta\phi_I^\bsg\delta({\bf r}-{\bf r}^{(0)}),
\eea
becomes deconfined excitations in the system. Since in a Chern-Simons theory (\ref{U(1)^N Chern-Simons}) gauge charges are always accompanied by gauge fluxes by the following equation of motion
\bea\notag
j^\mu_I=\frac{\epsilon^{\mu\nu\lambda}}{2\pi}\sum_J{\bf K}_{I,J}\partial_\nu a^J_\lambda
\eea
clearly this quasiparticle carries gauge charge vector ${\bf K}\delta\vec\phi^\bsg/(2\pi)$. We coin such new excitations emerged after gauging unitary symmetry $\bsg$ as ``\emph{$\bsg$ symmetry fluxes}'', and denote them as $\{q_\bsg\}$. A generic $\bsg$ symmetry flux corresponds to gauge charge vector $\frac{{\bf K}\delta\vec\phi^\bsg}{2\pi}+{\bf l}$ where ${\bf l}\in\mbz^N$, since it can always combine with any gapped excitation ($\forall~{\bf l}\in\mbz^N$) in the original (ungauged) SET phase.

The new topological order ${\bf K}_g$ obtained by gauging symmetry must include these new excitations in its quasiparticle content. More precisely, the quasiparticle content of topological order ${\bf K}_g$ is expanded by all the integer vectors as well as multiples of vectors $\{{\bf K}\delta\vec\phi^\bsg/(2\pi)\}$ for ${q_\bsg}$ particles
\bea\label{qp contents:after gauging symmetry}
&{\bf l}^\prime={\bf l}+\sum_{\bsg}n_{\bsg}\frac{{\bf K}\delta\vec\phi^\bsg}{2\pi},~~~{\bf l}\in\mbz^N,~n_{\bsg}\in\mbz.
\eea
where $\{\bsg\}$ denote the generators of symmetry group $G_s$ that are gauged. And we can identify the new matrix ${\bf K}_g$ which contains all these quasiparticles in its spectrum.

The above procedures work for all discrete Abelian symmetries. Gauging continuous Abelian symmetries (i.e. direct product of $U(1)$ and discrete $Z_n$ groups) symmetry can also be done conveniently in the Chern-Simons approach. To be specific, to gauge each $U(1)$ subgroup we coupled the physical degrees of freedom to a dynamical $U(1)$ gauge field, which increases the dimension of ${\bf K}$ matrix by 1. However we'll restrict ourselves to discrete Abelian symmetries for all examples in this paper. 

In the following we work on one example to demonstrate this gauging procedure. We consider all four SET phases in TABLE \ref{tab:Z2SL:z2:conventional}, with ${\bf K}=\bpm0&2\\2&0\epm\oplus\bpm0&1\\1&0\epm$ and symmetry transformation ${\bf W}^\bsg=1_{4\times4}$, $\delta\vec\phi^\bsg=\pi(i_1/2,i_2/2,1,i_4)^T$ with $i_{1,2,4}=0,1$. According to (\ref{qp contents:after gauging symmetry}) we know here a generic quasiparticle is labeled by gauge charge vector
\bea\notag
&{\bf l}^\prime={\bf M}{\bf l};~~~~{\bf M}=\bpm i_2/2&1&0&0\\i_1/2&0&1&0\\i_4/2&0&0&1\\1/2&0&0&0\epm,~~{\bf l}\in\mbz^4.
\eea
Since this new Abelian topological order is determined by the statistics of its quasiparticles, we immediately obtain
\bea
&\notag({\bf l}_1^\prime)^T{\bf K}^{-1}{\bf l}_2^\prime={\bf l}_1^T{\bf K}_g^{-1}{\bf l}_2,~~~{\bf l}^\prime_\alpha={\bf M}{\bf l}_\alpha.
\eea
and therefore
\bea
&\notag{\bf K}_g^{-1}={\bf M}^T{\bf K}^{-1}{\bf M}=\bpm\frac{i_1i_2+2i_4}{4}& i_1/4&i_2/4&1/2\\i_1/4&0&1/2&0\\i_2/4&1/2&0&0\\1/2&0&0&0\epm,\\
&\label{K mat:Z2 spin liquid:gauing z2 sym}
\Rightarrow{\bf K}_g=\bpm0&0&0&2\\0&0&2&-i_2\\0&2&0&-i_1\\2&-i_2&-i_1&-2i_4\epm.
\eea
Clearly from (\ref{statistics:self})-(\ref{statistics:mutual}) we know the statistical angle of new quasiparticle $q_\bsg\equiv\frac{{\bf K}\delta\vec\phi^\bsg}{2\pi}+{\bf l}$ is
\bea\notag
&\theta_{q_\bsg}=\pi q_\bsg^T{\bf K}^{-1}q_{\bsg}=\pi\big(\frac{\delta\vec\phi^\bsg}{2\pi}\big)^T{\bf K}\frac{\delta\vec\phi^\bsg}{2\pi}+\pi{\bf l}^T{\bf K}^{-1}{\bf l}+{\bf l}^T\delta\vec\phi^\bsg\\
\label{new qp:z2 spin liquid:self}&=\pi\big(\frac{i_1i_2+2i_4}{4}+l_1l_2+\frac{i_1l_1+i_2l_2}2+l_3+i_4l_4\big).
\eea
as summarized in TABLE \ref{tab:Z2SL:z2:conventional}. In addition to their (self) statistics, another important character of these ``$\bsg$ symmetry fluxes'' $\{q_\bsg\}$ is their mutual statistics with the original quasiparticles in the (ungauged) SET phase. Here \eg a generic electric charge is represented by gauge charge vector $e\equiv (1,2e_2,e_3,e_4)^T$ with $e_i\in\mbz$, and its mutual statistics with $\bsg$ symmetry flux $q_\bsg\equiv\frac{{\bf K}\delta\vec\phi^\bsg}{2\pi}+{\bf l}$ is
\bea\label{new qp:z2 spin liquid:e}
&\tilde\theta_{q_\bsg,e}=(\frac{i_1}{2}+l_2+e_2i_2+e_3+e_4i_4)\pi.
\eea
Meanwhile, a generic magnetic vortex $m=(2m_1,1,m_3,m_4)^T$ has mutual statistics
\bea\label{new qp:z2 spin liquid:m}
&\tilde\theta_{q_\bsg,m}=(\frac{i_2}2+l_1+m_1i_1+m_3+m_4i_4)\pi.
\eea
with $\bsg$ symmetry flux $q_\bsg$.
%
%
%and its mutual statistics with original quasiparticles $e,m,f$ of the $Z_2$ spin liquid are
%\bea
%&\notag\tilde\theta_{q_\bsg,e}=(\frac{i_1}{2}+l_2)\pi,~~~\tilde\theta_{q_\bsg,m}=(\frac{i_2}{2}+l_1)\pi,\\
%&\tilde\theta_{q_\bsg,f}=\frac{(i_1+i_2)\pi}{2}.\label{new qp:z2 spin liquid:mutual}
%\eea
Topological spin\cite{Kitaev2006} $\exp(2\pi\imth h_q)$, the Berry phase obtained by adiabatically rotating a quasiparticle $q$ by $2\pi$, is an important character of a 2+1-D topological order. In Abelian topological orders, the topological spin $\exp(2\pi\imth h_{q_\bsg})$ has a one-to-one correspondence to the self-statistics (\ref{new qp:z2 spin liquid:self}) of a quasiparticle in unit of $2\pi$:
\bea\label{new qp:z2 spin liquid:spin}
&h_{q_\bsg}=\frac{\theta_{q_\bsg}}{2\pi}=\frac{i_1i_2+2i_4}{8}+\frac{i_1l_1+i_2l_2}4+\frac{l_1l_2+l_3+i_4l_4}2.
\eea
All these statistical properties are summarized in TABLE \ref{tab:Z2SL:z2:conventional}.

For the ``unconventional'' SET phases, \eg in our case with $G_s=Z_2$, the $Z_2$ symmetry would exchange quasiparticles that belong to different superselection sectors in Abelian topological order ${\bf K}$. Gauging this kind of $Z_2$ symmetry will in general lead to $U(1)^N\rtimes Z_2$ Chern-Simons theory\cite{Barkeshli2010b}, which describes non-Abelian topological orders in relation to $Z_2$ orbifold conformal field theory\cite{Dijkgraaf1989,Moore1989}.

\section{Classifying double semion theory with onsite $Z_2$ symmetry}\label{app:double semion}

Double semion theory\cite{Freedman2004,Levin2005} is a ``twisted'' $Z_2$ gauge theory in 2+1-D, with Abelian topological order described by ${\bf K}=\bpm2&0\\0&-2\epm$. Due to the presence of nontrivial bosonic $Z_2$-SPT phase in 2+1-D, again here we use a $4\times4$ matrix
\bea\label{K mat:double semion}
{\bf K}=\bpm2&0\\0&-2\epm\oplus\bpm0&1\\1&0\epm=\bpm2&0&0&0\\0&-2&0&0\\0&0&0&1\\0&0&1&0\epm.
\eea
to capture all the different $Z_2$-symmetry-enriched double semion theory. Such a theory has the following quasiparticle contents in its spectra
\bea\label{qp contents:double semion}
&1\simeq\bpm0\\0\\0\\0\epm\simeq\bpm2\\0\\0\\0\epm\simeq\bpm0\\2\\0\\0\epm\simeq\bpm0\\0\\1\\0\epm\simeq\bpm0\\0\\0\\1\epm,\\
&\notag s\simeq\bpm1\\0\\0\\0\epm,~~\bar{s}\simeq\bpm0\\1\\0\\0\epm,~~b\simeq\bpm1\\1\\0\\0\epm.
\eea
where $s$ and $\bar{s}$ represents semion and anti-semion respectively, and $b$ is the bound state of a semion and an anti-semion. $b$ has bosonic (self) statistics (\ref{statistics:self}) but mutual semion(anti-semion) statistics with $s$($\bar{s}$). Here $\{1,s,\bar{s},b\}$ represent the 4 superselection sectors of double semion theory. Any two quasiparticles differing by a local excitation $\simeq0$ belong to the same superselection sector.

Now let's consider the implementation of unitary $G_s=Z_2$ symmetry on double semion theory. We have group compatibility condition (\ref{constraint: sym transf Z2}) for symmetry transformation (\ref{sym transf}) on quasiparticles:
\bea
&\label{condition:double semion:z2}\big({\bf W}^\bsg\big)^2=1_{4\times 4},\\
&\notag\bpm2&0&0&0\\0&-2&0&0\\0&0&0&1\\0&0&1&0\epm=\big({\bf W}^\bsg\big)^T\bpm2&0&0&0\\0&-2&0&0\\0&0&0&1\\0&0&1&0\epm{\bf W}^\bsg,\\
&\notag\big(1_{4\times 4}+{\bf W}^\bsg\big)\delta\vec{\phi}^\bsg=\pi\bpm1&0&0&0\\0&-1&0&0\\0&0&0&2\\0&0&2&0\epm{\bf n},~~~{\bf n}\in\mbz^4.
\eea
We consider\footnote{We believe there is no ``unconventional'' implementation of onsite $Z_2$ symmetry in the double semion theory. This is because here the three types of anyons $s,\bar{s},b$ have different statistical angles. As a result exchanging any two of them shouldn't be a symmetry of the system.} the solutions to (\ref{condition:double semion:z2}) with ${\bf W}^\bsg=1_{4\times4}$. Due to Criterion I, naively there are 8 distinct solutions to to (\ref{condition:double semion:z2}): $\delta\vec\phi^\bsg=\pi(i_1/2,i_2/2,1,i_4)^T$ with $i_{1,2,4}=0,1$. However a careful analysis reveals the following gauge equivalence between certain solutions:
\bea
&\delta\vec\phi^\bsg_{(1)}=\bpm\pi/2\\0\\ \pi\\0\epm\simeq\bpm\pi/2\\0\\ \pi\\ \pi\epm\simeq\bpm-\pi/2\\ \pi\\ \pi\\ \pi\epm={\bf X}^{-1}\delta\vec\phi^\bsg_{(1)}\notag,\\
&\delta\vec\phi^\bsg_{(2)}=\bpm0\\ \pi/2\\ \pi\\0\epm\simeq\bpm0\\ \pi/2\\ \pi\\ \pi\epm\simeq\bpm\pi\\-\pi/2\\ \pi\\ \pi\epm={\bf X}^{-1}\delta\vec\phi^\bsg_{(2)}\notag,\\
&\notag{\bf X}=\bpm1&0&1&0\\0&1&-1&0\\0&0&1&0\\-2&-2&0&1\epm,~~~{\bf X}^T{\bf K}{\bf X}={\bf K}.
\eea
As a result there are only 6 gauge inequivalent solutions of $\delta\vec\phi^\bsg$ to (\ref{condition:double semion:z2}), as summarized in TABLE \ref{tab:double semion:z2:conventional}.

Following Appendix \ref{app:gauging}, we briefly discuss the consequence of gauging the unitary $Z_2$ symmetry in the double semion theory. Since the symmetry transformation is a $U(1)$ phase shift $\delta\vec\phi^\bsg=\pi(i_1/2,i_2/2,1,i_4)^T$ as shown in TABLE \ref{tab:double semion:z2:conventional}, the quasiparticle content in the new topological order obtained by gauging $Z_2$ symmetry is expanded by gauge charge vector:
\bea\notag
&{\bf l}^\prime={\bf M}{\bf l};~~~~{\bf M}=\bpm i_1/2&1&0&0\\-i_2/2&0&1&0\\i_4/2&0&0&1\\1/2&0&0&0\epm,~~{\bf l}\in\mbz^4.
\eea
and therefore
\bea
&\notag{\bf K}_g^{-1}={\bf M}^T{\bf K}^{-1}{\bf M}=\bpm\frac{i_1^2-i_2^2+4i_4}{8}& i_1/4&i_2/4&1/2\\i_1/4&1/2&0&0\\i_2/4&0&-1/2&0\\1/2&0&0&0\epm,\\
&\label{K mat:double semion:gauing z2 sym}
\Rightarrow{\bf K}_g=\bpm0&0&0&2\\0&2&0&-i_1\\0&0&-2&i_2\\2&-i_1&i_2&-2i_4\epm.
\eea
Take $\#5$ as an examples with $\delta\vec\phi^\bsg=\pi(1/2,1/2,1,0)^T$ (or $i_1=i_2=1,i_4=0$), the Abelian topological order obtained by gauging $Z_2$ symmetry is
\bea\notag
&{\bf K}_g\simeq{\bf X}^T{\bf K}_g{\bf X}=\bpm 0&4&0&0\\4&0&0&0\\0&0&0&1\\0&0&1&0\epm\simeq\bpm 0&4\\4&0\epm,\\
&\notag{\bf X}=\bpm1&0&0&0\\1&1&0&0\\-1&1&1&0\\-2&2&1&1\epm\in GL(4,\mbz).
\eea
Notice that
\bea\notag
&\bpm8&0\\0&-2\epm\simeq\bpm0&4\\4&6\epm,~~~\bpm2&0\\0&-8\epm\simeq\bpm0&4\\4&2\epm.
\eea
%therefore $\#3,\#5,\#7$ SET phases are all gauged into a $Z_4$ topological order.
Again from (\ref{statistics:self})-(\ref{statistics:mutual}) we can obtain the (self) statistics of $\bsg$ symmetry flux $q_\bsg=\frac{{\bf K}\delta\vec\phi^\bsg}{2\pi}+{\bf l},~{\bf l}\in\mbz^N$ as
\bea
\notag&\theta_{q_\bsg}=\pi q_\bsg^T{\bf K}^{-1}q_{\bsg}=\pi\big(\frac{\delta\vec\phi^\bsg}{2\pi}\big)^T{\bf K}\frac{\delta\vec\phi^\bsg}{2\pi}+\pi{\bf l}^T{\bf K}^{-1}{\bf l}+{\bf l}^T\delta\vec\phi^\bsg\\
&=\label{new qp:double semion:self}\pi(\frac{i_1^2-i_2^2}8+\frac{l_1^2-l_2^2+i_4+i_1l_1+i_2l_2}2+i_4l_4+l_3).
\eea
Its mutual statistics with original quasiparticles $s\equiv(1,2s_2,s_3,s_4)$ and $\bar s\equiv(2\bar s_1,1,\bar s_3,\bar s_4)$ are
\bea
&\notag\tilde\theta_{q_\bsg,s}=\pi(\frac{i_1}{2}+l_1+i_2s_2+s_3+i_4s_4),\\
&\tilde\theta_{q_\bsg,\bar{s}}=\pi(\frac{i_2}2+i_1\bar s_1-l_2+\bar s_3+i_4\bar s_4).\label{new qp:double semion:mutual}
\eea
The topological spin of new quasiparticle $q_\bsg$ is given by $\Theta_{q_\bsg}\equiv\exp(2\pi\imth h_{q_\bsg})$ where
\bea\label{new qp:double semion:spin}
&h_{q_\bsg}=\frac{\theta_{q_\bsg}}{2\pi}=\frac{i_1^2-i_2^2}{16}+\frac{l_1^2-l_2^2+i_4+i_1l_1+i_2l_2}4+\frac{i_4l_4+l_3}2.
\eea

Unlike others, for SET phases $\#3,\#4,\#6$ it's not easy to find a $GL(4,\mbz)$ transformation (\ref{GL(N,Z) transformation}) on ${\bf K}_g$ matrix (\ref{K mat:double semion:gauing z2 sym}) to reduce it to a simpler form. \eg one can only show for SET phase $\#3$
\bea
&{\bf K}_g\simeq{\bf X}^T{\bf K}_g{\bf X}=\bpm -2&0&0&0\\0&2&1&0\\0&1&2&2\\0&0&2&0\epm\notag,\\
&\notag{\bf X}=\bpm0&1&1&1\\0&1&0&0\\1&0&0&0\\0&0&1&0\epm\in GL(4,\mbz).
\eea
and for SET phase $\#4$
\bea
&{\bf K}_g\simeq{\bf X}^T{\bf K}_g{\bf X}=\bpm 2&0&0&0\\0&-2&1&0\\0&1&2&2\\0&0&2&0\epm\notag,\\
&\notag{\bf X}=\bpm0&0&1&1\\0&1&0&0\\1&0&0&0\\0&0&1&0\epm\in GL(4,\mbz).
\eea
However, a one-to-one correspondence between the quasiparticle contents of two seemingly different ${\bf K}_g$ matrices can be established. For example, there are 16 different superselection sectors (or 16 quasiparticle types) for Abelian topological order ${\bf K}_g$ in (\ref{K mat:double semion:gauing z2 sym}), obtained by gauging $Z_2$ symmetry in $\#3$ SET phase ($i_1=1=i_4,i_2=0$):
\bea\notag
&\bpm\gamma_1+4\gamma_2\\0\\ \gamma_1+\gamma_2\\0\epm~\text{in (\ref{K mat:double semion:gauing z2 sym})}\Leftrightarrow\bpm\gamma_1\\ \gamma_2\epm~\text{in}~\bpm8&0\\0&-2\epm,\\
&\notag\gamma_1=0,1,\cdots,7,~~~\gamma_2=0,1.
\eea
For $\#4$ SET phase ($i_2=1=i_4,i_1=0$):
\bea\notag
&\bpm4\gamma_1+3\gamma_2\\ \gamma_1+\gamma_2\\0\\0\epm~\text{in (\ref{K mat:double semion:gauing z2 sym})}\Leftrightarrow\bpm\gamma_1\\ \gamma_2\epm~\text{in}~\bpm2&0\\0&-8\epm,\\
&\notag\gamma_2=0,1,\cdots,7,~~~\gamma_1=0,1.
\eea

The Abelian topological order obtained by gauging $Z_2$ symmetry in $\#6$ SET phase ($i_2=i_4=i_1=1$) is characterized by the following matrix
\bea
&{\bf K}_g\simeq{\bf X}^T{\bf K}_g{\bf X}=\bpm0&2&0&0\\2&2&1&0\\0&1&2&2\\0&0&2&0\epm,\\
&\notag{\bf X}=\bpm0&0&1&1\\-1&-1&0&0\\-1&0&0&0\\0&0&1&0\epm,~~~\det{\bf X}=1.
\eea
And it has 16 different types of quasiparticles:
\bea\notag
&\vec\gamma\equiv\bpm\gamma_1\\ \gamma_2\\0\\0\epm~\text{in (\ref{K mat:double semion:gauing z2 sym})},~~~\gamma_{1,2}=0,1,2,3.
\eea
Among them four has bosonic self statistics ($\theta=0\mod2\pi$), six with semionic statistics ($\theta=\frac{\pi}2\mod2\pi$) and the other six with anti-semionic statistics ($\theta=-\frac{\pi}2\mod2\pi$). In the above basis, the $16\times16$ modular $\mathcal{S}$-matrix\cite{Kitaev2006,Nayak2008} of this Abelian topological order is given by
\bea
\mathcal{S}_{\vec\gamma,\vec\gamma^\prime}=\frac14\exp\Big[\frac{\pi\imth}{2}(2\sum_{a=1}^2\gamma_a\gamma_a^\prime+\gamma_1\gamma_2^\prime+\gamma_2\gamma_1^\prime)\Big].
\eea

\section{Discussions on $Z_2$-symmetry-enriched $Z_2$ gauge theories}\label{COMPARISON}

\begin{table*}[tb]
\centering
\begin{ruledtabular}
\begin{tabular}{ |c||c|c|c|c|c|c|}
\hline
\multicolumn{7}{|c|}{${\bf K}\simeq\bpm2&0\\0&-2\epm$ with unitary symmetry $G_s=Z_2=\{\bsg,\bse=\bsg^2\}$} \\
\hline
\multicolumn{7}{|c|}{Data set in (\ref{data of a SET phase}): $[{\bf K}=\bpm2&0\\0&-2\epm\oplus\bpm0&1\\1&0\epm,\{\eta^\bsg=+1,{\bf W}^\bsg=1_{4\times4},\delta\vec\phi^\bsg\}]$}\\
\hline
Label&$\#1$&$\#2$&$\#3$&$\#4$&$\#5$&$\#6$\\
\hline$\delta\vec\phi^\bsg$&$\bpm0\\0\\ \pi\\0\epm$&$\bpm0\\0\\ \pi\\ \pi\epm$&$\bpm\pi/2\\0\\ \pi\\0\epm\simeq\bpm\pi/2\\0\\ \pi\\ \pi\epm$&
$\bpm0\\ \pi/2\\ \pi\\0\epm\simeq\bpm0\\ \pi/2\\ \pi\\ \pi\epm$&$\bpm\pi/2\\ \pi/2\\ \pi\\0\epm$&$\bpm\pi/2\\ \pi/2\\ \pi\\ \pi\epm$\\
\hline Proj. Sym. ($s$)&No&No&Yes&No&Yes&Yes\\
\hline Proj. Sym. ($\bar{s}$)&No&No&No&Yes&Yes&Yes\\
\hline Proj. Sym. ($b$)&No&No&Yes&Yes&No&No\\
\hline Symmetry protected edge&No&No&Yes&Yes&No&Yes\\
\hline Central charge $c$&0&0&1&1&0&1\\
\hline \color{blue} After gauging symmetry $\bsg$&&&&&& \\${\bf K}_g\simeq$&$\bpm2&0&0&0\\0&-2&0&0\\0&0&0&2\\0&0&2&0\epm$&
$\bpm2&0&0&0\\0&-2&0&0\\0&0&2&0\\0&0&0&-2\epm$&
$\bpm8&0\\0&-2\epm$&
%$\bpm8&0\\0&-2\epm$&
$\bpm2&0\\0&-8\epm$&
%$\bpm2&0\\0&-8\epm$&
$\bpm0&4\\4&0\epm$&
$\bpm0&2&0&0\\2&2&1&0\\0&1&2&2\\0&0&2&0\epm$\\
\hline  \color{blue}$\theta_{q_\bsg}/2\pi\equiv h_{q_\bsg}\mod1$&$0,\pm\frac14,\frac12$&$0,\pm\frac14,\frac12$&$\frac1{16},-\frac3{16},\frac{5}{16},-\frac7{16}$&$-\frac1{16},\frac3{16},-\frac{5}{16},\frac7{16}$&$0,\frac12$&$\pm\frac14$\\
\hline  \color{blue} $\tilde\theta_{q_\bsg,s}/2\pi\mod1$&$0,1/2$&$0,1/2$&$\pm1/4$&$0,1/2$&$\pm1/4$&$\pm1/4$\\
\hline  \color{blue} $\tilde\theta_{q_\bsg,\bar{s}}/2\pi\mod1$&$0,1/2$&$0,1/2$&$0,1/2$&$\pm1/4$&$\pm1/4$&$\pm1/4$\\
%\hline  \color{blue} $\tilde\theta_{q_\bsg,b}\mod2\pi$&$0$&$0$&$\pi/2$&$\pi/2$&$\pi$&$\pi$\\
\hline \color{green} Notation in \Ref{Hung2013}&$(001)$&$(101)$&$m_1=3$&$m_1=1$&$(011)$&$(111)$\\
\hline
 \end{tabular}
\caption{Classification of double semion theory (\ref{K mat:double semion}) enriched by onsite (unitary) $G_s=Z_2$ symmetry, see Appendix \ref{app:double semion} for details. There are 6 different ``conventional'' SET phases, where under $Z_2$ symmetry all quasiparticles ($s,\bar{s},b$) merely obtain a $U(1)$ phase factor. The data set in the 2nd line completely characterizes these SET phases. ${\bf K}_g$ denotes the topological order, which is obtained by gauging the unitary $G_s=Z_2$ symmetry in the double semion theory. Some of these SET phases have $Z_2$ symmetry protected edge states, which will be gapless unless $Z_2$ symmetry is spontaneously broken. On gauging the $Z_2$ symmetry (blue entries) new quasiparticles $\{q_\bsg\}$ (coined ``$\bsg$ symmetry fluxes'') are obtained, as described in Appendix \ref{app:gauging}. Its statistics (\ref{new qp:double semion:self})-(\ref{new qp:double semion:mutual}) are also summarized in the table: its self statistics $\theta_{q_\bsg}=2\pi h_{q_\bsg}$ has a one-to-one correspondence with its topological spin $\Theta_{q_\bsg}\equiv\exp(2\pi\imth h_{q_\bsg})$. Note, there are no ``unconventional'' symmetry realizations of onsite $Z_2$ symmetry for this topological order.}
\label{tab:double semion:z2:conventional}
\end{ruledtabular}
\end{table*}

First we discuss the results for ``conventional'' SET phases with onsite $Z_2$ symmetry, and their relation to Dijkgraaf-Witten gauge theories\cite{Dijkgraaf1990} in 2+1-D. For $Z_2$ spin liquids ${\bf K}\simeq\bpm0&2\\2&0\epm$, with on-site $Z_2$ symmetry we obtain 4 different conventional $Z_2$-SET phases as summarized in TABLE \ref{tab:Z2SL:z2:conventional}. For double semion theory ${\bf K}\simeq\bpm2&0\\0&-2\epm$, with on-site $Z_2$ symmetry we obtain 6 different conventional $Z_2$-SET phases as summarized in TABLE \ref{tab:double semion:z2:conventional}. Here we make some connection between these SET phases and $Z_2$-symmetry-enriched $Z_2$ gauge theories obtained in previous studies\cite{Mesaros2013,Hung2013}.

In \Ref{Mesaros2013} the exact soluble lattice models for 8 different $G_s=Z_2$-symmetry-enriched $G_g=Z_2$ gauge theories are obtained. They correspond to group cohomology $\mathcal{H}^3(G_s\times G_g,U(1))=\mathcal{H}^3(Z_2\times Z_2,U(1))=\mbz_2^3$. Among these 8 different SET phases, 4 comes from $Z_2$ spin liquids with on-site $G_s=Z_2$ symmetry, and the other 4 from double semion theory with onsite $G_s=Z_2$ symmetry. They are nothing but $\#1,\#2,\#3$ in TABLE \ref{tab:Z2SL:z2:conventional}, together with $\#1,\#2,\#5,\#6$ in TABLE \ref{tab:double semion:z2:conventional}. In fact two different models constructed in \Ref{Mesaros2013}, labeled by $(010)$ and $(110)$ in TABLE \ref{tab:Z2SL:z2:conventional}, belong to the same SET phase ($\#3$ in TABLE \ref{tab:Z2SL:z2:conventional}).

On the other hand \Ref{Hung2013} claimed twelve different $Z_2$-symmetry-enriched $Z_2$ topological orders, among which six are $Z_2$ spin liquids and the others are double semion theories. It was conjectured that different $G_s$-symmetry-enriched $G_g$-gauge theory (with gauge group $G_g$) are classified by group cohomology $\mathcal{H}^{d+1}(G,U(1))$ or Dijkgraaf-Witten $G$-gauge theory\cite{Dijkgraaf1990} in $d$-spatial dimensions, where $G$ is an extension of symmetry group $G_s$ by gauge group $G_g$ (in other words $G/G_s=G_g$). When $G_s=G_g=Z_2$ we have $G=Z_2\times Z_2$ or $G=Z_4$. The number $12=2^3+4$ is associated with $\mathcal{H}^3(Z_2\times Z_2,U(1))\oplus\mathcal{H}^3(Z_4,U(1))=\mbz_2^3\oplus\mbz_4$. The proposed 8 different  SET phases from $\mathcal{H}^3(Z_2\times Z_2,U(1))$ are the same as those in \Ref{Mesaros2013}, which are discussed earlier. After gauging the $G_s=Z_2$ symmetry, these 8 different SET phases lead to Abelian topological orders described by a $4\times4$ matrix\cite{Hung2013}
\bea
&{\bf K}(n_1n_2n_3)=\bpm2n_1&2&n_2&0\\2&0&0&0\\n_2&0&2n_3&2\\0&0&2&0\epm\label{kmat:spt:Z2xZ2}
\eea
where $n_1,n_2,n_3=0,1$. It's not difficult to check that these 8 different SET phases labeled by $(n_1n_2n_3)$ have the following correspondence with our results:
\bea
&\notag{\bf K}(000)=\bpm0&2&0&0\\2&0&0&0\\0&0&0&2\\0&0&2&0\epm\Leftrightarrow\#1~\text{in TABLE \ref{tab:Z2SL:z2:conventional}},\\
&\notag{\bf K}(100)\simeq\bpm0&2&0&0\\2&0&0&0\\0&0&2&0\\0&0&0&-2\epm\Leftrightarrow\#2~\text{in TABLE \ref{tab:Z2SL:z2:conventional}},\\
&\notag{\bf K}(010)\simeq\bpm0&4\\4&0\epm\Leftrightarrow\#3~\text{in TABLE \ref{tab:Z2SL:z2:conventional}},\\
&\notag{\bf K}(110)\simeq\bpm0&4\\4&0\epm\Leftrightarrow\#3~\text{in TABLE \ref{tab:Z2SL:z2:conventional}}.
\eea
and
\bea
&\notag{\bf K}(001)\simeq \bpm2&0&0&0\\0&-2&0&0\\0&0&0&2\\0&0&2&0\epm\Leftrightarrow\#1~\text{in TABLE \ref{tab:double semion:z2:conventional}},\\
&\notag{\bf K}(101)\simeq \bpm2&0&0&0\\0&-2&0&0\\0&0&2&0\\0&0&0&-2\epm\Leftrightarrow\#2~\text{in TABLE \ref{tab:double semion:z2:conventional}},\\
&\notag{\bf K}(011)\simeq\bpm0&4\\4&0\epm\Leftrightarrow\#5~\text{in TABLE \ref{tab:double semion:z2:conventional}},\\
&\notag{\bf K}(111)\simeq\bpm0&2&0&0\\2&2&1&0\\0&1&2&2\\0&0&2&0\epm\Leftrightarrow\#6~\text{in TABLE \ref{tab:double semion:z2:conventional}}.
\eea
The other 4 SET phases proposed in \Ref{Hung2013} are associated to group cohomology $\mathcal{H}^3(Z_4,U(1))=\mbz_4$. \Ref{Hung2013} asserted that after gauging the $Z_2$ symmetry they lead to Abelian $Z_4$ topological orders described by
\bea
\notag{\bf K}(m_1)=\bpm2m_1&4\\4&0\epm,~~~m=0,1,2,3.
\eea
We found that these 4 different SET phases have overlap with the previous 7 SET phases associated to $\mathcal{H}^3(Z_2\times Z_2,U(1))=\mbz_2^3$: they turn out to be
\bea
&\notag{\bf K}(m_1=0)=\bpm0&4\\4&0\epm\Leftrightarrow\#3~\text{in TABLE \ref{tab:Z2SL:z2:conventional}},\\
&\notag{\bf K}(m_1=2)\simeq\bpm4&0\\0&-4\epm\Leftrightarrow\#4~\text{in TABLE \ref{tab:Z2SL:z2:conventional}}.
\eea
and
\bea
&\notag{\bf K}(m_1=3)\simeq\bpm8&0\\0&-2\epm\Leftrightarrow\#3~\text{in TABLE \ref{tab:double semion:z2:conventional}},\\
&\notag{\bf K}(m_1=1)\simeq\bpm2&0\\0&-8\epm\Leftrightarrow\#4~\text{in TABLE \ref{tab:double semion:z2:conventional}}.
\eea

We want to emphasize that when the on-site unitary $Z_2$ symmetry is gauged, a $Z_2$ spin liquid and another double semion theory could result in the same ``gauged'' topological order. For $Z_2$ spin liquid $\#2$ in TABLE \ref{tab:Z2SL:z2:conventional} and double semion theory $\#1$ in TABLE \ref{tab:double semion:z2:conventional}, it's straightforward to see ${\bf K}(100)\simeq{\bf K}(001)$. Similarly for $Z_2$ spin liquid $\#3$ in TABLE \ref{tab:Z2SL:z2:conventional} and double semion theory $\#5$ in TABLE \ref{tab:double semion:z2:conventional}, with ${\bf X}_{1,2}\in GL(4,\mbz)$ we have
\bea
&\notag{\bf X}_1^T\cdot{\bf K}(010)\cdot{\bf X}_1={\bf X}_2^T\cdot{\bf K}(110)\cdot{\bf X}_2=\bpm0&4&0&0\\4&0&0&0\\0&0&0&1\\0&0&1&0\epm,\\
&{\bf X}_1=\bpm2&0&0&-1\\0&1&0&0\\0&-2&-1&0\\-1&0&0&0\epm,~~{\bf X}_2=\bpm2&0&0&-1\\-1&1&0&0\\-2&-2&-1&1\\-1&0&0&0\epm.\notag
\eea
and one can further show ${\bf K}(110)\simeq{\bf K}(011)$. In fact the 8 different ${\bf K}$ matrices ${\bf K}(n_1n_2n_3)$ describe only 5 different Abelian topological orders. Since these 8 theories correspond to different Dijkgraaf-Witten \cite{Dijkgraaf1990} theories, \ie gauge theories with different topological terms specified by $\mathcal{H}^3(Z_2\times Z_2,U(1))=Z_2^3$, this also implies that \emph{different Dijkgraaf-Witten theories based on a particular gauge group can share the same topological order, and correspond to the same SET phase}. Further information regarding which particles comprise electric charges and magnetic vortices is required to uniquely define those phases.

\section{Vertex algebra approach to gauge a unitary symmetry}\label{app:vertex algebra}

The Chern-Simons approach to gauge a unitary symmetry, introduced in Appendix \ref{app:gauging}, applies to all cases where we obtain an Abelian topological order after gauging the symmetry. Thus for many ``conventional'' SET phases we can gauge its unitary symmetry and obtain an Abelian topological order in the Chern-Simons approach. For ``unconventional'' SET phases (and certain ``conventional'' ones \eg in section \ref{example:z2sl:Z2xZ2}), such as those summarized in TABLE \ref{tab:Z2SL:z2:unconventional}, gauging a unitary (\eg $Z_2$) symmetry will result in non-Abelian topological orders. In the case $G_s=Z_2$ as discussed in this work, these non-Abelian topological orders are described by $U(1)^N\rtimes Z_2$ Chern-Simons theory\cite{Barkeshli2010b}. In these ``unconventional'' cases the Chern-Simons approach introduced previously is not enough. In order to obtain the full structure (such as topological spin $\exp(2\pi\imth h)$ of quasiparticles and modular $\mathcal{S}$ matrix associated with quasiparticle statistics) of these non-Abelian topological orders, here we introduce a vertex algebra approach to gauge the unitary symmetry. It applies to both the ``conventional'' and ``unconventional'' SET phases and in the following we'll demonstrate its power by two examples: ``conventional'' and ``unconventional'' $Z_2$-symmetry-enriched $Z_2$ spin liquids.

\subsection{The vertex algebra formalism, and application to ``conventional'' SET phases}

The vertex algebra approach\cite{Moore1991,Wen1994,Lu2010} is based on the close connection\cite{Witten1989,Moore1989} between the bulk topological order (described by 2+1-D topological field theory) and its boundary excitations (described by 1+1-D conformal field theory) in two spatial dimensions. Let's take $Z_2$ spin liquid (\ref{K mat:Z2 spin liquid}) for an example. The edge effective theory (\ref{edge:right}) contains two branches of chiral bosons $\{\phi_{1,2}\}$, which could be reformulated by a $c=1$ $U(1)\times U(1)$ Gaussian model with a holomorphic and anti-holomorphic part:
\bea
&\notag\varphi(x+\imth\tau)\equiv\varphi(z)=\phi_1(x,t)+\phi_2(x,t),\\
&\bar\varphi(x-\imth\tau)\equiv\bar\varphi(\bar{z})=\phi_1(x,t)-\phi_2(x,t).\notag
\eea
The Gaussian model has Lagrangian density $\mathcal{L}_{Gaussian}=\frac1{2\pi}\partial\varphi(z)\bar\partial\varphi(z)=\frac1{8\pi}|\vec\nabla\varphi|^2$, yielding the following correlation function
\bea\label{OPE:gaussian}
&\langle\varphi(z)\varphi(w)\rangle=-\ln(z-w).
\eea
and $\langle\bar\varphi(\bar z)\bar\varphi(\bar w)\rangle=-\ln(\bar z-\bar w)$.
%Similarly for the anti-holomorphic part we have
%\bea
%&\langle\bar\varphi(\bar{z})\bar\varphi(\bar{w})\rangle=-\ln(\bar{z}-\bar{w}).
%\eea
The free boson field $\varphi$ has compactification radius $R=2$ for $Z_2$ spin liquid (\ref{K mat:Z2 spin liquid}) so that periodicity $\varphi\sim\varphi+2\pi R$ holds. In general for ${\bf K}=\bpm0&N\\N&0\epm$ in (\ref{edge:right}) the associated compactification radius of scalar boson
\bea
&\notag\sqrt{\frac2N}\varphi(z)=\phi_1(x,t)+\phi_2(x,t),\\
&\sqrt{\frac2N}\bar\varphi(\bar{z})=\phi_1(x,t)-\phi_2(x,t).\label{U(1)xU(1) Gaussian model<->chiral boson}
\eea
is $R=\sqrt{2N}$. The allowed physical excitations must be compatible with $2\pi R$ periodicity of bosons and they are\cite{Barkeshli2012}
\bea\label{vertex operator}
V_k(z)=e^{\imth k\varphi(z)/\sqrt{2N}},~~~k=0,1,\cdots,2N-1.
\eea
for holomorphic part (and similarly $\bar{V}_k$ for anti-holomorphic part). These $2N$ vertex operators are primary fields of the holomorphic $U(1)$ conformal field theory (CFT) and they form different representations of the conformal algebra. From (\ref{OPE:gaussian}) one can see they have the following (radial-ordered) operator product expansion\cite{Ginsparg1989,Francesco1997} (OPE):
\bea\label{OPE:vertex operator}
e^{\imth\alpha\varphi(z)}e^{\imth\beta\varphi(w)}=(z-w)^{\alpha\beta}e^{\imth(\alpha+\beta)\varphi(w)}+\cdots
\eea
for $\alpha+\beta\neq0$. There is an energy-momentum tensor $T=-\frac12(\partial\varphi)^2$ which generates the conformal transformation of the vertex algebra, so that any primary field $P(z)$ has the following OPE with energy-momentum tensor
\bea
T(z)P(w)=\frac{h_P}{(z-w)^2}P(w)+\frac{1}{z-w}\partial P(w)+\cdots
\eea
where $h_P$ is the scaling dimension of primary field $P$. Apparently the vertex operator $\exp\big[\imth\alpha\varphi(z)\big]$ has scaling dimension $h_{\alpha}=\frac12\alpha^2$. Another primary field is the current operator $j(z)\equiv\imth\partial\varphi(z)$ which have scaling dimension $h_j=1$. And we have
\bea
e^{\imth\alpha\varphi(z)}e^{-\imth\alpha\varphi(w)}=\frac{1}{(z-w)^{\alpha^2}}+\frac{\alpha j(w)}{(z-w)^{\alpha^2-1}}+\cdots.\notag
\eea
These OPEs imply the following \emph{fusion rules} of primary fields
\bea
&\notag e^{\imth\alpha\varphi}\times e^{\imth\alpha\varphi}=1+j,~~~(\alpha\neq0)\\
&e^{\imth\alpha\varphi}\times e^{\imth\beta\varphi}=e^{\imth(\alpha+\beta)\varphi},~~~(\alpha\neq-\beta).\notag
\eea
Similar results hold for anti-holomorphic $\bar\varphi(\bar{z})$ part, only that all scaling dimensions changes sign for their anti-holomorphic counterparts.

A natural question is among all these primary fields, which ones appear in the physical edge spectra of the topologically ordered phase? There are a few physical principles to follow. First of all every physical edge excitation ($e^{\imth\sum_il_i\phi_i},~l_i\in\mbz$) must be a primary field. Secondly, there are electron operators (or the microscopic local degrees of freedom $e^{\imth\sum_{i,j}l_i{\bf K}_{i,j}\phi_j},~l_i\in\mbz$) which is local with respect to all other edge excitations. In the context of vertex algebra, two operators $A$ and $B$ are local w.r.t. each other if and only if in OPE
\bea
&A(z)B(w)=\frac{f_{A,B}^C C(w)}{(z-w)^{\alpha_{A,B}}}+O\Big((z-w)^{1-\alpha_{A,B}}\Big),\label{OPE:A and B}\\
&\notag\alpha_{A,B}\in\mbz.
\eea
where $C$ is also a primary field and $f_{A,B}^C$ is a structure constant. For example before we gauge the $Z_2$ symmetry, for Abelian topological order ${\bf K}=\bpm0&N\\N&0\epm$ the electron operator is
\bea
&\notag e^{\imth N(l_1\phi_1+l_2\phi_2)}=\exp\Big[{\imth}\sqrt{\frac N2}\big((l_1+l_2)\varphi(z)\\
&+(l_1-l_2)\bar\varphi(\bar{z})\big)\Big],~~~~~~l_{1,2}\in\mbz.\label{electron:Zn gauge theory}
\eea
It's straightforward to check that all allowed quasiparticles (local w.r.t. the above electron operator) have the following form
\bea
&e^{\imth(l_1\phi_1+l_2\phi_2)}=\exp\Big[\imth\frac{(l_1+l_2)\varphi(z)+(l_1-l_2)\bar\varphi(\bar{z})}{\sqrt{2N}}\Big],~~~l_{1,2}\in\mbz.\notag
\eea
Lastly, any two primary fields differing by an electron operator are regarded as the same (or belong to the same superselection sector).\\

Now let's go back to $Z_2$ spin liquids with ${\bf K}=\bpm0&2\\2&0\epm\oplus\bpm0&1\\1&0\epm$, which have 4 branches of chiral bosons $\{\phi_i,1\leq i\leq4\}$. We can introduce
free bosons $\varphi_1(z),\bar\varphi_1(\bar z)$ for chiral bosons $\phi_{1,2}$ as in (\ref{U(1)xU(1) Gaussian model<->chiral boson}) with $N=2$, and free bosons $\varphi_2(z),\bar\varphi_2(\bar z)$ for chiral bosons $\phi_{3,4}$ as in (\ref{U(1)xU(1) Gaussian model<->chiral boson}) with $N=1$.
In other words we have
\bea
\notag\bpm\phi_1\\ \phi_2\\ \phi_3\\ \phi_4\epm=\bpm1/2&1/2&&\\1/2&-1/2&&\\&&1/\sqrt2&1/\sqrt2\\&&1/\sqrt2&-1/\sqrt2\epm\bpm\varphi_1\\ \bar\varphi_1\\ \varphi_2\\ \bar\varphi_2\epm.
\eea
Before gauging the unitary
$Z_2$ symmetry, the four superselection sectors (or 4 types of different quasiparticles) correspond to
\bea
&\notag1\sim \bar j_1(\bar z)\sim j_1(z)\sim e^{2\imth\varphi_1(z)}\sim e^{2\imth\bar\varphi_1(\bar{z})}\sim e^{\imth\big[\varphi_1(z)\pm\bar\varphi_1(\bar{z})\big]}\\
&\notag\sim \bar j_2(\bar z)\sim j_2(z)\sim e^{\imth\frac{\varphi_2(z)\pm\bar\varphi_2(\bar{z})}{\sqrt2}}\sim e^{\sqrt2\imth\varphi_2(z)}\sim e^{\sqrt2\imth\bar\varphi_2(\bar z)},\\
&e\sim e^{\imth\frac{\varphi_1(z)+\bar\varphi_1(\bar{z})}{2}},~m\sim e^{\imth\frac{\varphi_1(z)-\bar\varphi_1(\bar{z})}{2}},~f\sim e^{\imth\varphi_1(z)}\sim e^{\imth\bar\varphi_1(\bar{z})}.\notag\\
&\label{qp contents:Z2 spin liquid:vertex operator}
\eea
Now after gauging the ``conventional'' $Z_2$ symmetries in TABLE \ref{tab:Z2SL:z2:conventional}, as discussed in Appendix \ref{app:gauging}, a new type of quasiparticles $q_\bsg$ becomes deconfined
excitations:
\bea
&\notag q_\bsg\sim e^{\imth\sum_{I,J}\phi_I{\bf K}_{I,J}\delta\vec\phi^\bsg_J/2\pi}\sim\\
&\exp\Big[\imth\frac{(i_1+i_2)\varphi_1+(i_2-i_1)\bar\varphi_1+\sqrt2(1+i_4)\varphi_2+\sqrt2(i_4-1)\bar\varphi_2}4\Big].\notag
\eea
where $i_{1,2,4}=0,1$ in $\delta\vec\phi^\bsg$. Notice that when such a $Z_2$ symmetry flux $q_\bsg$ is deconfined, we have to modify the previous definition of electron operators $1$ in (\ref{qp contents:Z2 spin liquid:vertex operator}).
The new electron operator is defined as anything that is local w.r.t. quasiparticles $\{e,m,f,q_\bsg\}$. With this new definition for electron operators, we can track down all the
inequivalent quasiparticles (superselection sectors) and obtain the full structure of the topological order obtained by gauging $Z_2$ symmetry. One can easily check this approach indeed reproduces TABLE \ref{tab:Z2SL:z2:conventional},
consistent with the result of Chern-Simons approach.p

In the vertex algebra context, the scaling dimension $h$ of a quasiparticle determines its topological spin $\Theta\equiv\exp(2\pi\imth h)$, the Berry phase obtained by self-rotating
a quasiparticle adiabatically by $2\pi$. On the other hand, the mutual statistics of quasiparticle $A$ and $B$ is given by $\tilde\theta_{A,B}=-2\pi\alpha_{A,B}$ in OPE (\ref{OPE:A and B}).

\begin{table*}[tb]
\centering
\begin{ruledtabular}
\begin{tabular}{ |c|c|c|c|c|c|c|c|c|c|c|}
\hline
\multicolumn{3}{|c|}{``Unconventional'' SET phases}&\multicolumn{2}{|c|}{$\#5$}&\multicolumn{2}{|c|}{$\#6$}&\multicolumn{2}{|c|}{$\#5$}&\multicolumn{2}{|c|}{$\#6$}\\
\hline\multirow{2}{1.6cm}{$Z_2$ orbifold fields}&\multirow{2}{1.4cm}{Quantum dimension}&\multirow{2}{1cm}{Ising$^2$ fields}&\multicolumn{2}{|c|}{$\delta\vec\phi^\bsg=(0,0,\pi,0)^T$}&\multicolumn{2}{|c|}{$\delta\vec\phi^\bsg=(0,0,\pi,\pi)^T$}&\multicolumn{2}{|c|}{$\delta\vec\phi^\bsg=(\pi/2,\pi/2,\pi,0)^T$}&\multicolumn{2}{|c|}{$\delta\vec\phi^\bsg=(\pi/2,\pi/2,\pi,\pi)^T$}\\
\cline{4-11}&&& q.p. $q_a$&$h_{a}$& q.p. $q_a$&$h_a$& q.p. $q_a$&$h_a$& q.p. $q_a$&$h_a$\\
\hline$1\sim e^{2\imth\varphi_1}$&1&$1\otimes1$&1&0&1&0&1&0&1&0\\
\hline$j_1=\imth\partial\varphi_1$&1&$\psi\otimes\psi$&$\bar j_1$&1&$\bar j_1$&1&$\bar j_1$&1&$\bar j_1$&1\\
\hline$f^1\sim\cos\varphi_1$&1&$\psi\otimes1$&$\bar f^1$&$\frac12$&$\bar f^1$&$\frac12$&$\bar f^1$&$\frac12$&$\bar f^1$&$\frac12$\\
\hline$f^2\sim\sin\varphi_1$&1&$1\otimes\psi$&$\bar f^2$&$\frac12$&$\bar f^2$&$\frac12$&$\bar f^2$&$\frac12$&$\bar f^2$&$\frac12$\\
\hline$V_1\sim\cos\frac{\varphi_1}2$&2&$\sigma\otimes\sigma$&$\bar V_1 e^{\imth\varphi_1/2}$&0&$\bar V_1 e^{\imth\varphi_1/2}$&0&$\bar V_1 e^{\imth\varphi_1/2}$&0&$\bar V_1 e^{\imth\varphi_1/2}$&0\\
\hline$\sigma^1$&$\sqrt2$&$\sigma\otimes1$
&$\bal\bar\sigma^1e^{\imth\frac{\varphi_1}2}\cdot \\e^{\imth\frac{\varphi_2-\bar\varphi_2}{2\sqrt2}}\eal$&$\frac1{16}$
&$\bar\sigma^1e^{\imth(\frac{\varphi_1}2+\frac{\varphi_2}{\sqrt2})}$&$\frac5{16}$
&$\bar\sigma^1e^{\imth\frac{\varphi_2-\bar\varphi_2}{2\sqrt2}}$&$\frac{-1}{16}$
&$\bar\sigma^1e^{\imth\frac{\varphi_2}{\sqrt2}}$&$\frac3{16}$\\
\hline$\sigma^2$&$\sqrt2$&$1\otimes\sigma$
&$\bar\sigma^2e^{\imth\frac{\varphi_2-\bar\varphi_2}{2\sqrt2}}$&$\frac{-1}{16}$
&$\bar\sigma^2e^{\imth\frac{\varphi_2}{\sqrt2}}$&$\frac3{16}$
&$\bal\bar\sigma^2e^{\imth\frac{\varphi_1}2}\cdot \\e^{\imth\frac{\varphi_2-\bar\varphi_2}{2\sqrt2}}\eal$&$\frac1{16}$
&$\bar\sigma^2e^{\imth(\frac{\varphi_1}2+\frac{\varphi_2}{\sqrt2})}$&$\frac5{16}$\\
\hline$\tau^1$&$\sqrt2$&$\sigma\otimes\psi$
&$\bal\bar\tau^1e^{\imth\frac{\varphi_1}2}\cdot \\e^{\imth\frac{\varphi_2-\bar\varphi_2}{2\sqrt2}}\eal$&$\frac9{16}$
&$\bar\tau^1e^{\imth(\frac{\varphi_1}2+\frac{\varphi_2}{\sqrt2})}$&$\frac{-3}{16}$
&$\bar\tau^1e^{\imth\frac{\varphi_2-\bar\varphi_2}{2\sqrt2}}$&$\frac{-9}{16}$
&$\bar\tau^1e^{\imth\frac{\varphi_2}{\sqrt2}}$&$\frac{-5}{16}$\\
\hline$\tau^2$&$\sqrt2$&$\psi\otimes\sigma$
&$\bar\tau^2e^{\imth\frac{\varphi_2-\bar\varphi_2}{2\sqrt2}}$&$\frac{-9}{16}$
&$\bar\tau^2e^{\imth\frac{\varphi_2}{\sqrt2}}$&$\frac{-5}{16}$
&$\bal\bar\tau^2e^{\imth\frac{\varphi_1}2}\cdot \\e^{\imth\frac{\varphi_2-\bar\varphi_2}{2\sqrt2}}\eal$&$\frac9{16}$
&$\bar\tau^2e^{\imth(\frac{\varphi_1}2+\frac{\varphi_2}{\sqrt2})}$&$\frac{-3}{16}$\\
\hline
\multicolumn{3}{|c|}{$\begin{aligned}\text{Electron operator}:~1\sim \\e^{2\imth\varphi_1}\sim e^{\sqrt2\imth(\varphi_2\pm\bar\varphi_2)}\sim\end{aligned}$}
&\multicolumn{2}{|c|}{$\bal\bar f^1 e^{\imth\varphi_1}\sim e^{\imth\frac{\varphi_2-\bar\varphi_2}{\sqrt2}}\\ \sim \bar{j}_1e^{\imth\frac{\varphi_2+\bar\varphi_2}{\sqrt2}}\eal$}
&\multicolumn{2}{|c|}{$\bal\bar f^1 e^{\imth\varphi_1}\sim e^{\imth\sqrt{2}\bar\varphi_2}\\ \sim \bar j_1e^{\imth\frac{\varphi_2+\bar\varphi_2}{\sqrt2}}\eal$}
&\multicolumn{2}{|c|}{$\bal\bar f^2 e^{\imth\varphi_1}\sim e^{\imth\frac{\varphi_2-\bar\varphi_2}{\sqrt2}}\\ \sim \bar{j}_1e^{\imth\frac{\varphi_2+\bar\varphi_2}{\sqrt2}}\eal$}
&\multicolumn{2}{|c|}{$\bal\bar f^2 e^{\imth\varphi_1}\sim e^{\imth\sqrt{2}\bar\varphi_2}\\ \sim \bar j_1e^{\imth\frac{\varphi_2+\bar\varphi_2}{\sqrt2}}\eal$}\\
\hline
\multicolumn{3}{|c|}{Relation to Kitaev's 16-fold way\cite{Kitaev2006}}&\multicolumn{2}{|c|}{$(\nu=1)\otimes(\nu=15)$}&\multicolumn{2}{|c|}{$(\nu=5)\otimes(\nu=11)$}
&\multicolumn{2}{|c|}{$(\nu=7)\otimes(\nu=9)$}&\multicolumn{2}{|c|}{$(\nu=3)\otimes(\nu=13)$}\\
\hline
 \end{tabular}
\caption{Quasiparticle (q.p.) contents of non-Abelian topological orders obtained by gauging the $Z_2$ symmetry in ``unconventional'' SET phases as summarized in TABLE \ref{tab:Z2SL:z2:unconventional}. They are related to $Z_2$ orbifold CFT compactified at radius $R=2$. The fusion rules of non-Abelian quasiparticles have a one-to-one correspondence to the two copies of Ising CFT (\ie Ising$^2$ theory). Each quasiparticle $q_a$ correspond to a vertex operator (a primary field) in the vertex algebra (which are CFTs) defined through operator product expansion (OPE), and its (conformal) scaling dimension $h_a~(\mod1)$ physically relates to the topological spin $\exp(2\pi\imth h_a)$ of the quasiparticle. The modular $\mathcal{S}$ matrix of such non-Abelian topological orders is also determined by the OPEs between vertex operators. Allowed quasiparticles must be local w.r.t. any electron operators $\sim1$. Any two quasiparticles differing by an electron operator are considered as the same. The scaling dimensions in the Ising$^2$ CFT (or $Z_2$ orbifold model) are $h_1=0$,~$h_j=1$,~$h_{f^1}=h_{f^2}=1/2$,~$h_{V_1}=1/8$, $h_{\sigma^1}=h_{\sigma^2}=1/16$ and $h_{\tau^1}=h_{\tau^2}=9/16$. We label these 9 different quasiparticles (or superselection sectors) as $q_a,~0\leq a\leq8$, which is shown in the $(a+2)$-th row of this table. All non-Abelian topological orders in this table have 9-fold GSD on a torus, corresponding to 9 different superselection sectors.}
\label{tab:Z2SL:z2:unconventional:vertex algebra}
\end{ruledtabular}
\end{table*}

\subsection{Application to ``unconventional'' SET phases}

For a ``unconventional'' SET phase, \eg where two inequivalent quasiparticles ($e$ and $m$ in $Z_2$ spin liquid) are exchanged under $Z_2$ symmetry operation as summarized in TABLE \ref{tab:Z2SL:z2:unconventional}, a non-Abelian topological order is obtained by gauging the $Z_2$ symmetry. Here we apply the vertex algebra approach to extract the full structure of these non-Abelian topological orders.

First let's review some known results, discussed in detail in \Ref{Barkeshli2010b,Barkeshli2012}. When the ``unconventional'' $Z_2$ symmetry $\{{\bf W}^\bsg=\bpm0&1\\1&0\epm,\delta\vec\phi^\bsg=0\}$ is gauged for Abelian topological order ${\bf K}=\bpm0&N\\N&0\epm$, the resultant topological order is described by $U(1)\times U(1)\rtimes Z_2$ Chern-Simons theory (coined ``twisted'' $Z_N$ gauge theory in \Ref{Barkeshli2012}), which has GSD$=(N^g/2)[N^g+1+(2^{2g}-1)(N^{g-1}+1)]$ on a genus-$g$ Riemann surface. It contains $2N$ different quasiparticles with quantum dimension $d=1$, another $2N$ quasiparticles with $d=\sqrt{N}$ and $N(N-1)/2$ quasiparticles with $d=2$. Under the unconventional $Z_2$ symmetry operation two superselection sectors $e\leftrightarrow m$ exchanges and so does chiral bosons $\phi_1\leftrightarrow\phi_2$. Therefore in the context of vertex algebra (\ref{U(1)xU(1) Gaussian model<->chiral boson}) the anti-holomorphic free boson $\bar\varphi\rightarrow-\bar\varphi$ under $Z_2$ symmetry operation! After this ``unconventional'' $Z_2$ symmetry is gauged for $Z_2$ spin liquids ($N=2$), we obtain an non-Abelian topological order whose quasiparticle content has an antiholomorphic part (from $\bar\varphi$) given by $Z_2$ orbifold CFT\cite{Moore1989,Dijkgraaf1989} with compactification radius $R=2$. It has been shown that $Z_2$ orbifold CFT is equivalent to Ising$\times$Ising (or Ising$^2$) CFT\cite{Dijkgraaf1989}. In each Ising CFT there are 3 different quasiparticles: vacuum (or boson) $1$, fermion $\psi$ and the ``disorder'' field\cite{Zamolodchikov1985a} $\sigma$ with the following fusions rules:
\bea\label{fusion rule:Ising}
&\psi\times\psi=1,~~\psi\times\sigma=\sigma,~~\sigma\times\sigma=1+\psi.
\eea
Both $1$ and $\psi$ have quantum dimension 1 while disorder operator $\sigma$ has quantum dimension $\sqrt2$. Their scaling dimensions are $0,~1/2$ and $1/16$. Therefore the $Z_2$ orbifold CFT, equivalent to the direct product of two copies of Ising CFTs, contains $9=3\times3$ inequivalent quasiparticles (superselection sectors). The quasiparticle contents of the $Z_2$ orbifold CFT are summarized in the first 3 columns of TABLE \ref{tab:Z2SL:z2:unconventional:vertex algebra}.\\

Now let's get back to our cases of $Z_2$ spin liquids with unconventional on-site $Z_2$ symmetry. There are two such SET phases as summarized in TABLE \ref{tab:Z2SL:z2:unconventional}. After gauging the unitary $Z_2$ symmetry, they both lead to non-Abelian topological orders with 9 inequivalent quasiparticles (superselection sectors). In the vertex algebra context, they share the same antiholomorphic ($\bar\varphi_1$) part which gives rise to the non-Abelian quasiparticles. However, their different holomorphic parts discriminates these two SET phases. A key issue in determining the quasiparticle contents is: which quasiparticles are identical (or belong to the same superselection sector), after the symmetry is gauged?

In the vertex algebra context, once we fix the electron operator $1\sim~?$ (or the vacuum/trivial sector) which is local w.r.t. all quasiparticles, the full structure of inequivalent quasiparticles is determined. So the above issue becomes the following question: how to determine the electron operators in the vertex algebra, once we gauge the unitary symmetry? The answer lies in the following physical principle:

\emph{If in the original SET phase, two quasiparticles belong to the same superselection sector (\ie they are equivalent) and transform in the same way under a unitary symmetry, then they belong to the same superselection sector after the unitary symmetry is gauged}.

To be specific, if two quasiparticles $q_A$ and $q_B$ belong the same superselection sector and transform in the same way under unitary symmetry, then after gauging the symmetry, quasiparticle $q_Aq_B^\dagger\sim1$ ($q_B^\dagger$ is the anti-particle of $q_B$) belong to the trivial sector. For instance, in SET phase $\#1$ in TABLE \ref{tab:Z2SL:z2:unconventional} and \ref{tab:Z2SL:z2:unconventional:vertex algebra}, the following two quasiparticles belong the the trivial sector and are both odd under $Z_2$ symmetry $\bsg$:
\bea
\bar j_1\sim e^{\imth\phi_3}=e^{\imth\frac{\varphi_2+\bar\varphi_2}{\sqrt2}}\notag
\eea
and they are both their own anti-particles. Besides the following two fermions also belong the same superselection sector and are both even under $Z_2$ symmetry:
\bea
\bar f^1=\cos(\phi_1-\phi_2)=\cos(\bar\varphi_1)\sim e^{\imth(\phi_1+\phi_2)}=e^{\imth\varphi_1}.\notag
\eea
Both of them are also their own anti-particles. Therefore we have the following definitions of electron operators (or trivial sector) as shown in TABLE \ref{tab:Z2SL:z2:unconventional:vertex algebra}:
\bea
\notag1\sim\bar j_1 e^{\imth\frac{\varphi_2+\bar\varphi_2}{\sqrt2}}\sim\bar f^1e^{\imth\varphi_1}.
\eea
This enables us to obtain all the 9 inequivalent quasiparticles (superselection sectors) as summarized in TABLE \ref{tab:Z2SL:z2:unconventional:vertex algebra}, for the non-Abelian topological order acquired by gauging $Z_2$ symmetry in these SET phases.

As discussed earlier, in the vertex algebra approach, the mutual statistics of two quasiparticles $A$ and $B$ is given in their OPE (\ref{OPE:A and B}) by $\mathcal{S}_{A,B}=\exp(\imth\tilde\theta_{A,B})=\exp(-2\pi\imth\alpha_{A,B})$. If quasiparticles $A$ and $B$ leads to more than one fusion channels, the corresponding entry $\mathcal{S}_{A,B}=0$ vanishes in the modular $\mathcal{S}$ matrix. Besides, scaling dimensions $\{h_q\}$ of quasiparticles $\{q\}$ determine their topological spins $\mathcal{T}_{A,B}=\delta_{A,B}\exp(2\pi\imth h_A)$, which corresponds to the modular $\mathcal{T}$ matrix. So we can extract all the topological properties of the non-Abelian topological orders, obtained by gauging $Z_2$ symmetry in SET phases.

The modular $\mathcal{S}$ matrix in the basis $q_a~(0\leq a\leq8,~\text{see TABLE \ref{tab:Z2SL:z2:unconventional:vertex algebra}})$ of the 9 different quasiparticles (superselection sectors) is $\mathcal{S}_{\#5}=$
\bea
&\notag\frac14\bpm1&1&1&1&2&\sqrt2&\sqrt2&\sqrt2&\sqrt2\\1&1&1&1&2&-\sqrt2&-\sqrt2&-\sqrt2&-\sqrt2\\1&1&1&1&-2&-\sqrt2&\sqrt2&-\sqrt2&\sqrt2
\\1&1&1&1&-2&\sqrt2&-\sqrt2&\sqrt2&-\sqrt2\\2&2&-2&-2&0&0&0&0&0\\ \sqrt2&-\sqrt2&-\sqrt2&\sqrt2&0&0&2&0&-2
\\ \sqrt2&-\sqrt2&\sqrt2&-\sqrt2&0&2&0&-2&0\\ \sqrt2&-\sqrt2&-\sqrt2&\sqrt2&0&0&-2&0&2\\ \sqrt2&-\sqrt2&\sqrt2&-\sqrt2&0&-2&0&2&0
\epm
\eea
for gauged ``unconventional'' SET phase $\#5$. Meanwhile the $\mathcal{T}$ matrix is~%given by $\mathcal{T}_{\#1}=$
%\bea
%\bpm1&&&&&&&&\\&1&&&&&&&\\&&-1&&&&&&\\&&&-1&&&&&\\&&&&1&&&&\\&&&&&e^{\imth\frac{\pi}8}&&&\\&&&&&&e^{-\imth\frac{\pi}8}&&\\&&&&&&&-e^{\imth\frac{\pi}8}&\\&&&&&&&&-e^{-\imth\frac{\pi}8}\epm
%\eea
a diagonal unitary matrix $\mathcal{T}_{a,b}=\delta_{a,b}\exp(2\pi\imth h_a)$, where $h_a$ gives the topological spin $\Theta_a=\exp(2\pi\imth h_a)$ of quasiparticle $q_a$ shown in TABLE \ref{tab:Z2SL:z2:unconventional:vertex algebra}. To be specific we have
\bea\notag
{\mathcal{T}}_{\#5}=\bpm1&&&&&&&&\\&1&&&&&&&\\&&-1&&&&&&\\&&&-1&&&&&\\&&&&1&&&&\\&&&&&e^{\imth\pi/8}&&&\\&&&&&&e^{-\imth\pi/8}&&\\&&&&&&&-e^{\imth\pi/8}&\\&&&&&&&&-e^{-\imth\pi/8}\epm
\eea

For SET phase $\#6$, after gauging the unitary $Z_2$ symmetry we have its modular $\mathcal{S}$ matrix as $\mathcal{S}_{\#6}=$
\bea
&\notag\frac14\bpm1&1&1&1&2&\sqrt2&\sqrt2&\sqrt2&\sqrt2\\1&1&1&1&2&-\sqrt2&-\sqrt2&-\sqrt2&-\sqrt2\\1&1&1&1&-2&-\sqrt2&\sqrt2&-\sqrt2&\sqrt2
\\1&1&1&1&-2&\sqrt2&-\sqrt2&\sqrt2&-\sqrt2\\2&2&-2&-2&0&0&0&0&0\\ \sqrt2&-\sqrt2&-\sqrt2&\sqrt2&0&0&-2&0&2
\\ \sqrt2&-\sqrt2&\sqrt2&-\sqrt2&0&-2&0&2&0\\ \sqrt2&-\sqrt2&-\sqrt2&\sqrt2&0&0&2&0&-2\\ \sqrt2&-\sqrt2&\sqrt2&-\sqrt2&0&2&0&-2&0
\epm
\eea
and its $\mathcal{T}$ matrix as
\bea\notag
{\mathcal{T}}_{\#6}=\bpm1&&&&&&&&\\&1&&&&&&&\\&&-1&&&&&&\\&&&-1&&&&&\\&&&&1&&&&\\&&&&&e^{\imth5\pi/8}&&&\\&&&&&&e^{\imth3\pi/8}&&\\&&&&&&&-e^{\imth5\pi/8}&\\&&&&&&&&-e^{3\imth\pi/8}\epm
\eea

Clearly after gauging the unconventional $Z_2$ symmetry, $\delta\vec\phi^\bsg=(0,0,\pi,0)^T$ and $\delta\vec\phi^\bsg=(\pi/2,\pi/2,\pi,0)^T$ lead to the same non-Abelian topological order, since they belong to the same SET phase. They share the same $\mathcal{S}$ and $\mathcal{T}$ matrices, differing by a relabel of quasiparticles in TABLE \ref{tab:Z2SL:z2:unconventional:vertex algebra}. For example quasiparticle $q_5$ in $\delta\vec\phi^\bsg=(0,0,\pi,0)^T$ case corresponds to quasiparticle $q_6$ in $\delta\vec\phi^\bsg=(\pi/2,\pi/2,\pi,0)^T$ case. Similarly two cases $\delta\vec\phi^\bsg=(0,0,\pi,\pi)^T$ and $\delta\vec\phi^\bsg=(\pi/2,\pi/2,\pi,\pi)^T$ lead to the same non-Abelian topological order, by gauging the unconventional $Z_2$ symmetry.

It's easy to verify that they satisfy the following consistency conditions\cite{Kitaev2006} for modular transformations:
\bea
(\mathcal{S}\mathcal{T})^3=\Theta\cdot\mathcal{S}^2,~~~\mathcal{S}^4=1.
\eea
where the $U(1)$ phase factor $\Theta$ is defined as
\bea
\Theta\equiv d_a^2\cdot e^{2\pi\imth h_a}/\sqrt{\sum_ad_a^2}=e^{2\pi\imth c_-/8}.
\eea
$d_a$ and $h_a$ corresponds to the quantum dimension and topological spin $\exp(2\pi\imth h_a)$ of quasiparticle $q_a$ respectively. $c_-$ is the chiral central charge of the edge excitations of the topological ordered phase. Both non-Abelian topological orders in TABLE \ref{tab:Z2SL:z2:unconventional:vertex algebra} have $c_-=0$ and hence $\Theta=1$. In fact for both non-Abelian topological orders ($\#5-\#6$) summarized in TABLE \ref{tab:Z2SL:z2:unconventional:vertex algebra}, their modular $\mathcal{S}$ and $\mathcal{T}$ matrices satisfy ${\mathcal{S}}^2=(\mathcal{S}\mathcal{T})^3=1_{9\times9}$.

Starting from a $Z_2$ gauge theory ($Z_2$ spin liquid or double semion theory) with unitary $Z_2$ symmetry, once the symmetry is gauged, a resultant $Z_2\times Z_2$ gauge theory is expected\cite{Dijkgraaf1990,Hung2013}. The above non-Abelian topological orders can be regarded as ``unconventional'' $Z_2\times Z_2$ gauge theories, related to Kitaev's 16-fold way classification\cite{Kitaev2006} of $Z_2$ gauge theories in 2+1-D. In particular, they are associated with $Z_2$ gauge theories where fermions having an odd Chern number ($\nu=$~odd) couple to $Z_2$ gauge fields. Notice that before gauging the symmetry, all SET phases have non-chiral edge excitations with chiral central charge $c_-=0$. As a result, we expect that after gauging the $Z_2$ symmetry their edge states remain non-chiral and should be gapped due to backscattering in a generic situation. Indeed in all the ``gauged'' non-Abelian topological orders in TABLE \ref{tab:Z2SL:z2:unconventional:vertex algebra}, a $Z_2$ gauge theory with fermion Chern number $\nu$ is always accompanied by its time-reversal counterpart $\bar{\nu}\equiv16-\nu\mod16$ through a direct product.

Specifically, in \Ref{Kitaev2006} Kitaev introduced a 16-fold way classification of 2+1-D $Z_2$ gauge theories, describing fermions coupled to a $Z_2$ gauge field. When the Chern number $\nu$ of fermions changes by 16, one ends up with the same $Z_2$ gauge theory. Specifically when $\nu=$odd, associated $Z_2$ gauge theory contains 3 inequivalent quasiparticles: vaccum (or boson) $1$, fermion $\psi$ ($\varepsilon$ in Kitaev's notation\cite{Kitaev2006}) and vortex $\sigma$. Their fusions rules are the same as (\ref{fusion rule:Ising}), \ie those in Ising anyon theory\cite{Nayak2008}. Their quantum dimensions are
\bea
\notag d_1=d_\psi=1,~~~d_\sigma=\sqrt2.
\eea
The topological spin $\exp(2\pi\imth h)$ of these quasiparticles are given by
\bea
\notag h_1=0,~~~h_\psi=\frac12,~~~h_\sigma=\frac{\nu}{16}.
\eea
Therefore when $\nu=1$ this corresponds to the Ising anyon theory. When a direct product of a $Z_2$ gauge theory with Chern number $\nu$ (we denote this $Z_2$ gauge theory by $\nu$) and its time reversal counterpart $\bar{\nu}=16-\nu$ is made, one can combine the fermion $\psi$ in $\nu$ and the vortex $\bar{\sigma}$ in $\bar\nu$ to form a new vortex operator, which have scaling dimension $\frac12-\frac{\nu}{16}=\frac{8-\nu}{16}$. Therefore one can clearly see the following two seemingly different direct products
\bea
\nu\otimes(16-\nu)\simeq(8-\nu)\otimes(8+\nu).
\eea
lead to the same topological order. As a result SET phases $\#5$ and $\#6$ in TABLE \ref{tab:Z2SL:z2:unconventional} lead to two distinct non-Abelian topological orders ($\nu=1,7$ and $\nu=3,5$), by gauging the unitary $Z_2$ symmetry.

%\bibliographystyle{naturemag_no_url}
%\bibliography{bibs}
%%\bibliographystyle{C://CTEX//MiKTeX//bibtex//bst//nature//naturemag_no_url}
%%\bibliographystyle{C://CTEX//MiKTeX//bibtex//bst//revtex4//apsrev_nurl}
%%\bibliography{E://Dropbox//notes//bibs}
%\bibliography{F://Research//Ctex//Bib//mybib//bibs}

\begin{thebibliography}{100}
\expandafter\ifx\csname url\endcsname\relax
  \def\url#1{\texttt{#1}}\fi
\expandafter\ifx\csname urlprefix\endcsname\relax\def\urlprefix{URL }\fi
\providecommand{\bibinfo}[2]{#2}
\providecommand{\eprint}[2][]{\url{#2}}

\bibitem{Landau1937}
\bibinfo{author}{Landau, L.~D.}
\newblock \bibinfo{title}{Theory of phase transformations. i}.
\newblock \emph{\bibinfo{journal}{Phys. Z. Sowjetunion}}
  \textbf{\bibinfo{volume}{11}}, \bibinfo{pages}{26} (\bibinfo{year}{1937}).

\bibitem{Landau1937a}
\bibinfo{author}{Landau, L.~D.}
\newblock \bibinfo{title}{Theory of phase transformations. ii}.
\newblock \emph{\bibinfo{journal}{Phys. Z. Sowjetunion}}
  \textbf{\bibinfo{volume}{11}}, \bibinfo{pages}{545} (\bibinfo{year}{1937}).

\bibitem{Klitzing1980}
\bibinfo{author}{Klitzing, K.~v.}, \bibinfo{author}{Dorda, G.} \&
  \bibinfo{author}{Pepper, M.}
\newblock \bibinfo{title}{New method for high-accuracy determination of the
  fine-structure constant based on quantized hall resistance}.
\newblock \emph{\bibinfo{journal}{Phys. Rev. Lett.}}
  \textbf{\bibinfo{volume}{45}}, \bibinfo{pages}{494--} (\bibinfo{year}{1980}).

\bibitem{Tsui1982}
\bibinfo{author}{Tsui, D.~C.}, \bibinfo{author}{Stormer, H.~L.} \&
  \bibinfo{author}{Gossard, A.~C.}
\newblock \bibinfo{title}{Two-dimensional magnetotransport in the extreme
  quantum limit}.
\newblock \emph{\bibinfo{journal}{Phys. Rev. Lett.}}
  \textbf{\bibinfo{volume}{48}}, \bibinfo{pages}{1559--}
  (\bibinfo{year}{1982}).

\bibitem{Laughlin1983}
\bibinfo{author}{Laughlin, R.~B.}
\newblock \bibinfo{title}{Anomalous quantum hall effect: An incompressible
  quantum fluid with fractionally charged excitations}.
\newblock \emph{\bibinfo{journal}{Phys. Rev. Lett.}}
  \textbf{\bibinfo{volume}{50}}, \bibinfo{pages}{1395--}
  (\bibinfo{year}{1983}).

\bibitem{Wen2004B}
\bibinfo{author}{Wen, X.-G.}
\newblock \emph{\bibinfo{title}{Quantum Field Theory Of Many-body Systems: From
  The Origin Of Sound To An Origin Of Light And Electrons}}
  (\bibinfo{publisher}{Oxford University Press, New York},
  \bibinfo{year}{2004}).

\bibitem{Wilczek1990B}
\bibinfo{author}{Wilczek, F.}
\newblock \emph{\bibinfo{title}{Fractional Statistics and Anyon
  Superconductivity}} (\bibinfo{publisher}{World Scientific Pub Co Inc,
  Singapore}, \bibinfo{year}{1990}).

\bibitem{Yan2011}
\bibinfo{author}{Yan, S.}, \bibinfo{author}{Huse, D.~A.} \&
  \bibinfo{author}{White, S.~R.}
\newblock \bibinfo{title}{Spin-liquid ground state of the s = 1/2 kagome
  heisenberg antiferromagnet}.
\newblock \emph{\bibinfo{journal}{Science}} \textbf{\bibinfo{volume}{332}},
  \bibinfo{pages}{1173--1176} (\bibinfo{year}{2011}).

\bibitem{Jiang2012}
\bibinfo{author}{Jiang, H.-C.}, \bibinfo{author}{Yao, H.} \&
  \bibinfo{author}{Balents, L.}
\newblock \bibinfo{title}{Spin liquid ground state of the spin-1/2 square
  j$_{1}$-j$_{2}$ heisenberg model}.
\newblock \emph{\bibinfo{journal}{Phys. Rev. B}} \textbf{\bibinfo{volume}{86}},
  \bibinfo{pages}{024424--} (\bibinfo{year}{2012}).

\bibitem{Wang2011a}
\bibinfo{author}{Wang, L.}, \bibinfo{author}{Gu, Z.-C.},
  \bibinfo{author}{Verstraete, F.} \& \bibinfo{author}{Wen, X.-G.}
\newblock \bibinfo{title}{Spin-liquid phase in spin-1/2 square $j_1$-$j_2$
  heisenberg model: A tensor product state approach}.
\newblock \emph{\bibinfo{journal}{ArXiv e-prints 1112.3331}}
  (\bibinfo{year}{2011}).

\bibitem{Kitaev2006a}
\bibinfo{author}{Kitaev, A.} \& \bibinfo{author}{Preskill, J.}
\newblock \bibinfo{title}{Topological entanglement entropy}.
\newblock \emph{\bibinfo{journal}{Phys. Rev. Lett.}}
  \textbf{\bibinfo{volume}{96}}, \bibinfo{pages}{110404--}
  (\bibinfo{year}{2006}).

\bibitem{Levin2006}
\bibinfo{author}{Levin, M.} \& \bibinfo{author}{Wen, X.-G.}
\newblock \bibinfo{title}{Detecting topological order in a ground state wave
  function}.
\newblock \emph{\bibinfo{journal}{Phys. Rev. Lett.}}
  \textbf{\bibinfo{volume}{96}}, \bibinfo{pages}{110405--}
  (\bibinfo{year}{2006}).

\bibitem{Jiang2012a}
\bibinfo{author}{Jiang, H.-C.}, \bibinfo{author}{Wang, Z.} \&
  \bibinfo{author}{Balents, L.}
\newblock \bibinfo{title}{Identifying topological order by entanglement
  entropy}.
\newblock \emph{\bibinfo{journal}{Nat Phys}} \textbf{\bibinfo{volume}{8}},
  \bibinfo{pages}{902--905} (\bibinfo{year}{2012}).

\bibitem{Depenbrock2012}
\bibinfo{author}{Depenbrock, S.}, \bibinfo{author}{McCulloch, I.~P.} \&
  \bibinfo{author}{Schollw$\ddot{o}$ck, U.}
\newblock \bibinfo{title}{Nature of the spin-liquid ground state of the s=1/2
  heisenberg model on the kagome lattice}.
\newblock \emph{\bibinfo{journal}{Phys. Rev. Lett.}}
  \textbf{\bibinfo{volume}{109}}, \bibinfo{pages}{067201--}
  (\bibinfo{year}{2012}).

\bibitem{Wen2002}
\bibinfo{author}{Wen, X.-G.}
\newblock \bibinfo{title}{Quantum orders and symmetric spin liquids}.
\newblock \emph{\bibinfo{journal}{Phys. Rev. B}} \textbf{\bibinfo{volume}{65}},
  \bibinfo{pages}{165113} (\bibinfo{year}{2002}).

\bibitem{Wang2006}
\bibinfo{author}{Wang, F.} \& \bibinfo{author}{Vishwanath, A.}
\newblock \bibinfo{title}{Spin-liquid states on the triangular and
  kagom\&eacute; lattices: A projective-symmetry-group analysis of schwinger
  boson states}.
\newblock \emph{\bibinfo{journal}{Phys. Rev. B}} \textbf{\bibinfo{volume}{74}},
  \bibinfo{pages}{174423--} (\bibinfo{year}{2006}).

\bibitem{Levin2012a}
\bibinfo{author}{Levin, M.} \& \bibinfo{author}{Stern, A.}
\newblock \bibinfo{title}{Classification and analysis of two-dimensional
  abelian fractional topological insulators}.
\newblock \emph{\bibinfo{journal}{Phys. Rev. B}} \textbf{\bibinfo{volume}{86}},
  \bibinfo{pages}{115131--} (\bibinfo{year}{2012}).

\bibitem{Essin2013}
\bibinfo{author}{Essin, A.~M.} \& \bibinfo{author}{Hermele, M.}
\newblock \bibinfo{title}{Classifying fractionalization: Symmetry
  classification of gapped $z_{2}$ spin liquids in two dimensions}.
\newblock \emph{\bibinfo{journal}{Phys. Rev. B}} \textbf{\bibinfo{volume}{87}},
  \bibinfo{pages}{104406--} (\bibinfo{year}{2013}).

\bibitem{Mesaros2013}
\bibinfo{author}{Mesaros, A.} \& \bibinfo{author}{Ran, Y.}
\newblock \bibinfo{title}{Classification of symmetry enriched topological
  phases with exactly solvable models}.
\newblock \emph{\bibinfo{journal}{Phys. Rev. B}} \textbf{\bibinfo{volume}{87}},
  \bibinfo{pages}{155115--} (\bibinfo{year}{2013}).

\bibitem{Hung2013a}
\bibinfo{author}{Hung, L.-Y.} \& \bibinfo{author}{Wan, Y.}
\newblock \bibinfo{title}{K matrix construction of symmetry-enriched phases of
  matter}.
\newblock \emph{\bibinfo{journal}{Phys. Rev. B}} \textbf{\bibinfo{volume}{87}},
  \bibinfo{pages}{195103--} (\bibinfo{year}{2013}).

\bibitem{Hung2013}
\bibinfo{author}{Hung, L.-Y.} \& \bibinfo{author}{Wen, X.-G.}
\newblock \bibinfo{title}{Quantized topological terms in weak-coupling gauge
  theories with a global symmetry and their connection to symmetry-enriched
  topological phases}.
\newblock \emph{\bibinfo{journal}{Phys. Rev. B}} \textbf{\bibinfo{volume}{87}},
  \bibinfo{pages}{165107--} (\bibinfo{year}{2013}).

\bibitem{Yao2010}
\bibinfo{author}{Yao, H.}, \bibinfo{author}{Fu, L.} \& \bibinfo{author}{Qi,
  X.-L.}
\newblock \bibinfo{title}{{Symmetry fractional quantization in two
  dimensions}}.
\newblock \emph{\bibinfo{journal}{ArXiv e-prints 1012.4470}}
  (\bibinfo{year}{2010}).
\newblock \eprint{1012.4470}.

\bibitem{Chen2013}
\bibinfo{author}{Chen, X.}, \bibinfo{author}{Gu, Z.-C.}, \bibinfo{author}{Liu,
  Z.-X.} \& \bibinfo{author}{Wen, X.-G.}
\newblock \bibinfo{title}{Symmetry protected topological orders and the group
  cohomology of their symmetry group}.
\newblock \emph{\bibinfo{journal}{Phys. Rev. B}} \textbf{\bibinfo{volume}{87}},
  \bibinfo{pages}{155114--} (\bibinfo{year}{2013}).

\bibitem{Hasan2010}
\bibinfo{author}{Hasan, M.~Z.} \& \bibinfo{author}{Kane, C.~L.}
\newblock \bibinfo{title}{Colloquium: Topological insulators}.
\newblock \emph{\bibinfo{journal}{Rev. Mod. Phys.}}
  \textbf{\bibinfo{volume}{82}}, \bibinfo{pages}{3045--}
  (\bibinfo{year}{2010}).

\bibitem{Hasan2011}
\bibinfo{author}{Hasan, M.~Z.} \& \bibinfo{author}{Moore, J.~E.}
\newblock \bibinfo{title}{Three-dimensional topological insulators}.
\newblock \emph{\bibinfo{journal}{Annu. Rev. Condens. Matter Phys.}}
  \textbf{\bibinfo{volume}{2}}, \bibinfo{pages}{55--78} (\bibinfo{year}{2011}).

\bibitem{Qi2011}
\bibinfo{author}{Qi, X.-L.} \& \bibinfo{author}{Zhang, S.-C.}
\newblock \bibinfo{title}{Topological insulators and superconductors}.
\newblock \emph{\bibinfo{journal}{Rev. Mod. Phys.}}
  \textbf{\bibinfo{volume}{83}}, \bibinfo{pages}{1057--1110}
  (\bibinfo{year}{2011}).

\bibitem{Chen2011b}
\bibinfo{author}{Chen, X.}, \bibinfo{author}{Liu, Z.-X.} \&
  \bibinfo{author}{Wen, X.-G.}
\newblock \bibinfo{title}{Two-dimensional symmetry-protected topological orders
  and their protected gapless edge excitations}.
\newblock \emph{\bibinfo{journal}{Phys. Rev. B}} \textbf{\bibinfo{volume}{84}},
  \bibinfo{pages}{235141--} (\bibinfo{year}{2011}).

\bibitem{Levin2012}
\bibinfo{author}{Levin, M.} \& \bibinfo{author}{Gu, Z.-C.}
\newblock \bibinfo{title}{Braiding statistics approach to symmetry-protected
  topological phases}.
\newblock \emph{\bibinfo{journal}{Phys. Rev. B}} \textbf{\bibinfo{volume}{86}},
  \bibinfo{pages}{115109--} (\bibinfo{year}{2012}).

\bibitem{Lu2012a}
\bibinfo{author}{Lu, Y.-M.} \& \bibinfo{author}{Vishwanath, A.}
\newblock \bibinfo{title}{Theory and classification of interacting integer
  topological phases in two dimensions: A chern-simons approach}.
\newblock \emph{\bibinfo{journal}{Phys. Rev. B}} \textbf{\bibinfo{volume}{86}},
  \bibinfo{pages}{125119--} (\bibinfo{year}{2012}).

\bibitem{Read1990}
\bibinfo{author}{Read, N.}
\newblock \bibinfo{title}{Excitation structure of the hierarchy scheme in the
  fractional quantum hall effect}.
\newblock \emph{\bibinfo{journal}{Phys. Rev. Lett.}}
  \textbf{\bibinfo{volume}{65}}, \bibinfo{pages}{1502--}
  (\bibinfo{year}{1990}).

\bibitem{Wen1992}
\bibinfo{author}{Wen, X.~G.} \& \bibinfo{author}{Zee, A.}
\newblock \bibinfo{title}{Classification of abelian quantum hall states and
  matrix formulation of topological fluids}.
\newblock \emph{\bibinfo{journal}{Phys. Rev. B}} \textbf{\bibinfo{volume}{46}},
  \bibinfo{pages}{2290--} (\bibinfo{year}{1992}).

\bibitem{Frohlich1991}
\bibinfo{author}{Frohlich, J.} \& \bibinfo{author}{Zee, A.}
\newblock \bibinfo{title}{Large scale physics of the quantum hall fluid}.
\newblock \emph{\bibinfo{journal}{Nuclear Physics B}}
  \textbf{\bibinfo{volume}{364}}, \bibinfo{pages}{517--540}
  (\bibinfo{year}{1991}).

\bibitem{Witten1988}
\bibinfo{author}{Witten, E.}
\newblock \bibinfo{title}{{Topological quantum field theory.}}
\newblock \emph{\bibinfo{journal}{Commun. Math. Phys.}}
  \textbf{\bibinfo{volume}{117}}, \bibinfo{pages}{353--386}
  (\bibinfo{year}{1988}).

\bibitem{Wen1995}
\bibinfo{author}{Wen, X.-G.}
\newblock \bibinfo{title}{Topological orders and edge excitations in fractional
  quantum hall states}.
\newblock \emph{\bibinfo{journal}{Advances in Physics}}
  \textbf{\bibinfo{volume}{44}}, \bibinfo{pages}{405--473}
  (\bibinfo{year}{1995}).

\bibitem{Kitaev2003}
\bibinfo{author}{Kitaev, A.~Y.}
\newblock \bibinfo{title}{Fault-tolerant quantum computation by anyons}.
\newblock \emph{\bibinfo{journal}{Annals of Physics}}
  \textbf{\bibinfo{volume}{303}}, \bibinfo{pages}{2--30}
  (\bibinfo{year}{2003}).

\bibitem{Freedman2004}
\bibinfo{author}{Freedman, M.}, \bibinfo{author}{Nayak, C.},
  \bibinfo{author}{Shtengel, K.}, \bibinfo{author}{Walker, K.} \&
  \bibinfo{author}{Wang, Z.}
\newblock \bibinfo{title}{A class of p,t-invariant topological phases of
  interacting electrons}.
\newblock \emph{\bibinfo{journal}{Annals of Physics}}
  \textbf{\bibinfo{volume}{310}}, \bibinfo{pages}{428--492}
  (\bibinfo{year}{2004}).

\bibitem{Levin2005}
\bibinfo{author}{Levin, M.~A.} \& \bibinfo{author}{Wen, X.-G.}
\newblock \bibinfo{title}{String-net condensation: A physical mechanism for
  topological phases}.
\newblock \emph{\bibinfo{journal}{Phys. Rev. B}} \textbf{\bibinfo{volume}{71}},
  \bibinfo{pages}{045110--} (\bibinfo{year}{2005}).

\bibitem{Wen2003}
\bibinfo{author}{Wen, X.-G.}
\newblock \bibinfo{title}{Quantum orders in an exact soluble model}.
\newblock \emph{\bibinfo{journal}{Phys. Rev. Lett.}}
  \textbf{\bibinfo{volume}{90}}, \bibinfo{pages}{016803--}
  (\bibinfo{year}{2003}).

\bibitem{Kou2008}
\bibinfo{author}{Kou, S.-P.}, \bibinfo{author}{Levin, M.} \&
  \bibinfo{author}{Wen, X.-G.}
\newblock \bibinfo{title}{Mutual chern-simons theory for $z_{2}$ topological
  order}.
\newblock \emph{\bibinfo{journal}{Phys. Rev. B}} \textbf{\bibinfo{volume}{78}},
  \bibinfo{pages}{155134--} (\bibinfo{year}{2008}).

\bibitem{Cho2012}
\bibinfo{author}{Cho, G.~Y.}, \bibinfo{author}{Lu, Y.-M.} \&
  \bibinfo{author}{Moore, J.~E.}
\newblock \bibinfo{title}{Gapless edge states of background field theory and
  translation-symmetric $z_{2}$ spin liquids}.
\newblock \emph{\bibinfo{journal}{Phys. Rev. B}} \textbf{\bibinfo{volume}{86}},
  \bibinfo{pages}{125101--} (\bibinfo{year}{2012}).

\bibitem{Dijkgraaf1990}
\bibinfo{author}{Dijkgraaf, R.} \& \bibinfo{author}{Witten, E.}
\newblock \bibinfo{title}{{Topological gauge theories and group cohomology.}}
\newblock \emph{\bibinfo{journal}{Commun. Math. Phys.}}
  \textbf{\bibinfo{volume}{129}}, \bibinfo{pages}{393--429}
  (\bibinfo{year}{1990}).

\bibitem{Nayak2008}
\bibinfo{author}{Nayak, C.}, \bibinfo{author}{Simon, S.~H.},
  \bibinfo{author}{Stern, A.}, \bibinfo{author}{Freedman, M.} \&
  \bibinfo{author}{Das~Sarma, S.}
\newblock \bibinfo{title}{Non-abelian anyons and topological quantum
  computation}.
\newblock \emph{\bibinfo{journal}{Rev. Mod. Phys.}}
  \textbf{\bibinfo{volume}{80}}, \bibinfo{pages}{1083--}
  (\bibinfo{year}{2008}).

\bibitem{Kitaev2006}
\bibinfo{author}{Kitaev, A.}
\newblock \bibinfo{title}{Anyons in an exactly solved model and beyond}.
\newblock \emph{\bibinfo{journal}{Annals of Physics}}
  \textbf{\bibinfo{volume}{321}}, \bibinfo{pages}{2--111}
  (\bibinfo{year}{2006}).

\bibitem{Lu2010}
\bibinfo{author}{Lu, Y.-M.}, \bibinfo{author}{Wen, X.-G.},
  \bibinfo{author}{Wang, Z.} \& \bibinfo{author}{Wang, Z.}
\newblock \bibinfo{title}{Non-abelian quantum hall states and their
  quasiparticles: From the pattern of zeros to vertex algebra}.
\newblock \emph{\bibinfo{journal}{Phys. Rev. B}} \textbf{\bibinfo{volume}{81}},
  \bibinfo{pages}{115124--} (\bibinfo{year}{2010}).

\bibitem{Vishwanath2013}
\bibinfo{author}{Vishwanath, A.} \& \bibinfo{author}{Senthil, T.}
\newblock \bibinfo{title}{Physics of three-dimensional bosonic topological
  insulators: Surface-deconfined criticality and quantized magnetoelectric
  effect}.
\newblock \emph{\bibinfo{journal}{Phys. Rev. X}} \textbf{\bibinfo{volume}{3}},
  \bibinfo{pages}{011016--} (\bibinfo{year}{2013}).

\bibitem{Fradkin1998}
\bibinfo{author}{Fradkin, E.}, \bibinfo{author}{Nayak, C.},
  \bibinfo{author}{Tsvelik, A.} \& \bibinfo{author}{Wilczek, F.}
\newblock \bibinfo{title}{A chern-simons effective field theory for the
  pfaffian quantum hall state}.
\newblock \emph{\bibinfo{journal}{Nuclear Physics B}}
  \textbf{\bibinfo{volume}{516}}, \bibinfo{pages}{704--718}
  (\bibinfo{year}{1998}).

\bibitem{Kalmeyer1987}
\bibinfo{author}{Kalmeyer, V.} \& \bibinfo{author}{Laughlin, R.~B.}
\newblock \bibinfo{title}{Equivalence of the resonating-valence-bond and
  fractional quantum hall states}.
\newblock \emph{\bibinfo{journal}{Phys. Rev. Lett.}}
  \textbf{\bibinfo{volume}{59}}, \bibinfo{pages}{2095--2098}
  (\bibinfo{year}{1987}).

\bibitem{Hansson2004}
\bibinfo{author}{Hansson, T.~H.}, \bibinfo{author}{Oganesyan, V.} \&
  \bibinfo{author}{Sondhi, S.~L.}
\newblock \bibinfo{title}{Superconductors are topologically ordered}.
\newblock \emph{\bibinfo{journal}{Annals of Physics}}
  \textbf{\bibinfo{volume}{313}}, \bibinfo{pages}{497--538}
  (\bibinfo{year}{2004}).

\bibitem{Wilczek1983}
\bibinfo{author}{Wilczek, F.} \& \bibinfo{author}{Zee, A.}
\newblock \bibinfo{title}{Linking numbers, spin, and statistics of solitons}.
\newblock \emph{\bibinfo{journal}{Phys. Rev. Lett.}}
  \textbf{\bibinfo{volume}{51}}, \bibinfo{pages}{2250--2252}
  (\bibinfo{year}{1983}).

\bibitem{Frohlich1990}
\bibinfo{author}{Frohlich, J.} \& \bibinfo{author}{Gabbiani, F.}
\newblock \bibinfo{title}{Braid statistics in local quantum theory}.
\newblock \emph{\bibinfo{journal}{Reviews in Mathematical Physics}}
  \textbf{\bibinfo{volume}{2}}, \bibinfo{pages}{251--353}
  (\bibinfo{year}{1990}).

\bibitem{Witten1989}
\bibinfo{author}{Witten, E.}
\newblock \bibinfo{title}{{Quantum field theory and the Jones polynomial.}}
\newblock \emph{\bibinfo{journal}{Commun. Math. Phys.}}
  \textbf{\bibinfo{volume}{121}}, \bibinfo{pages}{351--399}
  (\bibinfo{year}{1989}).

\bibitem{Wen1990}
\bibinfo{author}{Wen, X.~G.}
\newblock \bibinfo{title}{Chiral luttinger liquid and the edge excitations in
  the fractional quantum hall states}.
\newblock \emph{\bibinfo{journal}{Phys. Rev. B}} \textbf{\bibinfo{volume}{41}},
  \bibinfo{pages}{12838--} (\bibinfo{year}{1990}).

\bibitem{Wen1990b}
\bibinfo{author}{Wen, X.~G.} \& \bibinfo{author}{Niu, Q.}
\newblock \bibinfo{title}{Ground-state degeneracy of the fractional quantum
  hall states in the presence of a random potential and on high-genus riemann
  surfaces}.
\newblock \emph{\bibinfo{journal}{Phys. Rev. B}} \textbf{\bibinfo{volume}{41}},
  \bibinfo{pages}{9377--} (\bibinfo{year}{1990}).

\bibitem{Lu2012}
\bibinfo{author}{Lu, Y.-M.} \& \bibinfo{author}{Ran, Y.}
\newblock \bibinfo{title}{Symmetry-protected fractional chern insulators and
  fractional topological insulators}.
\newblock \emph{\bibinfo{journal}{Phys. Rev. B}} \textbf{\bibinfo{volume}{85}},
  \bibinfo{pages}{165134--} (\bibinfo{year}{2012}).

\bibitem{Haldane1995}
\bibinfo{author}{Haldane, F. D.~M.}
\newblock \bibinfo{title}{Stability of chiral luttinger liquids and abelian
  quantum hall states}.
\newblock \emph{\bibinfo{journal}{Phys. Rev. Lett.}}
  \textbf{\bibinfo{volume}{74}}, \bibinfo{pages}{2090--}
  (\bibinfo{year}{1995}).

\bibitem{Kane1996}
\bibinfo{author}{Kane, C.~L.} \& \bibinfo{author}{Fisher, M. P.~A.}
\newblock \bibinfo{title}{Thermal transport in a luttinger liquid}.
\newblock \emph{\bibinfo{journal}{Phys. Rev. Lett.}}
  \textbf{\bibinfo{volume}{76}}, \bibinfo{pages}{3192--3195}
  (\bibinfo{year}{1996}).

\bibitem{Levin2013}
\bibinfo{author}{Levin, M.}
\newblock \bibinfo{title}{Protected edge modes without symmetry}.
\newblock \emph{\bibinfo{journal}{Phys. Rev. X}} \textbf{\bibinfo{volume}{3}},
  \bibinfo{pages}{021009--} (\bibinfo{year}{2013}).

\bibitem{Levin2009}
\bibinfo{author}{Levin, M.} \& \bibinfo{author}{Stern, A.}
\newblock \bibinfo{title}{Fractional topological insulators}.
\newblock \emph{\bibinfo{journal}{Phys. Rev. Lett.}}
  \textbf{\bibinfo{volume}{103}}, \bibinfo{pages}{196803--}
  (\bibinfo{year}{2009}).

\bibitem{Cano2014}
\bibinfo{author}{Cano, J.} \emph{et~al.}
\newblock \bibinfo{title}{Bulk-edge correspondence in (2 + 1)-dimensional
  abelian topological phases}.
\newblock \emph{\bibinfo{journal}{Phys. Rev. B}} \textbf{\bibinfo{volume}{89}},
  \bibinfo{pages}{115116--} (\bibinfo{year}{2014}).

\bibitem{Lu2014c}
\bibinfo{author}{Lu, Y.-M.} \& \bibinfo{author}{Lee, D.-H.}
\newblock \bibinfo{title}{Gapped symmetric edges of symmetry-protected
  topological phases}.
\newblock \emph{\bibinfo{journal}{Phys. Rev. B}} \textbf{\bibinfo{volume}{89}},
  \bibinfo{pages}{205117--} (\bibinfo{year}{2014}).

\bibitem{Wang2013b}
\bibinfo{author}{Wang, C.} \& \bibinfo{author}{Levin, M.}
\newblock \bibinfo{title}{Weak symmetry breaking in two-dimensional topological
  insulators}.
\newblock \emph{\bibinfo{journal}{Phys. Rev. B}} \textbf{\bibinfo{volume}{88}},
  \bibinfo{pages}{245136--} (\bibinfo{year}{2013}).

\bibitem{Barkeshli2010b}
\bibinfo{author}{Barkeshli, M.} \& \bibinfo{author}{Wen, X.-G.}
\newblock \bibinfo{title}{$u(1)\times u(1)\rtimes z_{2}$ chern-simons theory
  and $z_{4}$ parafermion fractional quantum hall states}.
\newblock \emph{\bibinfo{journal}{Phys. Rev. B}} \textbf{\bibinfo{volume}{81}},
  \bibinfo{pages}{045323--} (\bibinfo{year}{2010}).

\bibitem{Abrikosov1965}
\bibinfo{author}{Abrikosov, A.~A.}
\newblock \bibinfo{title}{Electron scattering on magnetic impurities in metals
  and anomalous resistivity effects}.
\newblock \emph{\bibinfo{journal}{Physics}} \textbf{\bibinfo{volume}{2}},
  \bibinfo{pages}{5--20} (\bibinfo{year}{1965}).

\bibitem{Affleck1988b}
\bibinfo{author}{Affleck, I.}, \bibinfo{author}{Zou, Z.}, \bibinfo{author}{Hsu,
  T.} \& \bibinfo{author}{Anderson, P.~W.}
\newblock \bibinfo{title}{Su(2) gauge symmetry of the large-u limit of the
  hubbard model}.
\newblock \emph{\bibinfo{journal}{Phys. Rev. B}} \textbf{\bibinfo{volume}{38}},
  \bibinfo{pages}{745--} (\bibinfo{year}{1988}).

\bibitem{Baskaran1988}
\bibinfo{author}{Baskaran, G.} \& \bibinfo{author}{Anderson, P.~W.}
\newblock \bibinfo{title}{Gauge theory of high-temperature superconductors and
  strongly correlated fermi systems}.
\newblock \emph{\bibinfo{journal}{Phys. Rev. B}} \textbf{\bibinfo{volume}{37}},
  \bibinfo{pages}{580--} (\bibinfo{year}{1988}).

\bibitem{Schnyder2008}
\bibinfo{author}{Schnyder, A.~P.}, \bibinfo{author}{Ryu, S.},
  \bibinfo{author}{Furusaki, A.} \& \bibinfo{author}{Ludwig, A. W.~W.}
\newblock \bibinfo{title}{Classification of topological insulators and
  superconductors in three spatial dimensions}.
\newblock \emph{\bibinfo{journal}{Phys. Rev. B}} \textbf{\bibinfo{volume}{78}},
  \bibinfo{pages}{195125--} (\bibinfo{year}{2008}).

\bibitem{Kitaev2009}
\bibinfo{author}{Kitaev, A.}
\newblock \bibinfo{title}{Periodic table for topological insulators and
  superconductors}.
\newblock \emph{\bibinfo{journal}{AIP Conf. Proc.}}
  \textbf{\bibinfo{volume}{1134}}, \bibinfo{pages}{22--30}
  (\bibinfo{year}{2009}).

\bibitem{Liu2010a}
\bibinfo{author}{Liu, Z.-X.}, \bibinfo{author}{Zhou, Y.} \&
  \bibinfo{author}{Ng, T.-K.}
\newblock \bibinfo{title}{Fermionic theory for quantum antiferromagnets with
  spin $s>1/2$}.
\newblock \emph{\bibinfo{journal}{Phys. Rev. B}} \textbf{\bibinfo{volume}{82}},
  \bibinfo{pages}{144422--} (\bibinfo{year}{2010}).

\bibitem{Read2000}
\bibinfo{author}{Read, N.} \& \bibinfo{author}{Green, D.}
\newblock \bibinfo{title}{Paired states of fermions in two dimensions with
  breaking of parity and time-reversal symmetries and the fractional quantum
  hall effect}.
\newblock \emph{\bibinfo{journal}{Phys. Rev. B}} \textbf{\bibinfo{volume}{61}},
  \bibinfo{pages}{10267--} (\bibinfo{year}{2000}).

\bibitem{Zhang2012}
\bibinfo{author}{Zhang, Y.}, \bibinfo{author}{Grover, T.},
  \bibinfo{author}{Turner, A.}, \bibinfo{author}{Oshikawa, M.} \&
  \bibinfo{author}{Vishwanath, A.}
\newblock \bibinfo{title}{Quasiparticle statistics and braiding from
  ground-state entanglement}.
\newblock \emph{\bibinfo{journal}{Phys. Rev. B}} \textbf{\bibinfo{volume}{85}},
  \bibinfo{pages}{235151--} (\bibinfo{year}{2012}).

\bibitem{Eisert2010}
\bibinfo{author}{Eisert, J.}, \bibinfo{author}{Cramer, M.} \&
  \bibinfo{author}{Plenio, M.~B.}
\newblock \bibinfo{title}{Colloquium: Area laws for the entanglement entropy}.
\newblock \emph{\bibinfo{journal}{Rev. Mod. Phys.}}
  \textbf{\bibinfo{volume}{82}}, \bibinfo{pages}{277--306}
  (\bibinfo{year}{2010}).

\bibitem{Dong2008}
\bibinfo{author}{Dong, S.}, \bibinfo{author}{Fradkin, E.},
  \bibinfo{author}{Leigh, R.~G.} \& \bibinfo{author}{Nowling, S.}
\newblock \bibinfo{title}{Topological entanglement entropy in chern-simons
  theories and quantum hall fluids}.
\newblock \emph{\bibinfo{journal}{Journal of High Energy Physics}}
  \textbf{\bibinfo{volume}{2008}}, \bibinfo{pages}{016--}
  (\bibinfo{year}{2008}).

\bibitem{Bombin2010}
\bibinfo{author}{Bombin, H.}
\newblock \bibinfo{title}{Topological order with a twist: Ising anyons from an
  abelian model}.
\newblock \emph{\bibinfo{journal}{Phys. Rev. Lett.}}
  \textbf{\bibinfo{volume}{105}}, \bibinfo{pages}{030403--}
  (\bibinfo{year}{2010}).

\bibitem{Fu2009}
\bibinfo{author}{Fu, L.} \& \bibinfo{author}{Kane, C.~L.}
\newblock \bibinfo{title}{Josephson current and noise at a
  superconductor/quantum-spin-hall-insulator/superconductor junction}.
\newblock \emph{\bibinfo{journal}{Phys. Rev. B}} \textbf{\bibinfo{volume}{79}},
  \bibinfo{pages}{161408--} (\bibinfo{year}{2009}).

\bibitem{Wen1989}
\bibinfo{author}{Wen, X.~G.}, \bibinfo{author}{Wilczek, F.} \&
  \bibinfo{author}{Zee, A.}
\newblock \bibinfo{title}{Chiral spin states and superconductivity}.
\newblock \emph{\bibinfo{journal}{Phys. Rev. B}} \textbf{\bibinfo{volume}{39}},
  \bibinfo{pages}{11413--11423} (\bibinfo{year}{1989}).

\bibitem{Lu2014b}
\bibinfo{author}{Lu, Y.-M.} \& \bibinfo{author}{Fidkowski, L.}
\newblock \bibinfo{title}{Symmetry-induced anyon breeding in fractional quantum
  hall states}.
\newblock \emph{\bibinfo{journal}{Phys. Rev. B}} \textbf{\bibinfo{volume}{89}},
  \bibinfo{pages}{115321--} (\bibinfo{year}{2014}).

\bibitem{Gu2012}
\bibinfo{author}{Gu, Z.} \& \bibinfo{author}{Wen, X.~G.}
\newblock \bibinfo{title}{Symmetry-protected topological orders for interacting
  fermions -- fermionic topological non-linear sigma-models and a group
  super-cohomology theory}.
\newblock \emph{\bibinfo{journal}{arXiv:1201.2648v1 [cond-mat.str-el]}}
  (\bibinfo{year}{2012}).

\bibitem{Auerbach1994B}
\bibinfo{author}{Auerbach, A.}
\newblock \emph{\bibinfo{title}{Interacting electrons and quantum magnetism}}.
\newblock Graduate Texts in Contemporary Physics (\bibinfo{publisher}{Springer,
  New York}, \bibinfo{year}{1994}).

\bibitem{Sachdev1992}
\bibinfo{author}{Sachdev, S.}
\newblock \bibinfo{title}{Kagome and triangular-lattice heisenberg
  antiferromagnets: Ordering from quantum fluctuations and quantum-disordered
  ground states with unconfined bosonic spinons}.
\newblock \emph{\bibinfo{journal}{Phys. Rev. B}} \textbf{\bibinfo{volume}{45}},
  \bibinfo{pages}{12377--} (\bibinfo{year}{1992}).

\bibitem{Wang2012a}
\bibinfo{author}{{Wang}, J.} \& \bibinfo{author}{{Wen}, X.-G.}
\newblock \bibinfo{title}{{Boundary Degeneracy of Topological Order}}.
\newblock \emph{\bibinfo{journal}{ArXiv e-prints}}  (\bibinfo{year}{2012}).
\newblock \eprint{1212.4863}.

\bibitem{Dijkgraaf1989}
\bibinfo{author}{Dijkgraaf, R.}, \bibinfo{author}{Vafa, C.},
  \bibinfo{author}{Verlinde, E.} \& \bibinfo{author}{Verlinde, H.}
\newblock \bibinfo{title}{{The operator algebra of orbifold models.}}
\newblock \emph{\bibinfo{journal}{Commun. Math. Phys.}}
  \textbf{\bibinfo{volume}{123}}, \bibinfo{pages}{485--526}
  (\bibinfo{year}{1989}).

\bibitem{Moore1989}
\bibinfo{author}{Moore, G.} \& \bibinfo{author}{Seiberg, N.}
\newblock \bibinfo{title}{Taming the conformal zoo}.
\newblock \emph{\bibinfo{journal}{Physics Letters B}}
  \textbf{\bibinfo{volume}{220}}, \bibinfo{pages}{422--430}
  (\bibinfo{year}{1989}).

\bibitem{Moore1991}
\bibinfo{author}{Moore, G.} \& \bibinfo{author}{Read, N.}
\newblock \bibinfo{title}{Nonabelions in the fractional quantum hall effect}.
\newblock \emph{\bibinfo{journal}{Nuclear Physics B}}
  \textbf{\bibinfo{volume}{360}}, \bibinfo{pages}{362--396}
  (\bibinfo{year}{1991}).

\bibitem{Wen1994}
\bibinfo{author}{Wen, X.-G.} \& \bibinfo{author}{Wu, Y.-S.}
\newblock \bibinfo{title}{Chiral operator product algebra hidden in certain
  fractional quantum hall wave functions}.
\newblock \emph{\bibinfo{journal}{Nuclear Physics B}}
  \textbf{\bibinfo{volume}{419}}, \bibinfo{pages}{455--479}
  (\bibinfo{year}{1994}).

\bibitem{Barkeshli2012}
\bibinfo{author}{Barkeshli, M.} \& \bibinfo{author}{Wen, X.-G.}
\newblock \bibinfo{title}{Phase transitions in $z_{N}$ gauge theory and twisted
  $z_{N}$ topological phases}.
\newblock \emph{\bibinfo{journal}{Phys. Rev. B}} \textbf{\bibinfo{volume}{86}},
  \bibinfo{pages}{085114--} (\bibinfo{year}{2012}).

\bibitem{Ginsparg1989}
\bibinfo{author}{Ginsparg, P.~H.}
\newblock \bibinfo{title}{Applied conformal field theory}.
\newblock \emph{\bibinfo{journal}{Fields, Strings and Critical Phenomena, (Les
  Houches, Session XLIX, 1988) ed. by E. Br\'ezin and J. Zinn Justin}}
  (\bibinfo{year}{1989}).

\bibitem{Francesco1997}
\bibinfo{author}{Di~Francesco, P.}, \bibinfo{author}{Mathieu, P.} \&
  \bibinfo{author}{Senechal, D.}
\newblock \emph{\bibinfo{title}{Conformal field theory}}.
\newblock Graduate texts in contemporary physics. (\bibinfo{publisher}{New York
  : Springer}, \bibinfo{year}{1997}).

\bibitem{Zamolodchikov1985a}
\bibinfo{author}{Zamolodchikov, A.~B.} \& \bibinfo{author}{Fateev, V.~A.}
\newblock \bibinfo{title}{Disorder fields in two-dimensional conformal
  quantum-field theory and n=2 extended supersymmetry}.
\newblock \emph{\bibinfo{journal}{Sov. Phys. JETP}}
  \textbf{\bibinfo{volume}{63}}, \bibinfo{pages}{913} (\bibinfo{year}{1985}).

\end{thebibliography}

\end{document}